\documentclass[11pt]{article}
\usepackage{setspace}
\usepackage{graphicx}
\usepackage{lscape}
\usepackage{pdflscape}
\usepackage{afterpage}
\usepackage{pdfpages}
\usepackage{float}
\usepackage{booktabs}
\usepackage[utf8]{inputenc}
\usepackage{indentfirst}
\usepackage{arydshln}
\usepackage{arydshln}
\usepackage{tikz}
\usepackage{lipsum}
\usepackage{pgffor}
\usepackage{dsfont}
\usepackage{amsfonts}
\usepackage{amsmath}
\allowdisplaybreaks
\usepackage{mathtools}
\newcommand\independent{\protect\mathpalette{\protect\independenT}{\perp}}
\def\independenT#1#2{\mathrel{\rlap{$#1#2$}\mkern2mu{#1#2}}}

\usepackage{xfrac}
\usepackage{amssymb}
\usepackage{algorithm}
\usepackage{bigints}
\usepackage{graphicx}
\usepackage{url}				
\usepackage{caption}
\usepackage{subcaption} 
\usepackage{enumitem}
\usepackage{relsize,exscale}
\usepackage{multirow}
\usepackage{natbib}
\setcitestyle{authoryear, open={(},close={)}}
\usepackage[flushleft]{threeparttable}

\usepackage{datetime}

\newdateformat{monthyeardate}{%
	\monthname[\THEMONTH] \THEYEAR}

\usepackage{titlesec}
\titleformat*{\section}{\large\bfseries}
\titleformat*{\subsection}{\large\bfseries}


\newcounter{parentnumber}
\usepackage{theorem}

\newtheorem{assumption}{Assumption}

\newtheorem{proposition}{Proposition}

\newenvironment{proof}[1][Proof]{\noindent \textbf{#1.} }{\  \rule{0.5em}{0.5em}}

\newtheorem{npassumption}{Assumption}

\newtheorem{spassumption}{Assumption}

\newcommand{\ffrac}[2]{\ensuremath{\frac{\displaystyle #1}{\displaystyle #2}}}

\providecommand{\U}[1]{\protect\rule{.1in}{.1in}}

\usepackage{hyperref} 
\hypersetup{
	colorlinks=true, breaklinks=true, bookmarks=true,bookmarksnumbered,
	urlcolor=blue, linkcolor=blue, citecolor=blue, 
	pdftitle={}, 
	pdfauthor={\textcopyright}, 
	pdfsubject={}, 
	pdfkeywords={}, 
	pdfcreator={pdfLaTeX}, 
	pdfproducer={LaTeX with hyperref and ClassicThesis} 
}

\begin{document}
	\setstretch{1}
	\title{{\LARGE Nonlinear Treatment Effects in Shift-Share Designs\thanks{We thank Arthur Botinha, Caio Castro, Giovanni di Pietra, Rafael Dix-Carneiro, Leon Eliezer, Bruno Ferman, Nícolas Goulart, Toru Kitagawa, Lucas Mariano, Emanuel Ornelas, João Paulo Pessoa, Pedro Ogeda, Pedro Picchetti, Pedro Sant'Anna, and Otávio Tecchio for insightful thoughts. We also thank seminar participants at the Sao Paulo School of Economics - FGV for their valuable comments.}}}
	
	\author{
		Luigi Garzon\thanks{Sao Paulo School of Economics - FGV. Email: \href{mailto:luigi.garzon@fgv.br}{luigi.garzon@fgv.br}}  \and Vitor Possebom\thanks{Sao Paulo School of Economics - FGV. Email: \href{mailto:vitor.possebom@fgv.br}{vitor.possebom@fgv.br}. This study was financed, in part, by the São Paulo Research Foundation (FAPESP), Brazil. Process Number \#2025/04857-0.}
	}
	\date{}
	
	\maketitle

	\newsavebox{\tablebox} \newlength{\tableboxwidth}
	

	\begin{center}
		
		First Draft: July 2025; This Draft: \monthyeardate\today
		
		
		%
		%
		\href{https://sites.google.com/site/vitorapossebom/working-papers}{Please click here for the most recent version}
		
		\

		\large{\textbf{Abstract}}
	\end{center}
	
	We analyze heterogenous, nonlinear treatment effects in shift-share designs with exogenous shares. We employ a triangular model and correct for treatment endogeneity using a control function. Our tools identify four target parameters. Two of them capture the observable heterogeneity of treatment effects, while one summarizes this heterogeneity in a single measure. The last parameter analyzes counterfactual, policy-relevant treatment assignment mechanisms. We propose flexible parametric estimators for these parameters and apply them to reevaluate the impact of Chinese imports on U.S. manufacturing employment. Our results highlight substantial treatment effect heterogeneity, which is not captured by commonly used shift-share tools.
	
	\
	
	
	\
	

	\newpage
	
	\doublespacing
	
	\section{Introduction}\label{SecIntro}
	
	Shift-share designs have become a widely used tool in many fields, including trade \citep{autor2013}, political economy \citep{Dippel2021,Campante2023}, development economics \citep{topalova2010}, and labor economics \citep{Acemoglu2020}. The empirical importance of this tool motivated many methodological articles to discuss its identification and inferential aspects \citep{adao2019,goldsmith2020,borusyak2021}. However, those papers rely on linearity or homogeneity assumptions, which may be potentially strong in many applications \citep{hahn2024}. For instance, in the China shock setting \citep{autor2013}, opening trade relations with China could have a large effect, whereas increasing import exposure when trade relations with China are already well-established may have little effect. Moreover, allowing for nonlinear heterogeneity in the context of two-stage least squares may lead to negative weighting problems, affecting the causal interpretation of the estimand \citep{heckman2006,blandhol2022,słoczyński2024,alvarez2024,hahn2024}.
	
	To overcome these two limitations, we propose a nonparametric model for shift-share designs. We adapt the triangular equation model proposed by \cite{imbens2009}, allowing the outcome to be a nonseparable structural function of the treatment and a (possibly infinitely dimensional) shock, and the treatment to be a nonseparable structural function of the instrument and a scalar idiosyncratic term. The instrument combines common shocks (shifts) and individual measures of exposure to those shocks (shares). We leverage variation in the exogenous shares to construct a variable that controls for the endogenous part of the outcome shock.
	
	By doing so, we identify both Average and Policy Effects. For Average Effects, we nonparametrically identify the Local Average Response \citep{altonji2005}, the Average Derivative \citep{imbens2009}, and the Average Structural Function \citep{blundell2003}. First, the Local Average Response (LAR) function is the effect of the treatment on the outcome for a given value of the treatment, capturing the observed heterogeneity of our treatment effects. Second, the Average Derivative (AD) summarizes this heterogeneity into a single parameter that does not suffer from negative weighting issues. Third, the Average Structural Function (ASF) represents the average outcome at a fixed treatment value, capturing the observed heterogeneity in the levels of the outcome variable. Lastly, the Policy Effect (PE) captures the effect of a given treatment policy on the outcome variable.\footnote{An important limitation of shift-share designs, which also applies to our framework, is that they generally do not allow the identification of in-level effects, only in-changes. This drawback stems from the nature of the identifying variation, which relies on differential changes in exposure to common shocks.}
	
	The Policy Effect is, potentially, a very interesting parameter to explore in Shift-Share applications. For example, in the China shock application, we may be interested in understanding how a new tariff policy that reduces imports from China affects the U.S. labor market. Interestingly, this parameter enables us to study counterfactual policies that have not yet been implemented.
	
	In addition to identifying different parameters of interest, we provide a brief discussion of the two-stage least squares estimand (2SLS). By connecting this estimand to our nonlinear setting, we show that the 2SLS estimand may lack a causal interpretation without strong assumptions regarding the instrument-treatment relationship. Moreover, even if those assumptions are met, the 2SLS estimand captures a hard-to-interpret convex combination of local effects. 
	
	Since estimating those parameters nonparametrically is complicated by the large number of instruments (i.e., the size of the vector of shares), we also identify them using a semiparametric method. This semiparametric control function approach adapts the method proposed by \cite{chernozhukov2020} to the shift-share setting. It also provides a semiparametric estimator of the control function and flexibly parametric estimators of the target parameters. Moreover, we propose a uniform inference procedure using the weighted bootstrap.
	
	Lastly, we reevaluate the impact of increasing Chinese imports on the growth of manufacturing employment in the U.S. between 1990-2000 and 2000-2007. To do so, we use commuting zone data from \cite{autor2013} and flexibly estimate our four target parameters. We also compare them against estimates based on a linear 2SLS estimator, finding that allowing for nonlinear heterogeneous treatment effects is fundamental to understanding the China shock’s impact on the U.S. economy.
	
	Our Average Derivative estimates are small and do not reject the null hypothesis of zero average effects. In contrast, the 2SLS estimates are negative and significant. We find that these differences are explained by the negative weights within the 2SLS estimand in our dataset.
	
	Our estimates of the Local Average Response function find strong evidence of nonlinear effects between 2000 and 2007. In particular, for regions that faced lower exposure to growth in Chinese imports, a marginal increase in exposure to growth in Chinese imports leads to a greater intertemporal difference in manufacturing employment. In other words, for lower values of exposure to Chinese imports, employment in manufacturing decreases less than it would without the marginal increase in exposure.
	
	Our estimates of the Average Structural Function also find evidence of nonlinear effects between 2000 and 2007. They suggest that, for lower values of growth in exposure to Chinese imports, the reduction in the manufacturing employment rate was smaller than for those that faced higher values
	of growth in exposure to Chinese imports. However, in this case, the results of our nonlinear estimator are similar to the estimates based on linear 2SLS regressions.
	
	For the Policy Effect, we evaluate the effects of counterfactually increasing U.S. import tariffs during the analyzed period. To do so, we connect the exposure to Chinese imports with import tariffs by using the elasticity of substitution estimated by \cite{fajgelbaum2019}. We find that larger tariffs would not be sufficient to compensate for the loss of manufacturing employment caused by the increased exposure to Chinese imports. Importantly, the 2SLS estimates for the policy effect are larger and fall outside the confidence band of our nonlinear method. This result highlights how the linearity imposed by the 2SLS specification may lead researchers to overestimate the effect of tariffs. Consequently, allowing for nonlinear heterogeneous treatment effects is essential to our understanding of the impact of Chinese imports on the U.S. economy.
	
	\textbf{Related Literature.} This article contributes to three distinct strands of literature. Concerning its contribution to shift-share designs, there exists a growing literature on identification and estimation \citep{bartik1991,goldsmith2020,borusyak2021,dechaisemartin2022,hahn2024} and inference \citep{adao2019,alvarez2022}. Our contribution is focused on identification. While these articles clarify the identifying variation in shift-share designs, they rely on assumptions of linearity or homogeneity.
	
	We depart from this framework by allowing for heterogeneous and nonlinear effects in a nonparametric setting. By doing so, we broaden the applicability of shift-share designs to contexts where effect heterogeneity is empirically and theoretically plausible. Such an extension is methodologically relevant because \citet[Corollaries 3.1 and 3.2]{hahn2024} show that causally interpreting the 2SLS estimand in shift-share designs implies implausibly strong necessary conditions.
	
	Concerning its contribution to control function approaches, our work is inserted in the literature about identifying treatment effect parameters when the structural functions follow a triangular nonseparable model \citep{chesher2003,imbens2009,blundell2014}. We adapt the triangular model proposed by \cite{imbens2009} and the semiparametric tools proposed by \cite{chernozhukov2020} to the shift-share setting. These strategies complement existing work on control function models, providing a practical solution for applied researchers using shift-share instruments.
	
	Lastly, we speak directly to the empirical literature on the China shock and its labor market consequences \citep{autor2013,costa2016}. Our paper revisits the influential work of \cite{autor2013} through a nonparametric lens and shows that the standard 2SLS estimates may obscure meaningful heterogeneity in the impact of Chinese import exposure. By doing so, we contribute to a more nuanced understanding of the consequences of trade shocks and the limits of protectionist policy responses.
	
	\textbf{Paper Organization.} This paper is organized as follows. Section \ref{SecFramework} describes the econometric framework, presents the target parameters, and discusses our identifying assumptions. Section \ref{identification} presents our identification results and analyzes the 2SLS estimand in a nonlinear setting. Section \ref{estimation} explains our estimation and inferential procedures. Section \ref{empirical} discusses our empirical results with data from \cite{autor2013}, while Section \ref{conclusion} concludes.

	\section{Econometric Framework}\label{SecFramework}
	In this section, we explain our econometric framework. Section \ref{model} starts by describing our model’s setting. Then, Section \ref{SecParameters} defines our target parameters. Lastly, Section \ref{assumptions} states the model’s assumptions. In all these sections, we use our empirical application as an example to provide intuition.
	
	\subsection{Model's Setting}\label{model}
	
	We analyze the following triangular system with nonparametric, nonseparable equations:
	\begin{align}
		Y & = h_2(X, D, \varepsilon)\label{2nd_stage}\\
		X & = h_1(Z, D, \eta)\label{1st_stage} \\
		Z & = h_0(S,W),\label{EqIV}
	\end{align}
	where $Y$ is an outcome variable (e.g., growth in the manufacturing employment rate in a commuting zone in the U.S.) with support in $\mathcal{Y} \subseteq \mathbb{R}$, $X$ is a continuous treatment variable (e.g., temporal change in exposure to Chinese imports in a commuting zone) with support in $\mathcal{X} \subseteq \mathbb{R}$, $D$ is a vector of covariates (e.g., commuting zone's demographic characteristics) with support in $\mathcal{D} \subseteq \mathbb{R}^{dim(D)}$, $Z$ is a shift-share variable (e.g., $Z \coloneqq S \cdot W$) with support in $\mathcal{Z} \subseteq \mathbb{R}$, $S$ is a vector of shifts (e.g., sectoral growth of Chinese imports in high-income countries that are not the U.S.) with support in $\mathcal{S} \subseteq \mathbb{R}^J$, and $W$ is a vector of shares (e.g., sectoral employment shares in a commuting zone) with support in $\mathcal{W} \subseteq \mathbb{R}^J$. The natural number $J$ is the dimension of the vector of shifts and the vector of shares.
	
	This model also has two latent variables. First, $\varepsilon$ is the outcome-equation error with a (possibly) infinite-dimensional support $\mathcal{E}_2$. In our empirical application, it can be interpreted as local policies that impact the growth of the manufacturing employment rate. Second, $\eta$ is a scalar unobserved variable in the treatment assignment equation with support $\mathcal{E}_1 \subseteq \mathbb{R}$. In our empirical application, it can be interpreted as local shocks in the demand for Chinese imports that are not captured by shocks in other high-income countries.
	
	Moreover, this model has three key functions. First, $h_2: \mathcal{X} \times \mathcal{D} \times \mathcal{E}_2 \rightarrow \mathcal{Y}$ is the outcome function. Second, $h_1: \mathcal{Z} \times \mathcal{D} \times \mathcal{E}_1 \rightarrow \mathcal{X}$ is the treatment assignment function.\footnote{Our model allows for simultaneity between X and Y under additional assumptions. \cite{blundell2014} discuss which structural assumptions are necessary to write a simultaneous model as a triangular model.} Third, $h_0: \mathcal{S} \times \mathcal{W} \rightarrow \mathcal{Z}$ is the shift-share function, which is chosen by the researcher. As we see further in this paper, there is no need for the researcher to specify a function $h_0$, since she will only need the vector $W$ as the instrument.
	
	Although the structural equations above are presented in levels for generality and clarity, these variables are typically expressed in first differences in most empirical applications of shift-share designs. Our framework is flexible enough to accommodate this specification. In particular, when the empirical setting identifies causal effects from changes in exposure to aggregate shocks—rather than from levels—our model can be reinterpreted with $Y$, $X$, and $Z$ representing temporal changes rather than levels. For instance, in \cite{autor2013}, the identifying variation arises from differential trends across commuting zones, rather than static levels of exposure. For this reason, the outcome variable is the ten-year change in manufacturing employment, the treatment variable is the ten-year change in exposure to Chinese imports, and the structural functions $h_1$ and $h_2$ are defined directly for differenced variables. Consequently, we can only identify the effect of differential trade shocks on the growth rate of employment in each commuting zone. We cannot identify the effect of trade on employment levels.
	
	\subsection{Target Parameters}\label{SecParameters}
	
	In this section, we define our four parameters of interest. All objects are defined conditioning in $S = s$. In our empirical application, we treat each time period as a separate dataset. Consequently, it is as if we observed a single draw of the distribution of $S$, implying that conditioning on $S = s$ is basically conditioning on the available population.
	
	The first target parameter is the Local Average Response (LAR) function, studied in \cite{altonji2005}. It is defined as
	\begin{equation}\label{betax}
		\beta(x) := \mathbb{E}\left[\left.\frac{\partial h_2(X,D,\varepsilon)}{\partial x}\,\, \right\vert\, X = x, S = s \right] \text{ for } x \in \mathcal{X}.
	\end{equation}
	It summarizes the marginal effect of $x$ on $Y$ over the population of $D$ and $\varepsilon$ for a given value of $X = x$. It captures the observable heterogeneity from the model and can be interpreted as a generalization of the Conditional Average Treatment Effect (CATE). In our empirical application, it captures the effect of marginally increasing the change in exposure to Chinese imports on the temporal change in the manufacturing employment rate. 
	
	The second target parameter is a single summary measure of the marginal effect of $x$ on $Y$: the Average Derivative (AD). It is studied by \cite{imbens2009} and is defined as
	\begin{equation}\label{beta}
		\beta := \mathbb{E}\left[\left.\frac{\partial h_2(X,D,\varepsilon)}{\partial x}\,\,\right\vert \, S = s\right].
	\end{equation}
	The Average Derivative integrates the Local Average Response function over the population of $X$ and summarizes the observable heterogeneity into a single object.\footnote{In Section \ref{2slsrepresentation}, we will relate the LAR and the AD to the 2SLS estimand.} It can be interpreted as a generalization of the Average Treatment Effect (ATE). In our empirical application, it captures the average effect of marginally increasing the change in exposure to Chinese imports on the temporal change in the manufacturing employment rate. 
	
	The third target parameter is the Average Structural Function (ASF), studied by \cite{blundell2003}. It is defined as
	\begin{equation}\label{asf}
		\mu(x) := \mathbb{E}[h_2(x,D,\varepsilon) \mid S = s]  \text{ for } x \in \mathcal{X}.
	\end{equation}
	Similar to the Local Average Response function, this parameter captures the observable heterogeneity of the model, but at the level of the outcome variable. In our empirical application, it captures how the percentage point change in manufacturing employment rate differs, on average, for different values of change in the Chinese import exposure. In other words, the Average Structural Function describes the average intertemporal change in manufacturing employment rate for a commuting zone that faced a growth in Chinese import exposure of $X = x$.
	
	Our fourth target parameter is the Policy Effect, studied in \cite{imbens2009}. It is defined as
	\begin{equation}\label{policyeffect}
		\gamma := \mathbb{E}[h_2(\ell(X),D,\varepsilon) - Y \mid S = s],
	\end{equation}
	where $\ell:\mathcal{X} \rightarrow \mathcal{X}$ is policy function chosen by the researcher. This parameter captures the average effect of introducing policy $l$ on the outcome $Y$. In our empirical application, the researcher could be interested in the effects of a policy $\ell$ that imposes an upper bound on exposure to Chinese imports, $X$. It could be through an import restriction on some specific sector (e.g., a policy that bans imports of cars from China) or aggregated in terms of exposure to all Chinese imports. This parameter is interesting, as it allows the researcher to capture the causal effects of policies that have never been introduced in real life. In Section \ref{result:pe}, we provide a detailed discuss about this parameter.
	
	Lastly, note that, when the model is implemented using first-differenced variables (as is common in shift-share applications), the target parameters should be interpreted as marginal or average effects in changes, rather than in levels. For example, in our empirical application, the LAR, AD, ASF, and Policy Effect capture how increases in import exposure affect the decline in manufacturing employment over time, rather than the level of employment itself. This distinction is crucial for interpreting the results appropriately.
	
	\subsection{Assumptions}\label{assumptions}
	
	To identify the parameters described in the last section, we impose five assumptions. The first two assumptions allow us to identify a control function that will be used to identify all target parameters. Then, when we impose our third assumption, we can identify the Average Structural Function. Lastly, the addition of the fourth assumption allows us to identity the Local Average Response and the Average Derivative, while the addition of the fifth assumption allows us to identify policy effects.
	
	Our primary assumption imposes the exogeneity of the vector of shares. It is closely connected to the identification assumption used by \cite{goldsmith2020} and \citet[Proposition 2.1]{hahn2024}. Formally, it imposes the following restriction on our data-generating process. 
	\begin{assumption}[Exogenous Shares]\label{exogeneity}
		$(\varepsilon, \eta) \independent W \mid S, D$
	\end{assumption}
	Assumption \ref{exogeneity} says that the vector of shares, $W$, is independent of the treatment-assignment and outcome-equation errors, $\eta$ and $\varepsilon$, given the vector of shifts, $S$, and the vector of covariates, $D$.\footnote{An alternative identification strategy would impose exogenous shifts as done by \cite{borusyak2021}. Challengingly, the shifts are the same for every region, implying that we cannot find exogenous variation at the regional level in a nonparametric setting. To circumvent this issue, \cite{borusyak2021} rely on linearity restrictions to derive an equivalence result between a region-level model and an industry-level model. Using the latter model, a researcher can explore ``shift'' variation across industries to identify the linear effect of interest. However, such an equivalence result is not trivial in a nonlinear setting such as ours. For this reason, shift-share designs with exogenous shifts are outside the scope of this paper.} In our empirical application, it imposes that employment in manufacturing would have trended similarly for regions that were more vs. less exposed to a possible shock in the previous period if there were no changes in exposure to Chinese imports.\footnote{\cite{goldsmith2020} and \cite{hahn2024} find that, when combined in a Bartik instrument, share variation does not seem to be exogenous in our China shock application. However, both groups of authors argue that it is still possible to analyze this empirical setting by directly using the shares as instruments without combining them into a Bartik instrument. Importantly, our identifying assumption adopts this approach of directly using the shares as instruments.} As noted by \cite{goldsmith2020}, the Exogenous Shares assumption could be interpreted as a set of parallel trend conditions when the outcome is measured in changes.
	
	Our second assumption imposes monotonicity of the treatment-assignment error, $\eta$.
	\begin{assumption}[Monotonicity]\label{increasing}
		The random variable $\eta$ is continuously distributed given $D$ and $S$, and the function $h_1$ is strictly monotonic in $\eta$.
	\end{assumption}
	Assumption \ref{increasing} is a generalization of the common IV monotonicity assumption \citep{imbens2009}. In our setting, it requires that the function $h_1$ is strictly increasing or strictly decreasing in the unobserved variable of the treatment assignment equation, $\eta$. Combined with the fact that this treatment-assignment error is a scalar, it allows us to invert the function $h_1$ with respect to $\eta$. Similarly to the work of \cite{imbens2009}, this step is essential to identifying the control function in our setting.
	
	Our third assumption is necessary to connect our model with the shift-share structure present in our empirical application and restricts the number of observed shocks. 
	\begin{assumption}[Common Shocks]\label{common}
		We observe a unique draw $\Tilde{s}$ from the distribution of $S$. 
	\end{assumption}
	Assumption \ref{common} says that the vector of shifts $S$ is common across the entire population. Consequently, conditioning on $S$ is equivalent to condition on the observed population. To the best of our knowledge, all shift-share applications assume that there is only one common vector of shifts. For example, in our empirical application, the shift is a vector of changes in imports from China to high-income countries. Each entry of the vector corresponds to an industry sector, but the vector is common to all regions.
	
	To state our next two assumptions, we need to define the following variable:
	\begin{equation}
		V:= F_{X \mid W,S,D}(X \mid W,S,D).
	\end{equation}
	The random variable $V$ is based entirely on observable variables and, later, works as our control function. This result is shown in Proposition \ref{controlfunction}.
	
	Our fourth assumption imposes three regularity conditions and is connected to the assumptions in Theorem 6 by \cite{imbens2009}.
	\begin{npassumption}[Regularity Conditions]\label{rcond}
		\phantom{a}
		\begin{enumerate}
			\item\label{Cond1} $h_2$ is continuously differentiable in $x$.
			
			\item\label{Cond2} $X$ has a continuous distribution given $V$, $S$ and $D$.
			
			\item\label{Cond3} For all $x \in \mathcal{X}$, and for some $\Delta > 0$, there exists $$\mathbb{E}\left[\int \sup_{\| x- X \| \leq \Delta} \left\|\frac{\partial h_2(x,d,e)}{\partial x}\right\| dF_{\varepsilon \mid V, S, D}(e \mid V, S, D) \right].$$
		\end{enumerate}
	\end{npassumption}
	Assumption \ref{rcond} is necessary to identify the Local Average Response (LAR) function and the Average Derivative (AD). Condition \ref{rcond}.\ref{Cond1} requires that the outcome function, $h_2$, to be continuously differentiable in the treatment variable, $x$, since this derivative appears in the definitions of the LAR function and the AD. Condition \ref{rcond}.\ref{Cond2} allows us to identify the LAR function for the entire support of $X$ and, then, integrate the LAR function over the distribution of $X$ to identify the Average Derivative. Finally, Condition \ref{rcond}.\ref{Cond3} is the weakest possible restriction that allows us to change the order of the derivative and the integral.
	
	Lastly, our fifth assumption is a common support assumption.
	\begin{npassumption}[Common Support]\label{csupp}
		$\text{supp}(\ell(X),D,V) \subseteq \text{supp}(X,D,V)$
	\end{npassumption}
	Assumption \ref{csupp} imposes that the support of the treatment variable after the policy $\ell$ is imposed is contained in the observed support of the treatment variable. When combined with Assumptions \ref{exogeneity}-\ref{common}, Assumption \ref{csupp} is sufficient to identify the Policy Effect.
	
	\section{Identification Results}\label{identification}
	In this section, we present the identification results for the target parameters listed in the previous section. Section \ref{npidentification} provides nonparametric identification results, while Section \ref{2slsrepresentation} relates our target parameters to the 2SLS estimand. Lastly, \ref{spidentification} presents semiparametric identification results that connect directly with our proposed estimation and inference procedures in Section \ref{estimation}.
	
	\subsection{Nonparametric Identification}\label{npidentification}
	
	Similarly to \cite{imbens2009}, we adopt the control function approach to identification and estimation. We begin by identifying the control function variable. Then, we identify the average structural function, the local average response function, and the average derivative. Lastly, we identify the policy effect.
	
	Our first proposition identifies our control function variable.
	\begin{proposition}[Control Function]\label{controlfunction}
		Under Assumptions \ref{exogeneity}-\ref{increasing}, we have that $$
		V:= F_{X\mid W,S,D}(X \mid W,S,D) = F_{\eta \mid S,D}(\eta \mid S,D),$$ where $V$ has a uniform distribution. Moreover, $ X \independent \varepsilon \mid V, S, D$.
	\end{proposition}
	\textit{Proof.} See Appendix \ref{cfproof}.
	
	The intuition behind Proposition \ref{controlfunction} is that we want to clean out the unobserved endogenous variation of $X$ that is driven by $\eta$. To do that, we construct a proxy for $\eta$. When controlling for this proxy and the other observed variables ($D$ and $S$), we isolate the exogenous variation in $X$ that is driven by $W$. Consequently, we can identify its effects on $Y$. This reasoning is formalized by the last statement in Proposition \ref{controlfunction}.
	
	Before identifying our target parameters, we define the following function:
	\begin{equation}\label{mfunction}
		\begin{split}
			m(x,d,v) &:= \mathbb{E}[Y \mid X = x, D = d, V = v, S = \Tilde{s}]\\
			& = \int_{\mathcal{E}_2} h_2(x,d,e)\, dF_{\varepsilon \mid X,D,V,S}(e\mid x,d,v,\Tilde{s})\\
			& = \int_{\mathcal{E}_2} h_2(x,d,e)\, dF_{\varepsilon \mid D,V,S}(e\mid d,v,\Tilde{s}),
		\end{split}
	\end{equation}
	where the second equality follows from Equation \eqref{2nd_stage} and the third equality follows from Proposition \ref{controlfunction}. Defining the function $m(x,d,v)$ simplifies our notation significantly because it appears in most of our identification results. Note that the function $m(x,d,v)$ is defined using observable variables only.
	
	Our second proposition identifies the Average Structural Function (Equation \eqref{asf}).
	\begin{proposition}[Average Structural Function]\label{asfidentification}
		Under Assumptions \ref{exogeneity}-\ref{common}, for any $x \in \mathcal{X}$, the Average Structural Function is given by $$
		\mu(x) = \mathbb{E}[m(x,D,V) \mid  S = \Tilde{s}].$$
	\end{proposition}
	\textit{Proof.} See Appendix \ref{asfproof}.
	
	Proposition \ref{asfidentification} states that the conditional expectation of the structural function $h_2$ evaluated at point $x \in \mathcal{X}$ (i.e., $\mu(x)$)  is captured by the conditional expectation of the function $m$, defined in Equation \eqref{mfunction}. This result is closely related to \citet[Equation (2.47)]{blundell2003}. However, we impose a slightly different exogeneity assumption.\footnote{In \cite{blundell2003}, they use the conditional independence assumption $\varepsilon\mid X,Z \sim \varepsilon\mid X, \eta$.}
	
	Our third proposition identifies the Local Average Response function (Equation \eqref{betax}) and the Average Derivative (Equation \eqref{beta}).
	\begin{proposition}[Average Effects]\label{propAE}
		Under Assumptions \ref{exogeneity}-\ref{common} and \ref{rcond}, the Local Average Response function and the Average Derivative are given by $$\beta(x) = \mathbb{E}\left[\left. \frac{\partial m(X,D,V)}{\partial x} \right\vert X = x, S = \Tilde{s} \right] \quad \text{and} \quad \beta = \mathbb{E}\left[\left. \frac{\partial m(X,D,V)}{\partial x} \right\vert S = \Tilde{s}\right]$$ respectively.
	\end{proposition}
	\textit{Proof.} See Appendix \ref{aeproof}.
	
	The results in Proposition \ref{propAE} state that the conditional expectation of the derivative of the structural function $h_2$ in $x$ is captured by the conditional expectation of the derivative of the function $m$ in $x$.
	
	Our fourth proposition identifies the Policy Effect (Equation \eqref{policyeffect}).
	\begin{proposition}[Policy Effect]\label{propPE}
		Under Assumptions \ref{exogeneity}-\ref{common} and \ref{csupp}, the Policy Effect is identified by  $$\gamma = \mathbb{E}[m(\ell(X), D, V) \mid S = \Tilde{s} ] - \mathbb{E}[Y \mid S = \Tilde{s}].$$
	\end{proposition}
	\textit{Proof.} See Appendix \ref{peproof}.
	
	Proposition \ref{propPE} says that the conditional expectation of the structural function $h_2$ when applying the policy transformation $\ell$ in the random variable $X$ is captured by the conditional expectation of the function $m$ when applying the same policy transformation.
	
	Importantly, the expectations in Proposition \ref{propPE} are properly defined only when Assumption \ref{csupp} holds. This common support assumption is potentially a strong restriction, limiting our choice of policy functions in a fully nonparametric setting. To avoid this type of restriction in our empirical application, we use semiparametric assumptions as explained in Section \ref{spidentification}.
	
	\subsection{Two-Stage Least Squares Representation}\label{2slsrepresentation}
	
	In this section, we discuss the causal interpretation of the Two-Stage Least Squares (2SLS) estimand in our model and compare this estimand against our target parameters. For brevity, we omit the extra covariates $D$ from the model. In this case, the 2SLS estimand is given by
	\begin{equation}
		\beta^{\text{2SLS}} := \frac{\mathrm{Cov}(Y,Z \mid S = \Tilde{s})}{\mathrm{Cov}(X,Z\mid S = \Tilde{s})}
	\end{equation}
	
	Our fifth proposition connects the 2SLS estimand with the structural functions in Equations \eqref{2nd_stage} and \eqref{1st_stage}.
	\begin{proposition}\label{thm2sls}
		Under Assumptions \ref{exogeneity} and \ref{common}, we have that $$\beta^{\text{2SLS}} = \mathbb{E}\left[\left.\int_{\mathcal{Z}} \frac{\partial h_2(h_1(\omega,\eta),\varepsilon)}{\partial x} \lambda(\omega,\eta) d\omega\,\, \right \vert S = \Tilde{s}\right],$$ where $$\lambda(z,\eta) := \ffrac{\frac{\partial h_1(z,\eta)}{\partial z} \int_{z_0}^z(\zeta - \mathbb{E}[Z]) f_{Z\mid S}(\zeta) d\zeta}{\mathbb{E}\left[\left.\int_{\mathcal{Z}}  \frac{\partial h_1(\omega,\eta)}{\partial z} \int_{z_0}^\omega(\zeta - \mathbb{E}[Z]) f_{Z\mid S}(\zeta) d\zeta d\omega\,\,\right \vert S = \Tilde{s}\right]}$$ and $z_0 := \mathrm{inf}(\mathcal{Z})$.
	\end{proposition}
	\textit{Proof.} See Appendix \ref{2slsproof}.
	
	Proposition \ref{thm2sls} states that the 2SLS estimand identifies an average of the derivative of the outcome function (Equation \eqref{2nd_stage}) weighted by $\lambda(z,\eta)$.\footnote{This result adapts the result in \citet[Theorem 4]{angrist2000} for the case of a triangular model. It is also related to results derived by \citet[Proposition 1]{adao2019}, \citet[Proposition A1]{borusyak2021}, \citet[Theorem 3]{dechaisemartin2022}, and \citet[Proposition 3.1]{hahn2024}. Most of these authors analyze how to interpret the 2SLS in a linear, heterogeneous shift-share model, while we use a non-linear shift-share model. Importantly, \citet[Proposition A1]{borusyak2021} analyze a partially linear model, but they impose that our function $h_{0}$ has an inner product structure while we left it unrestricted (Equation \eqref{EqIV}).} This estimand identifies a convex combination of causal effects when $\lambda(z,\eta) \geq 0 $ for all $(z,\eta) \in \mathcal{Z}\times \mathcal{E}_1$. These weights are proportional to the first-stage effect, $\partial h_1(z,\eta)/\partial z$, and they will be nonnegative if, for all $(z,\eta) \in \mathcal{Z}\times \mathcal{E}_1$, either $\partial h_1(z,\eta)/\partial z \geq 0$ or $\partial h_1(z,\eta)/\partial z \leq 0$.\footnote{We identify the first-stage effect in Appendix \ref{fseffect}.}\textsuperscript{,}\footnote{\citet[Corollaries 3.1 and 3.2]{hahn2024} derive necessary conditions for the 2SLS estimand to be a positively weighted average of causal effects in a linear heterogeneous treatment effects model under either the exogenous shares assumption or the exogenous shifts assumption. They argue that these necessary conditions are implausible in many empirical contexts. Consequently, analyzing nonlinear heterogeneous models like ours is methodologically relevant.} If the first-stage effect function has different signs for a positive mass of points in the support of $Z$ and $\eta$, then the 2SLS estimand faces negative weighting problems and is not weakly causal according to \cite{blandhol2022}.
	
	Even without negative weighting problems, the 2SLS estimand lacks a straightforward causal interpretation. Note that, instead of using the distribution of observable and unobservable variables like the LAR function and the AD parameter (Equations \eqref{betax} and \eqref{beta}), the 2SLS estimand places more weight on points where $\partial h_1(z,\eta)/\partial z$ is greater. Moreover, the weights used by the 2SLS estimand are not connected with policies motivated by economic theory, such as our policy effect (Equation \eqref{policyeffect}).
	
	
	\subsection{Semiparametric Identification}\label{spidentification}
	
	Although the nonparametric identification results are valid for a wide class of structural functions, they have two main drawbacks. First, when connecting them to nonparametric estimators, we must estimate a complex function with many covariates (e.g., there are 397 industry sectors in our empirical application). Consequently, these estimators would suffer greatly from the curse of dimensionality. Second, the common support assumption significantly limits our choice of policy functions. To avoid these issues, this section describes a semiparametric identification strategy based on the methods proposed by \cite{chernozhukov2020}.
	
	Before stating the required assumptions for semiparametric identification, we must introduce some notation. Let $q_A$ be a vector of transformations, such as powers, splines, and interactions, referring to a random variable $A$.\footnote{\cite{Chen2007} provides a detailed review of sieve estimators, explaining power, spline and other series that may be used in our vector of transformations.} Now, define $$K_1(W,D) := q_W(W) \otimes q_D(D) \qquad \text{and} \qquad K_2(X,D,V):= q_X(X) \otimes q_D(D) \otimes q_V(V)$$ where $\otimes$ denotes the Kronecker product. Moreover, let $Q_{A}\left(\left. \tau \right\vert B\right)$ denote the $\tau$-th quantile of variable $A$ conditional on variable $B$.
	
	Our first semiparametric assumption restricts the functional form of our structural functions (Equations \eqref{2nd_stage} and \eqref{1st_stage}).
	\begin{spassumption}[Quantile Regression Baseline]\label{qrmodel}
		The treatment variable, $X$, conditional on  $W, D$ and $S = \Tilde{s}$ follows a Quantile Regression (QR) model, $$X = Q_X(V \mid W,D,S = \Tilde{s}) = K_1(W,D)^\prime \pi_1(V)$$ and the outcome variable, $Y$, conditional on $X, D, V$ and $S = \Tilde{s}$ follows its own Quantile Regression (QR) model $$Y = Q_Y(U \mid X, D, V, S = \Tilde{s}) = K_2(X,D,V)^\prime \pi_2 (U).$$    
	\end{spassumption}
	Assumption \ref{qrmodel} imposes that our structural functions follow a semiparametric quantile regression model.\footnote{We chose a Quantile Regression model, but one could opt for other semiparametric models. For example, \cite{chernozhukov2020} also derives identification results for Distribution Regression models.} According to \citet[p. 509]{chernozhukov2020}, the quantile regression model is valid when the structural functions (Equations \eqref{2nd_stage} and \eqref{1st_stage}) follow a restricted random coefficient model or a heteroskedastic normal system of equations.
	
	Our second semiparametric assumption is a rank condition.
	\begin{spassumption}[Positive Definiteness]\label{nonsing}
		The matrix $\mathbb{E}[K_2(X,D,V)K_2(X,D,V)^\prime]$ is nonsingular.
	\end{spassumption}
	Assumption \ref{nonsing} guarantees that the vector $\pi_2(U)$ is unique, as shown by \cite{chernozhukov2020}.
	
	Now, we briefly discuss our semiparametric identification strategy. Note that, under Assumptions \ref{qrmodel} and \ref{nonsing}, Equation \eqref{mfunction} implies that
	\begin{equation}\label{EqSemiParFuncM}
		m(x,d,v) = K_2(x,d,v)^\prime \pi_2
	\end{equation}
	where $\pi_{2} \coloneqq \int_{0}^{1} \pi_{2}\left(u\right) \, du$ according to \citet[Equation (2.6)]{chernozhukov2020}. This result, when combined with Propositions \ref{asfidentification} and \ref{propPE}, implies that we can semiparametrically identify the Average Structural Function and the Policy Effect as
	\begin{equation}
		\begin{split}
			\mu(x) & = \mathbb{E}[K_2(x,D,V)^\prime \pi_2(U) \mid  S = \Tilde{s}]\\
			\gamma & = \mathbb{E}[K_2(\ell(X), D, V)^\prime \pi_2(U) \mid S = \Tilde{s} ] - \mathbb{E}[Y \mid S = \Tilde{s}]
		\end{split}
	\end{equation}
	when we add Assumptions \ref{exogeneity}-\ref{common} only.  Importantly, we do not need the common support assumption to semiparametrically identify the Policy Effect because the Quantile Regression restrictions allow us to extrapolate outside the support of the treatment variable.
	
	Lastly, we combine Equation \eqref{EqSemiParFuncM} with Proposition \ref{propAE} to semiparametrically identify the Local Average Response function and the Average Derivative as
	\begin{equation}\label{EqSemiParBetas}
		\begin{split}
			\beta(x) & = \mathbb{E}\left[\left(\frac{\partial K_2(X,D,V)}{\partial x}\right)^\prime \pi_2(U) \mid X = x, S = \Tilde{s} \right]\\
			\beta & = \mathbb{E}\left[\left(\frac{\partial K_2(X,D,V)}{\partial x}\right)^\prime \pi_2(U) \mid S = \Tilde{s}\right]
		\end{split}
	\end{equation}
	when we add Assumption \ref{rcond}.
	
	Note that Equations \eqref{EqSemiParFuncM}-\eqref{EqSemiParBetas} are key results to understand the estimation method proposed in Section \ref{estimation}.

	\section{Estimation and Inference}\label{estimation}
	
	In this section, we propose a three-step estimation process to estimate the parameters of interest. Section \ref{1stage} uses a semiparametric estimator to estimate the control function, $V$, based on \cite{chernozhukov2020}. Section \ref{2stage} proposes a flexibly parametric procedure to estimate the function $m(x,d,v)$ and its derivative, while Section \ref{3stage} uses the objects estimated in Section \ref{2stage} to estimate the parameters of interest. Lastly, Section \ref{SecAlgorithm} describes a simple-to-implement estimation algorithm with a bootstrap inference procedure.
	
	Below, we assume that we observe a sample $\left\lbrace Y_{i}, X_{i}, W_{i}, D_{i}\right\rbrace_{i = 1}^{N}$ of size $N \in \mathbb{N}$. Moreover, our sample is exposed to a single common shift shock $S_{i} = \Tilde{s}$ for all $i \in \left\lbrace 1, \ldots, N \right\rbrace$. Consequently, conditioning on the vector of shifts is equivalent to conditioning on our dataset.
	
	\subsection{Semiparametric First Stage Estimation} \label{1stage}
	
	Our first step is to estimate the values of the control function, $V_i = F_{X \mid W,D,S}(X_i \mid W_i, D_i, \Tilde{s})$, for $i \in \{1,\dots,N\}$. Following in \cite{chernozhukov2020}, we estimate this distribution in a trimmed support $\overline{\mathcal{X}}$ to avoid far tails. We use bars to denote trimmed supports with respect to $X$, such as $\overline{\mathcal{X}\mathcal{W}\mathcal{D}} := \{(x,w,d) \in \mathcal{X}\times \mathcal{W}\times \mathcal{D}: x \in \overline{\mathcal{X}}\}$. Moreover, we denote the usual check function by $\rho_v(a) = (v - \mathds{1}\{a < 0\})\cdot a$.
	
	Now, we can estimate the first stage as
	\begin{equation}\label{cfestimamtion}
		\hat{F}_{X\mid W,D,S}(x \mid w,d, \Tilde{s}) = \epsilon + \int^{1-\epsilon}_\epsilon \mathds{1}\{K_1(w,d)^\prime \hat{\pi}_1(v) \leq x\}dv, \qquad (x,w,d) \in \overline{\mathcal{X}\mathcal{W}\mathcal{D}}
	\end{equation}
	\begin{equation}\label{cfpi}
		\hat{\pi}_1(v) \in \mathrm{arg\,min}_{\pi_1} \sum_{i=1}^N \rho_v(X_i - K_1(W_i, D_i)^\prime \pi_1 )
	\end{equation}
	for a small constant $\epsilon > 0$. \cite{chernozhukov2020} adjusts the limits of the integral in Equation \eqref{cfestimamtion} to avoid tail estimation of quantiles.
	
	Given Equation \eqref{cfestimamtion}, we estimate the control function variable as $$\hat{V}_i = \hat{F}_{X\mid W,D,S}(X_i \mid W_i, D_i, \Tilde{s}).$$
	
	\subsection{Parametric Second Stage Estimation} \label{2stage}
	
	Here, we provide a flexibly parametric procedure to estimate the function $m(x,d,v)$ (Equation \eqref{EqSemiParFuncM}) and its derivative with respect to $x$. To estimate the function $m(x,d,v)$, we perform a OLS regression of $Y_i$ on $K_2(X_i,D_i,\hat{V}_i)$. From this OLS regression, we have that
	\begin{equation}\label{EqMfunctionOLS}
		\hat{m}(x,d,v) = K_2(x,d,v)^\prime\hat{\pi}_2,
	\end{equation}
	where $\hat{\pi}_2$ is the vector of estimated parameters from the OLS regression.
	
	Treating the dimension of $K_2(x,d,v)$ as fixed, we know the functional form of $\hat{m}(x,d,v)$. Then, we can take the derivative of $K_2(x,d,v)$ with respect to $x$ in order to estimate the derivative of the function $m(x,d,v)$. To simplify notation, let $m_x(x,d,v) := \partial m(x,d,v)/\partial x$. Therefore, the estimator for $m_x(x,d,v)$ is
	
	\begin{equation}\label{EqMderivativeOLS}
		\hat{m}_x(x,d,v) = \left(\frac{\partial K_2(x,d,v)}{\partial x}\right)^\prime \hat{\pi}_2.
	\end{equation}
	
	\subsection{Third Stage Estimation} \label{3stage}
	
	For the final parametric step, we use the estimated functions from Section \ref{2stage} to construct estimators for the target parameters.
	
	We start by constructing an estimator for the Average Structural Function (ASF):
	\begin{equation}\label{asfestimator}
		\hat{\mu}(x) = \frac{1}{N}\sum_{i=1}^N \hat{m}(x,D_i,\hat{V}_i).
	\end{equation}
	In Equation \eqref{asfestimator}, the ASF estimator is a function of $x$, integrating over observed values of $D_i$ and $\hat{V}_i$.
	
	Next, we construct an estimator for the Policy Effect of a given policy $\ell(\cdot)$: 
	\begin{equation}
		\hat{\gamma} = \frac{1}{N}\sum_{i=1}^N \left(\hat{m}(\ell(X_i),D_i,\hat{V}_i) - Y_i\right).
	\end{equation}
	
	Furthermore, using the estimator for the derivative of the function $m(x,d,v)$, we can construct an estimator for the Average Derivative (AD):
	\begin{equation}
		\hat{\beta} = \frac{1}{N}\sum_{i=1}^N \hat{m}_x(X_i,D_i,\hat{V}_i).
	\end{equation}
	
	Lastly, to estimate the Local Average Response (LAR) function, we perform an OLS regression of $\hat{m}_x(X_i,D_i,\hat{V}_i)$ on $q_X(X_i)$. Then, our estimator for the LAR is
	\begin{equation}\label{larestimator}
		\hat{\beta}(x) = q_X(x)^\prime \hat{\pi}_X,
	\end{equation}
	where $\hat{\pi}_X$ is the vector of estimated parameters from the OLS regression of $\hat{m}_x(X_i,D_i,\hat{V}_i)$ on $q_X(X_i)$.
	
	\subsection{Estimation Algorithm and Bootstrap Inference Procedure}\label{SecAlgorithm}
	
	Here, we provide an algorithm for the three-stage estimation procedure and an algorithm to perform uniform inference for the target parameters using the weighted bootstrap to estimate the standard errors.
	
	Algorithm \ref{algorithmest} provides the three-stage estimation procedure. The first stage is identical to the first stage of the Quantile Regression specification proposed by \cite{chernozhukov2020}. The later stages are based on procedures described in Sections \ref{2stage} and \ref{3stage}. In the empirical application of \cite{chernozhukov2020}, they find that their estimates are not sensitive to values of $M_1$ and $\epsilon$. Similarly, we also perform a robustness analysis with respect to those parameters in our empirical section.
	\begin{algorithm}
		\caption{Three-stage estimation procedure} 
		\label{algorithmest}
		\onehalfspacing
		\textit{First Stage}. (Control Function estimation)
		
		\begin{enumerate}
			\item Choose $\epsilon \in (0, 0.5)$ and a fine mesh of $M_1$ values $\{\epsilon = v_1 < \dots < v_{M_1} = 1-\epsilon\}$.
			\item Estimate $\{\hat{\pi}_1(v_{m_1})\}_{m_1=1}^{M_1}$ by solving Equation \eqref{cfpi}.
			\item Estimate $\hat{V}_i = \hat{F}_{X\mid W,D,S}(X_i \mid W_i, D_i, \Tilde{s})$, for $i = 1, \dots, N$, as in Equation \eqref{cfestimamtion}.
		\end{enumerate}

		\textit{Second Stage}. (Function $m(x,d,v)$ and its derivative with respect to $x$)
		
		\begin{enumerate}
			\item Estimate $\hat{\pi}_2$ using the OLS regression of $Y_i$ on $K_2(X_i, D_i, \hat{V}_i)$, for $i = 1, \dots, N$.
			\item Take the derivative of $K_2(x,d,v)$ with respect to $x$.
			\item Construct the functions $\hat{m}(x,d,v)$ and $\hat{m}_x(x,d,v)$ according to Equations \eqref{EqMfunctionOLS} and \eqref{EqMderivativeOLS}.
		\end{enumerate}
		
		\textit{Third Stage}. (Estimation of parameters of interest)
		
		\begin{enumerate}
			\item Estimate $\hat{\pi}_X$ by the OLS regression of $\hat{m}_x(X_i,D_i,\hat{V}_i)$ on $q_X(X_i)$, for $i = 1, \dots, N$.
			\item Compute $\hat{\mu}(x)$, $\hat{\gamma}$, $\hat{\beta}$, and $\hat{\beta}(x)$ according to Equations \eqref{asfestimator}-\eqref{larestimator}.
		\end{enumerate}
	\end{algorithm}
	
	Next, we present the inference procedure in Algorithm \ref{inference}. This procedure is based on the uniform inference procedure proposed by \cite{chernozhukov2020}. We begin by performing a weighted bootstrap using the standard exponential distribution.\footnote{One could use any random variable satisfying $e \geq 0$, $\mathbb{E}[e] = 1$, $\mathrm{Var}(e) = 1$, and $\mathbb{E}\vert e \vert^{2+\delta} < \infty$ for some $\delta > 0$. See Assumption 3 in \cite{chernozhukov2020} for more details.} Next, we compute the standard errors using the interquartile range function. Then, for each bootstrap iteration, we compute the maximal $t$-statistics, so we can finally form $(1-\alpha)$-confidence bands in the last step.
	
	\begin{algorithm}[!hbt]
		\caption{Four-step inference procedure} 
		\label{inference}
		\onehalfspacing
		\textit{Step 1}. (Weighted Bootstrap)
		
		\begin{enumerate}
			\item Draw $e_b := \{e_{ib}\}_{i=1}^N$ i.i.d. from the standard exponential distribution, for $b = 1, \dots, B$.
			\item For $b = 1, \dots, B$, repeat Algorithm \ref{algorithmest} with the sample weighted by $e_b$.
		\end{enumerate}

		\textit{Step 2}. (Standard Errors)
		
		\begin{enumerate}
			\item For any estimator $\hat{\tau}(\cdot)$, given its $B$ bootstrap draws $\{\hat{\tau}_b(\cdot)\}_{b=1}^B$, compute its standard error as
			$$
			\hat{\sigma}_\tau(\cdot) = \mathrm{IQR}(\{\hat{\tau}_b(\cdot)\}_{b=1}^B)/1.349,
			$$ where IQR is the interquartile range function.
		\end{enumerate}
		
		\textit{Step 3}. (Maximal $t$-statistics)
		
		\begin{enumerate}
			\item For $b = 1,\dots, B$, compute the bootstrap draws of the maximal $t$-statistics for any estimator $\hat{\tau}(\cdot)$ as
			
			$$
			\Vert t_{\tau, b}(\cdot)\Vert = \mathrm{sup}\left\vert\frac{\hat{\tau}_b(\cdot) - \hat{\tau}(\cdot)}{\hat{\sigma}_\tau(\cdot)}\right\vert.
			$$
		\end{enumerate}
		
		\textit{Step 4}. (Confidence Interval)
		
		\begin{enumerate}
			\item Let $\hat{k}_\tau(1-\alpha)$ be the sample $(1-\alpha)$-quantile of $\{\Vert t_{\tau, b}(\cdot)\Vert: 1 \leq b \leq B\}$.
			\item Form $(1-\alpha)$-confidence bands for any estimator $\hat{\tau}(\cdot)$ as
			
			$$
			\{\hat{\tau}(\cdot) \pm \hat{k}_\tau(1-\alpha) \cdot \hat{\sigma}_\tau(\cdot)\}.
			$$
		\end{enumerate}
	\end{algorithm}

	\section{Effects of Chinese Imports on Manufacture Employment in the U.S.}\label{empirical}
	
	In this section, we estimate the effects of the time evolution of Chinese imports on the temporal change of manufacturing employment in the United States using data previously analyzed by \cite{autor2013}. In Section \ref{spec}, we provide the specification of our model.  Section \ref{result:ae} provides the results for both the Local Average Response (LAR) and the Average Derivative (AD), while Section \ref{result:asf} provides the results for the Average Structural Function (ASF). Lastly, Section \ref{result:pe} provides the results for the Policy Effect.
	
	When analyzing the same empirical application as ours, \cite{hahn2024} reject the null hypothesis of constant and linear effects in this setting. They find statistical evidence to reject this hypothesis using either the ``exogenous shares'' identification approach \citep{goldsmith2020} or the ``exogenous shift'' identification approach \citep{borusyak2021}. Their results highlight the importance of adopting a nonlinear heterogeneous treatment effect model such as the one we use in the following sections.
	
	Moreover, \cite{goldsmith2020} and \citet[p. 24]{hahn2024} argue that, when exploring share variation to analyze the China shock application, any researcher should directly use the shares as instruments without combining them into a Bartik instrument. Our identification strategy and estimation algorithm follow exactly this approach.
	
	\subsection{Specification}\label{spec}
	
	\cite{autor2013} studies the effects of the time evolution of Chinese imports on the temporal change of manufacturing employment in the United States. Our main specification relies on the specification in Column (6) in Table 3 of \cite{autor2013}, which includes the full set of covariates:
	\begin{align}
		Y_{it} & = h_2(X_{it},D_{it}, \varepsilon_{it}) \\
		X_{it} & = h_1(Z_{it}, D_{it}, \eta_{it}) = \sum_{j =1}^J W_{ijt} S_{jt}^{US}\\
		Z_{it} & = h_0(W_{it-1}, S_{t}^{HI}) = \sum_{j =1}^J W_{ijt-1} S_{jt}^{HI}
	\end{align}
	
	\noindent where $Y_{it}$ is the percentage point change in manufacturing employment rate for location $i$ and period $t$, $X_{it}$ is the change in  Chinese import exposure in the United States per worker in a region, and $Z_{it}$ is the change in Chinese import exposure of other high-income countries.\footnote{In Appendix \ref{dataappendix}, we plot the empirical distribution of $X_{it}$, and we plot a map of $X_{it}$ by commuting zone in the US.} Note that both $X_{it}$ and $Z_{it}$ are constructed as shift-share variables. $X_{it}$ is a linear combination of a normalized measure of the growth of imports from China to the US in industry $j$, $S_{jt}^{US}$, weighted by the contemporaneous start-of-period sector share of industry $j$ in each commuting zone $i$, $W_{ijt}$. Similarly, $Z_{it}$ is a linear combination of a normalized measure of the growth of imports from China to other high-income countries, $S_{jt}^{HI}$, weighted by the lagged sector share of industry $j$ in each commuting zone $i$, $W_{ijt-1}$. Finally, $D_{it}$ is a set of fifteen covariates.\footnote{The set of covariates includes start-of-decade labor force and demographic composition variables.} Importantly, when we connect this application to Assumption \ref{exogeneity}, we have that $(\varepsilon_{it},\eta_{it}) \independent W_{i,t-1} \mid D_{it}, S_{it}^{HI}$.
	
	In the data, we have a total of 722 locations ($N = 722$), 397 industries ($J = 397$), and two time periods: 1990-2000 and 2000-2007. \cite{autor2013} include a dummy for the second period as a covariate. However, including this indicator variable and using both time periods simultaneously would violate Assumption \ref{common}. To avoid this issue, we split the sample into two disjoint datasets, one for each period, and estimate the parameters of interest separately.
	
	Lastly, our estimator and inference procedures require choosing (i) sieves for shares, covariates, treatment variable, and control variable, (ii) trimming parameters, (iii) the number of bootstrap repetitions, and (iv) the confidence level. We present our choices below.
	
	\noindent \textit{First Stage Specification.} We choose $\epsilon = 0.01$, and $M_1 = 599$. We choose a linear specification for $K_1(W,D)$ because of the large number of shares and covariates.
	
	\noindent \textit{Second Stage Specification.} We choose a B-Spline bases for $q_X(X)$ and $q_V(\hat{V})$ with degree 3 and 4 knots. We interact those two basis as $q_X(X) \otimes q_V(\hat{V})$, and add $D$ linearly to construct $K_2(X,D,\hat{V})$. Consequently, the derivative of $K_2(x,d,v)$ with respect to $x$ will not depend on $d$.
	
	\noindent \textit{Inference.} We choose $B = 199$ for the weighted bootstrap procedure and $\alpha = 0.1$ to construct 90\%-confidence bands.
	
	\subsection{Results: Average Derivative (AD) and Local Average Response (LAR)}\label{result:ae}
	
	In this section, we show the results for the AD and LAR function (Equations \eqref{betax} and \eqref{beta}). Here, the LAR captures the effect of marginally increasing the change in exposure to Chinese imports on the temporal change in the manufacturing employment rate. In other words, we compare the percentage point change in the manufacturing employment rate when $X = x$ against its counterfactual when $X = x + dx$, where $dx \rightarrow 0$. The AD should be interpreted as the average of these effects.
	
	We start by comparing the AD results with the estimates obtained from 2SLS specifications. The first 2SLS specification is the one used by \cite{autor2013} and is given by
	\begin{equation}\label{2slsadh}
		Y_{it} = \alpha_0 + \alpha_1 X_{it}  + \alpha_2 Period_t + \alpha_3 D_{it} + u_{it} 
	\end{equation}
	where $Period_t$ is a dummy that equals 1 when the period $t$ is equal to 2000-2007, and 0 otherwise. This specification is reported in Column (3) in Table \ref{tab:ad}. We also report estimates for a non-pooled 2SLS specification:
	\begin{equation}\label{2slsseparate}
		Y_{it} = \alpha_{0,t} + \alpha_{1,t} X_{it}  + \alpha_{2,t} D_{it} + \Tilde{u}_{it} 
	\end{equation}
	We estimate this specification for both periods separately.
	
	Table \ref{tab:ad} reports the results for the AD and the 2SLS estimates. For both periods, the AD estimates are positive, but not statistically significant. In other words, we do not reject the null that, on average, marginally increasing the exposure to growth in Chinese imports in a region will not affect the change in employment in manufacturing. This finding contradicts the results obtained by the pooled 2SLS method used by \cite{autor2013}, as this specification yields a negative and statistically significant estimate of -0.303.\footnote{\cite{autor2013} estimate an effect of -0.596, but their specification includes regression weights accounting for the start-of-period commuting zones' shares of the national population. For simplicity, we do not use these weights in any estimates.} These differences can be interpreted using the results in Section \ref{2slsrepresentation}. According to Proposition \ref{thm2sls}, the derivatives of $h_1$ with respect to $z$ are likely larger in magnitude in points where the derivative of $h_2$ with respect to $x$ is more negative.
	
	\begin{table}[!htb]
		\centering
		\caption{Average Derivative}
		\begin{tabular}{lccc}
			\toprule
			& \multicolumn{3}{c}{Period}\\
			\cmidrule{2-4}
			& 1990-2000 & 2000-2007 & Pooled \\
			& (1) & (2) & (3)\\
			\midrule
			Average Derivative \hspace{43mm} & 0.106 & 0.021 & \textemdash \\
			& [-0.972, 1.185] & [-0.337, 0.378] & \textemdash \\
			& & & \\  
			2SLS & -0.087 & -0.209 & -0.303 \\
			& (0.091) & (0.076) & (0.102) \\
			& & & \\  
			\midrule
			1st stage & 0.964 & 0.669 & 0.746 \\
			Observations & 722 & 722 & 1,444 \\
			\bottomrule
		\end{tabular}
		\caption*{\textit{Note}: Table \ref{tab:ad} reports results for the Average Derivative and compares them with 2SLS estimates. Column (1) reports results for the period of 1990-2000, column (2) reports results for the period 2000-2007, and column (3) reports results pooling both periods. For the AD estimates, we report 90\%-confidence intervals in brackets. For the 2SLS estimates, robust standard errors in parentheses are clustered at the state level. We also report the main coefficient of the first stage of the 2SLS regressions.}
		\label{tab:ad}
	\end{table}
	
	When comparing the AD estimates with the 2SLS estimates for each period, the difference between the estimates gets smaller. For the period of 1990-2000, the 2SLS estimate is not statistically significant. However, for the period of 2000-2007, the 2SLS estimate is negative and statistically significant. Moreover, note that the pooled 2SLS estimate, reported in Column (3), is not a linear combination of the separate 2SLS estimates. These results suggests that the 2SLS specifications may suffer from negative weighting problems. 
	
	To better understand the weighting problems in the 2SLS specifications, we estimate the first-stage effects, which are a key component of the results in Proposition \ref{thm2sls}. This derivative is identified in Proposition \ref{fseffectprop} in Appendix \ref{fseffect} and is estimated similarly to the methods described in Section \ref{estimation}. First, we follow the semiparametric estimation procedure in \cite{chernozhukov2020} to estimate $\hat{\pi}_1(V)$, as in Section \ref{1stage}. Then, we choose the spline basis $K_1(Z,D)$. Here, we follow the main specification in Section \ref{spec}, with the basis for $Z$ as a spline of degree 3 and 4 knots, and the vector $D$ entering linearly. Consequently, the derivative of $K_1$ with respect to $Z$ does not depend on $D$.
	
\begin{figure}[!hbt]
		\centering
		\caption{First-Stage Effects, $\sfrac{\partial h_1(z,v)}{\partial z}$, with degree of 3 and 4 knots}
		\begin{subfigure}{0.49\textwidth}
			\includegraphics[width=\textwidth]{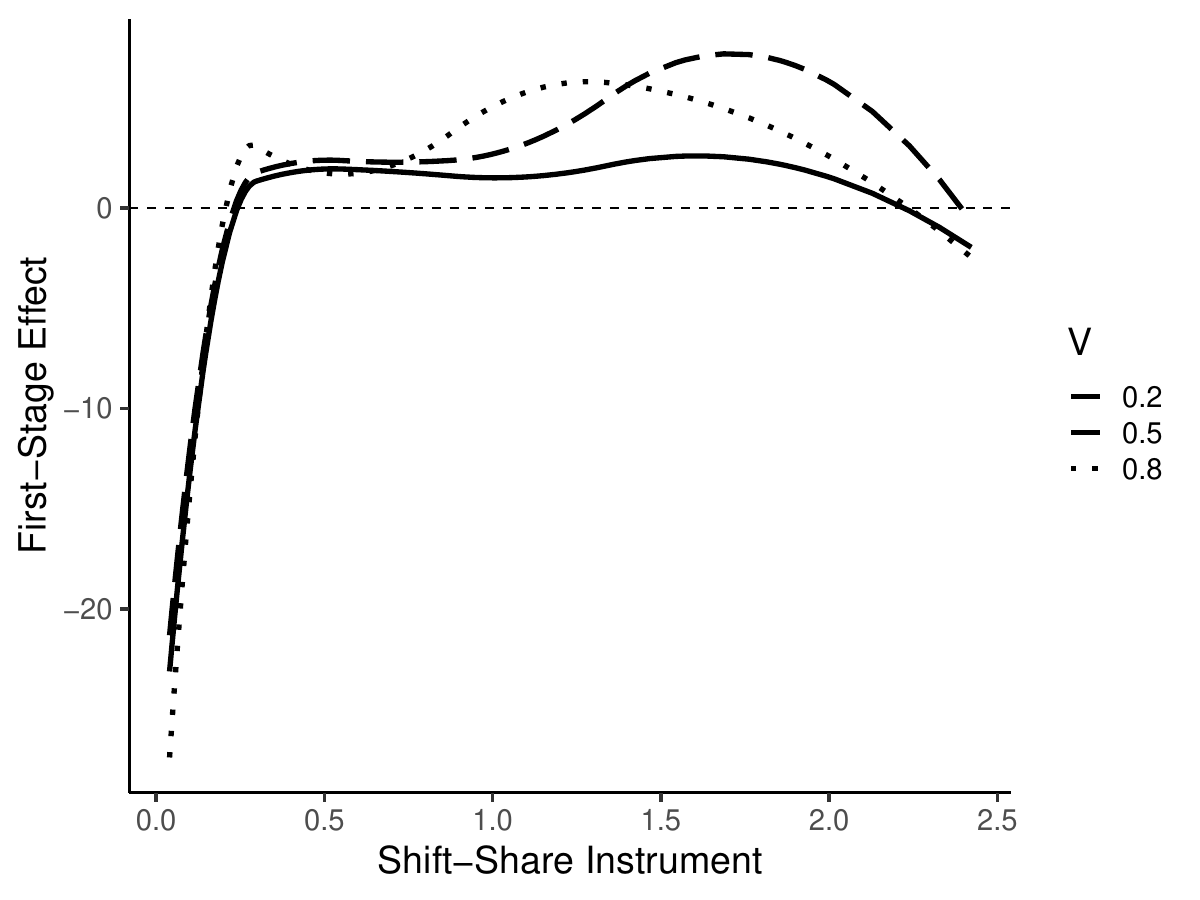}
			\caption{First-Stage Effects for 1990-2000}
		\end{subfigure}
		\begin{subfigure}{0.49\textwidth}
			\includegraphics[width=\textwidth]{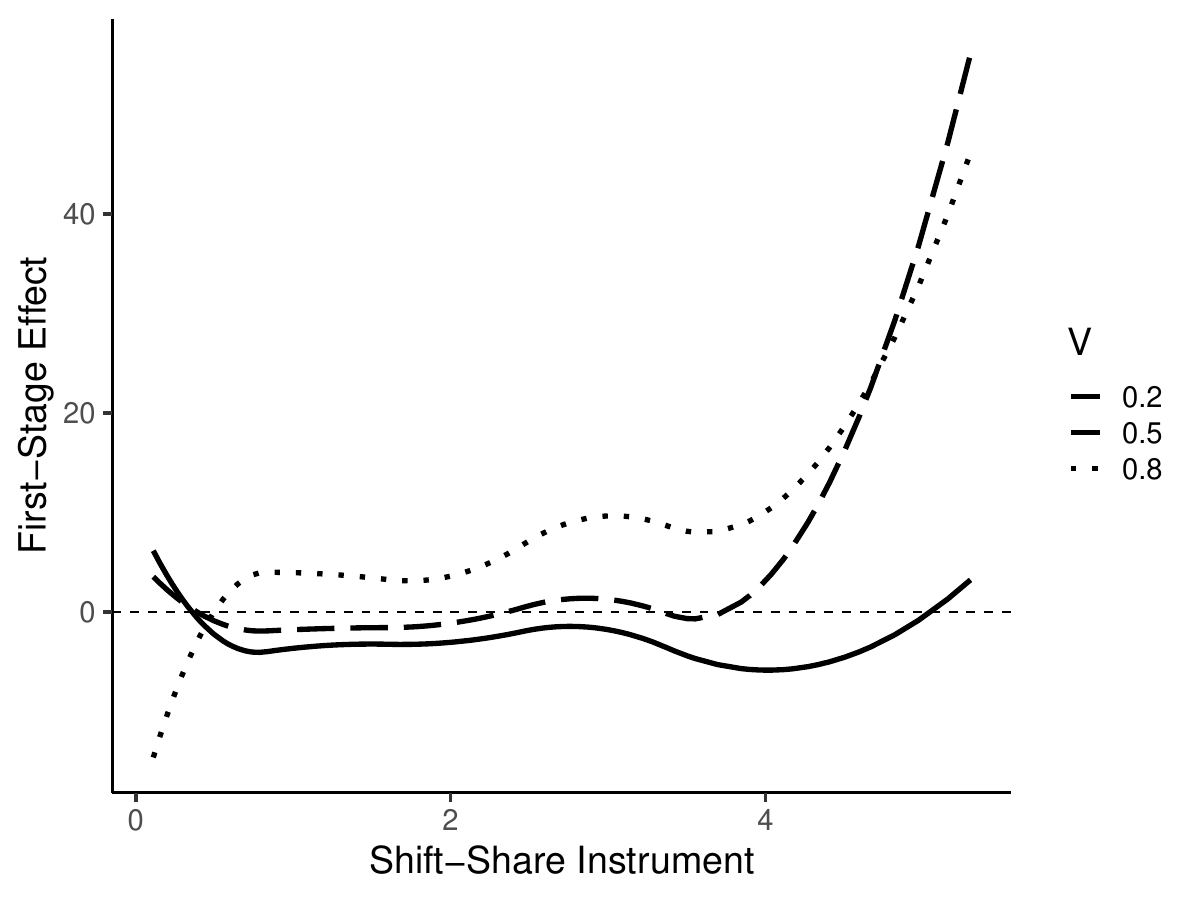}
			\caption{First-Stage Effects for 2000-2007}
		\end{subfigure}
		\caption*{\textit{Note:} Figure \ref{fig:fs34} plots the estimates for the First-Stage Effects,$\sfrac{\partial h_1(z,v)}{\partial z}$, for the periods of 1990-2000 and 2000-2007. Both estimates were produced according to the specification in Section \ref{result:ae}, which is justified based on the identification result in Appendix \ref{fseffect}. The solid line represents the First-Stage Effects for $V = 0.2$, the long-dashed line represents the First-Stage Effects for $V = 0.5$, and the dotted line represents the First-Stage Effects for $V = 0.8$.}
		\label{fig:fs34}
	\end{figure}

	Figure \ref{fig:fs34} shows the estimates of the first-stage effects associated with 2SLS regressions in the first two columns of Table \ref{tab:ad}. Importantly, the estimates of the $h_{1}$ function change sign depending on the value of $z$ and $v$. Combining these with Proposition \ref{thm2sls}, we have evidence that we cannot interpret the 2SLS estimates as weakly causal.\footnote{A similar conclusion is reached by \cite{hahn2024} using a different testing procedure.} Consequently, taking nonlinear treatment effects into consideration and focusing on the target parameters presented in Section \ref{SecParameters} are fundamental to understanding the impact of the China shock on the U.S. economy.
	
\begin{figure}[!hbt]
		\centering
		\caption{Local Average Effects}
		\begin{subfigure}{0.49\textwidth}
			\includegraphics[width=\textwidth]{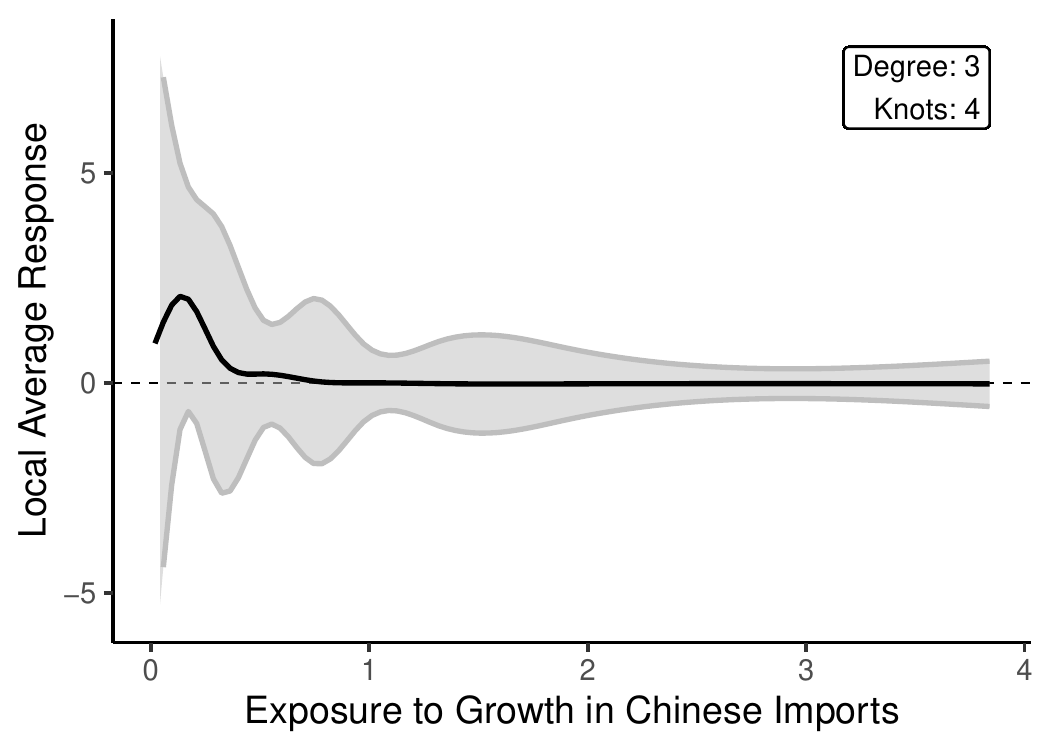}
			\caption{LAR for 1990-2000}
		\end{subfigure}
		\begin{subfigure}{0.49\textwidth}
			\includegraphics[width=\textwidth]{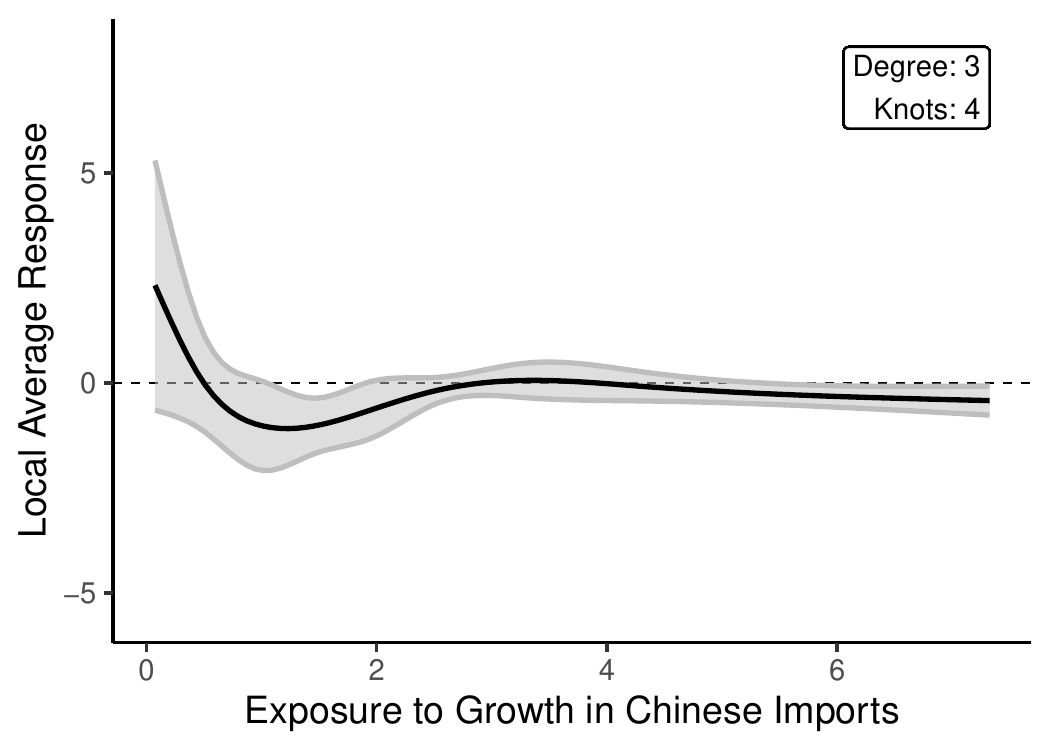}
			\caption{LAR for 2000-2007}
		\end{subfigure}
		\caption*{\textit{Note:} Figure \ref{fig:lar} plots the estimates for the Local Average Effect function for the periods 1990-2000 and 2000-2007. Both estimates were produced according to the specification in Section \ref{spec}. The gray areas represent bootstrapped uniform 90\%-confidence bands around the nonlinear LAR function. The values displayed on the x-axes represent the 5th to the 95th quantiles of the empirical distribution of the Exposure to Growth in Chinese Imports. }
		\label{fig:lar}
	\end{figure}

	To deepen our understanding of these nonlinear treatment effects, Figure \ref{fig:lar} reports the LAR estimates for the periods of 1990-2000 and 2000-2007. The estimates in Panel (a) indicate that the LAR function for the period 1990-2000  appears to be linear, as we can fit constant functions inside its confidence bands. In particular, a constant null effect is not rejected. On the other hand, the estimates in Panel (b) indicate that the LAR function for the period 2000-2007 is nonlinear. In particular, we cannot fit any constant function inside its confidence bands, since the maximum lower bound is greater than the minimum upper bound.\footnote{We note that the confidence bands are very sensitive to the chosen specification, specifically to the chosen number of knots. Appendix \ref{sens:ae} plots five alternative specifications of our estimator. When we lower the number of knots, we find tighter confidence bands. In particular, the positive values of the LAR function for the period of 2000-2007 become statistically significant when we choose 3 knots, no matter the chosen degree of the splines.}
	
	The point estimates in Panel (b) in Figure \ref{fig:lar} indicate that the effects for the period of 2000-2007 are positive for lower values of $X$, and they get negative for higher values. Specifically, for regions that faced lower exposure to growth in Chinese imports, a marginal increase in exposure to growth in Chinese imports leads to a greater intertemporal difference in manufacturing employment. In other words, for lower values of exposure to Chinese imports, employment in manufacturing decreases less than it would without the marginal increase in exposure. This type of nonlinearity underscores the importance of focusing on the target parameters presented in Section \ref{SecParameters} to understand the China shock's impact on the U.S. economy.
	
	\subsection{Results: Average Structural Function (ASF)}\label{result:asf}
	
	In this section, we present the results for the ASF (Equation \eqref{asf}). Since $Y$ and $X$ are defined as first differences, the ASF captures the relationship between a change in Chinese import exposure and the percentage point change in the manufacturing employment rate. Therefore, the ASF provides us with the average intertemporal change in the manufacturing employment rate for a commuting zone that experienced a growth in Chinese import exposure of $X = x$. 
	
	To illustrate the difference between nonlinear and linear models, we compare our specification with a linear estimator of the ASF. This linear estimator captures the ASF function if the true model (i.e., functions $h_{2}$ and $h_{1}$ in Equations \eqref{2nd_stage} and \eqref{1st_stage}) is given by the linear model associated with the non-pooled 2SLS regressions in Equation \eqref{2slsseparate}. The linear ASF estimator is given by
	\begin{equation}\label{asf2sls}
		\hat{\mu}_t^{2SLS}(x) = \hat{\alpha}_{0,t} + \hat{\alpha}_{1,t} \cdot x  + \hat{\alpha}_{2,t} \frac{1}{N}\sum_{i=1}^N D_{it}.
	\end{equation}
	
\begin{figure}[!hbt]
		\centering
		\caption{Average Structural Functions}
		\begin{subfigure}{0.49\textwidth}
			\includegraphics[width=\textwidth]{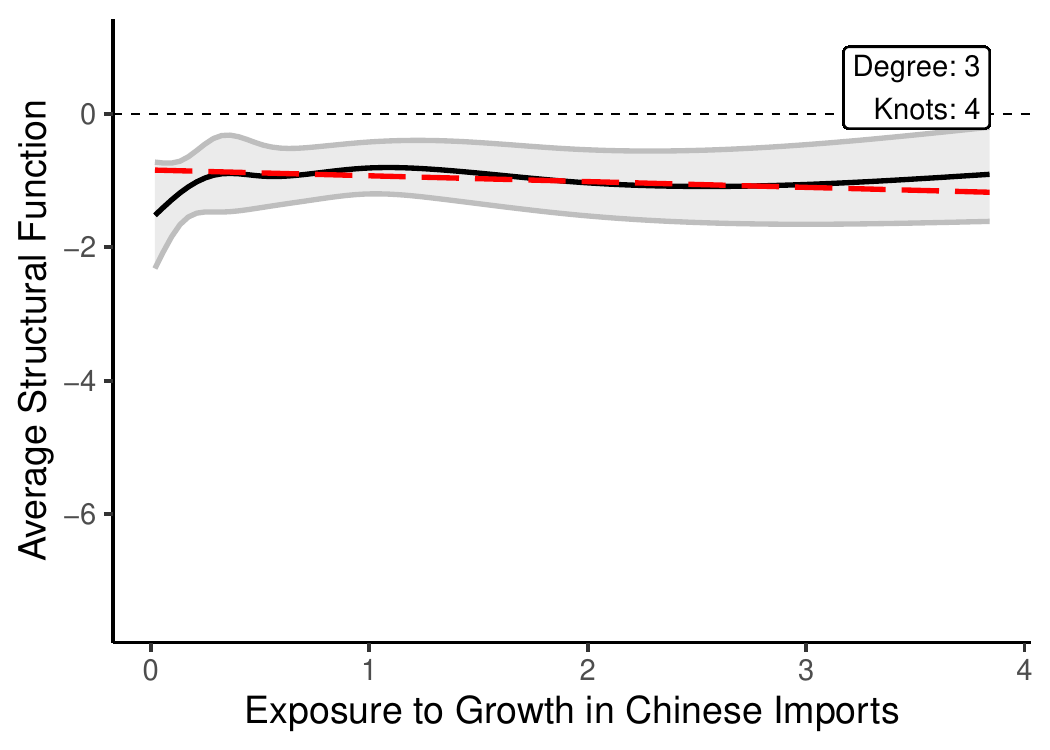}
			\caption{ASF for 1990-2000}
		\end{subfigure}
		\begin{subfigure}{0.49\textwidth}
			\includegraphics[width=\textwidth]{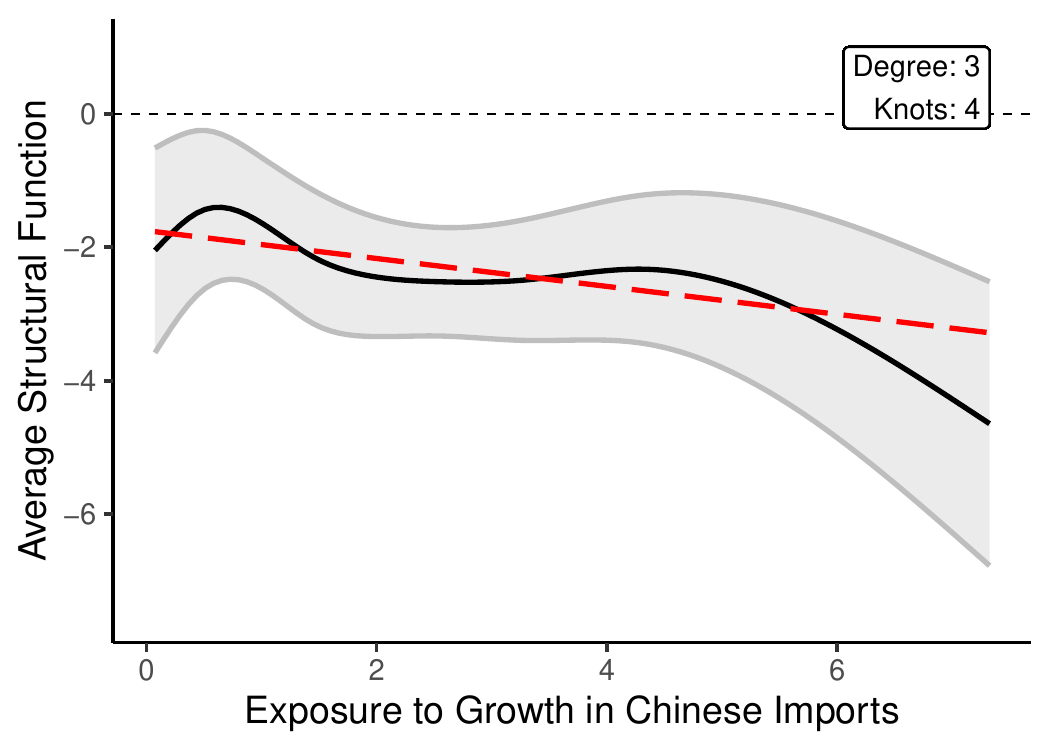}
			\caption{ASF for 2000-2007}
		\end{subfigure}
		\caption*{\textit{Note:} Figure \ref{fig:asf} plots the estimates for the Average Structural Function for the periods 1990-2000 and 2000-2007 in a black solid line. Both estimates were produced according to the specification in Section \ref{spec}. The gray areas represent bootstrapped uniform 90\%-confidence bands around the nonlinear ASF. The long-dashed lines in red represent the linear estimates of the ASF with the 2SLS parameters reported in Columns (1) and (2) of Table \ref{tab:ad}, and its estimator is described in Equation \eqref{asf2sls}.}
		\label{fig:asf}
	\end{figure}

	Figure \ref{fig:asf} plots the ASF estimates for our nonlinear specification (Section \ref{spec})  and the linear specification in Equation \eqref{asf2sls}. Moreover, we report uniform 90\%-confidence bands for our specification.
	
	Panel (a) plots the ASF estimates for the period of 1990-2000. The estimates provided by our specification suggest that, on average, a change in exposure to Chinese imports is associated with a reduction in the manufacturing employment rate that lies between 0 and 2 p.p. We observe little heterogeneity for that period, allowing us to fit constant functions within the confidence bands. The red line indicates the ASF produced by the 2SLS estimator in Equation \eqref{asf2sls}. It also fits inside the confidence bands, indicating that the 2SLS specification in Equation \eqref{2slsseparate} is flexible enough to estimate the ASF in this case. The point estimates for our specification also do not differ meaningfully from those produced by the 2SLS estimator.
	
	Panel (b) in Figure \ref{fig:asf} plots the ASF estimates for the period of 2000-2007. The ASF estimates suggest that, for lower values of growth in exposure to Chinese imports, the reduction in the manufacturing employment rate was smaller than for those that faced higher values of growth in exposure to Chinese imports. Similarly to the period 1990-2000, the red line, which indicates the estimates produced by the 2SLS estimator in Equation \eqref{asf2sls}, falls within the confidence bands for the period 2000-2007. On the other hand, the point estimates for our specification indicate that the reduction in the manufacturing employment rate was more negative than the estimates for the 2SLS specification for higher values of growth in exposure to Chinese imports.
	
	\subsection{Results: Policy Effects}\label{result:pe}
	
	In this section, we discuss the results for the Policy Effects (Equation \eqref{policyeffect}). We are interested in the effects of changes in import tariffs. Section \ref{SecConstructingPolicy} explains how we construct the policy of interest (function $\ell$ in Equation \eqref{policyeffect}) as a function of the counterfactual increase in import tariffs, while Section \ref{SecUnderstandingPolicy} interprets the estimated policy effects.
	
	\subsubsection{Constructing the Policy Function}\label{SecConstructingPolicy}
	To connect the changes in import tariffs with changes in our measure of exposure to growth in Chinese imports, we use the results discussed by \cite{fajgelbaum2019}. They estimate the elasticity of substitution between domestic goods and imports within a sector, $\hat{\kappa}$, of 1.19. They state the following relation between these variables:
	\begin{equation}\label{elasticity}
		\Delta \log (M_{jt}/D_{jt}) = (1-\hat{\kappa}) \Delta \log (1 + \phi_{jt}),
	\end{equation}
	where $M_{jt}$ is imports of products from sector $j$ in period $t$, $D_{jt}$ is expenditure of domestically produced goods, and $\phi_{jt}$ is the import tariff. Then, $\Delta \log (M_{jt}/D_{jt})$ is the log change in the ratio of imports from China to US in sector $j$ and period $t$ over domestically produced goods, driven by the log change in import tariffs $\Delta \log (1 + \phi_{jt})$. From Equation \eqref{elasticity}, we can derive the following relation:
	\begin{equation}\label{elasticity2}
		\frac{M_{jt}}{D_{jt}} = \frac{(1+\phi_{jt})^{1-\hat{\kappa}}}{(1+\phi_{jt-1})^{1-\hat{\kappa}}}\frac{M_{jt-1}}{D_{jt-1}},
	\end{equation}
	implying that we can construct a counterfactual $\frac{\Tilde{M}_{jt}}{\Tilde{D}_{jt}}$ from an increase of $\Delta\phi_j$ in the actual import tariff:
	\begin{equation}\label{elasticity3}
		\frac{\Tilde{M}_{jt}}{\Tilde{D}_{jt}} = \frac{(1+\phi_{jt} + \Delta\phi_j)^{1-\hat{\kappa}}}{(1+\phi_{jt-1})^{1-\hat{\kappa}}}\frac{M_{jt-1}}{D_{jt-1}}.
	\end{equation}
	
	We perform a partial equilibrium exercise, in which we assume that consumers do not change their consumption of domestically produced goods, \textit{i.e.}, $D_{jt} = \Tilde{D}_{jt}$. Combining Equations \eqref{elasticity2} and \eqref{elasticity3}, we have that
	\begin{equation}\label{elasiticity4}
		\left(\frac{\Tilde{M}_{jt}}{M_{jt}}\right)^{\frac{1}{1-\hat{\kappa}}} = 1 + \underbrace{\frac{1 + \phi_{jt} + \Delta\phi_j}{1 + \phi_{jt}} - 1}_{=: \Tilde{\phi}_{j}},
	\end{equation}
	where we define $\Tilde{\phi}_j$ as the percentage increase in import tariff related to the actual import tariff. By isolating $\Tilde{M}_{jt}$ in Equation \eqref{elasiticity4}, we find that
	\begin{equation}\label{elasticity5}
		\Tilde{M}_{jt} = (1 + \Tilde{\phi}_j)^{1-\hat{\kappa}} M_{jt}.
	\end{equation}
	
	Moreover, according to the definition of $S_{jt}^{US}$ used by \cite{autor2013}, we can construct a counterfactual $\Tilde{S}_{jt}^{US}$ using Equation \eqref{elasticity5} as
	\begin{equation}\label{shift2}
		\Tilde{S}_{jt}^{US} = \frac{\Tilde{M}_{jt} - M_{jt-1}}{L_{jt}} = \frac{(1 + \Tilde{\phi}_j)^{1-\hat{\kappa}} M_{jt} - M_{jt-1}}{L_{jt}},
	\end{equation}
	where $L_{jt}$ is the share of the workforce working in sector $j$.
	
	Furthermore, we are only interested in estimating the effects of a common increase in tariffs for all industries. For this reason, we impose that $\Tilde{\phi}_j = \Tilde{\phi}$ for all $j \in \{1,\dots,J\}$. Consequently, Equation \eqref{shift2} and the estimated elasticity in \cite{fajgelbaum2019} imply that our policy function is given by
	\begin{equation}\label{pfun1}
		\ell(X_{it}; \Tilde{\phi}) = \sum_{j=1}^J W_{ijt}\Tilde{S}_{jt}^{US} = \sum_{j=1}^J W_{ijt}\left(\frac{(1 + \Tilde{\phi})^{-0.19} M_{jt} - M_{jt-1}}{L_{jt}}\right).
	\end{equation}
	
	\subsubsection{Understanding the Policy Effect}\label{SecUnderstandingPolicy}
	
	Using Equation \eqref{pfun1} as our policy function, our Policy Effect (Equation \eqref{policyeffect}) captures the effect of a change in the actual exposure to growth in Chinese imports, driven by the percentage increase in the actual import tariff, on manufacturing employment at the end of the period. Consequently, we interpret this effect as a percentage point change in manufacturing employment rate at the end of the period driven by a percentage increase of $\Tilde{\phi}$ in the import tariff.\footnote{The factual world in Equation \eqref{policyeffect}, $\mathbb{E}[Y\mid S = \tilde{s}]$, contains the impacts of the actual growth in Chinese imports on the time change in the manufacturing employment rate. The counterfactual world in Equation \eqref{policyeffect}, $E[h_2(\ell(X),D, \varepsilon)\mid S = \tilde{s}]$,  contains the impact of changing the import tariffs through the term $\tilde{\phi}$ and the impacts of the actual growth in Chinese imports through the term $M_{jt} - M_{jt-1}$. Consequently, their difference---the policy effect---captures the effect of increasing import tariffs on the temporal change in the manufacturing employment rate.}
	
	To illustrate the difference between nonlinear and linear models, we compare our specification with a linear estimator of the policy effect. This linear estimator captures the policy effect if the true model (i.e., functions $h_{2}$ and $h_{1}$ in Equations \eqref{2nd_stage} and \eqref{1st_stage}) is given by the linear model associated with the non-pooled 2SLS regressions in Equation \eqref{2slsseparate}. The linear policy effect estimator is given by
	\begin{equation}\label{asf2sls_secondlabel}
		\hat{\gamma}_t^{2SLS}(\tilde{\phi}) = \hat{\alpha}_{0,t} + \hat{\alpha}_{1,t} \cdot \ell(X_{it}; \Tilde{\phi})  + \hat{\alpha}_{2,t} \frac{1}{N}\sum_{i=1}^N D_{it} - \frac{1}{N}\sum_{i=1}^N Y_{it}.
	\end{equation}
	
\begin{figure}[!hbt]
		\centering
		\caption{Policy Effects}
		\begin{subfigure}{0.49\textwidth}
			\includegraphics[width=\textwidth]{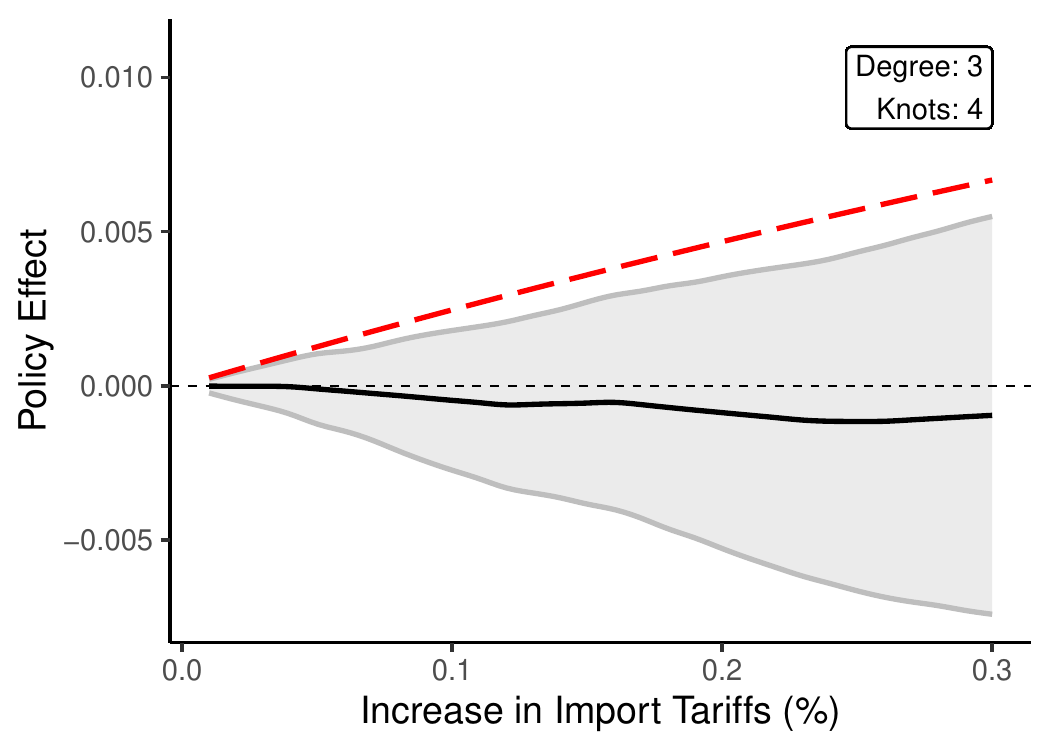}
			\caption{Policy Effect for 1990-2000}
		\end{subfigure}
		\begin{subfigure}{0.49\textwidth}
			\includegraphics[width=\textwidth]{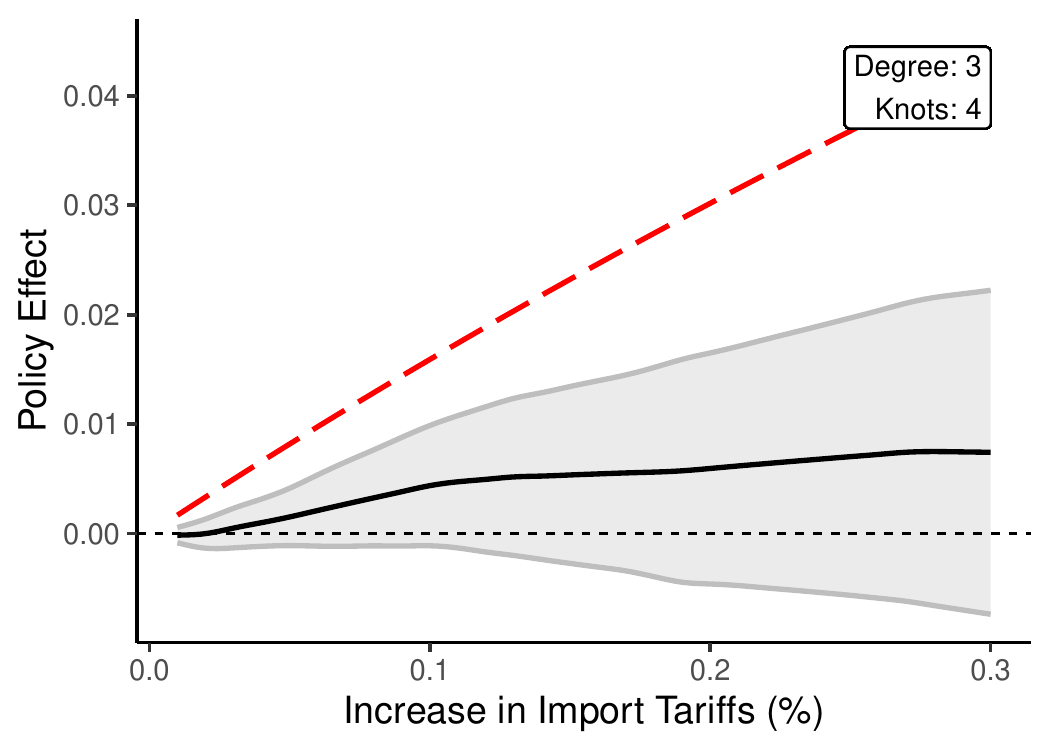}
			\caption{Policy Effect for 2000-2007}
		\end{subfigure}
		\caption*{\textit{Note:} Figure \ref{fig:pe} plots the estimates for the Policy Effect for the periods of 1990-2000 and 2000-2007 in a black solid line. The gray area represents bootstrapped uniform 90\%-confidence bands around the nonlinear policy effect. The red, long-dashed lines represent the linear policy effect estimates from a 2SLS specification.  Both estimates were produced according to the specification in Section \ref{spec}. The values displayed on the x-axes represent an increase in import tariff, going from a 1\% increase to a 30\% increase. }
		\label{fig:pe}
	\end{figure}

	We estimate the effects of the Policy Effect for values of $\Tilde{\phi}$ between 0.01 and 0.3, \textit{i.e.}, from an increase of 1\% to an increase of 30\% in import tariffs. Figure \ref{fig:pe} plots the estimates for the periods 1990-2000 and 2000-2007. In both periods, the estimated effects are not statistically significant.
	
	Panel (a) in Figure \ref{fig:pe} plots the estimates of the Policy Effect for the period of 1990-2000. The point estimates are negative and very close to zero, while the confidence band becomes wider for higher values of the increase in import tariffs. Those estimates suggest that increasing the import tariffs for the period of 1990-2000 would not affect the manufacturing employment rate at the end of the period. These results are consistent with the results in Sections \ref{result:ae} and \ref{result:asf}.\footnote{We note that the Policy Effect is sensitive to the chosen specification. Unlike our other objects of interest, not only the confidence bands, but also the point estimates of the Policy Effect are very sensitive to the choice of tuning parameters (e.g., the specification with degree of 3 and 3 knots). Appendix \ref{sens:pe} plots the Policy Effects for five alternative specifications.}
	
	Panel (b) in Figure \ref{fig:pe} plots the estimates of the Policy Effect for the period of 2000-2007. Although nonsignificant, the point estimates are now positive, and the area of the confidence band is more concentrated in the positive region of the plot. One interesting finding that our point estimates reveal is that the policy effect, as a function of the tariff increase, appears to become constant after a certain level of import tariffs is reached. For example, a 15\% increase in import tariffs has a similar effect to a 30\% increase.
	
	Importantly, the 2SLS estimates for the policy effect fall outside the confidence band of our nonlinear method. This result highlights how the linearity imposed by the 2SLS specification overlooks the heterogeneity found by our methods and may lead researchers to overestimate the effect of tariffs. Consequently, allowing for nonlinear heterogeneous treatment effects is fundamental to understanding the China shock's impact on the U.S. economy.
	
	\section{Conclusion}\label{conclusion}
	
	In this paper, we analyze nonlinear treatment effects in shift-share designs when shares are exogenous. To do so, we use a triangular model, similarly to \cite{imbens2009}, and identify four parameters of interest using a control function approach. Moreover, we propose a flexibly parametric estimation procedure to estimate these parameters. In this section, we discuss the contexts in which our proposed methodology can be applied and further develop our empirical discussion.
	
	Our methodology can be applied to any empirical problem that fits a shift-share design with exogenous shares. In our empirical application, we focus on the effects of the China shock on changes in manufacturing employment in the U.S., using data from \cite{autor2013}. Our results highlight substantial treatment effect heterogeneity, which commonly used tools in shift-share applications are not designed to capture.
	
	Regarding its empirical contribution, our work offers novel tools that answer policy-relevant questions in shift-share settings. For instance, we assess the impact of increasing import tariffs to compensate for the China shock, contributing to the literature on the effects of protectionist policies. Moreover, researchers can use our methodology to investigate the effects of different treatment assignment policies.

\singlespace

\bibliography{ref}


\pagebreak

\newpage

\pagebreak

\appendix

\begin{center}
	\huge
	Supporting Information

	(Online Appendix)

\end{center}

\doublespacing
\normalsize

\section{Proofs of the Main Results} \label{AppProof}

\setcounter{table}{0}
\renewcommand\thetable{A.\arabic{table}}

\setcounter{figure}{0}
\renewcommand\thefigure{A.\arabic{figure}}

\setcounter{equation}{0}
\renewcommand\theequation{A.\arabic{equation}}

\setcounter{theorem}{0}
\renewcommand\thetheorem{A.\arabic{theorem}}

\setcounter{proposition}{0}
\renewcommand\theproposition{A.\arabic{proposition}}

\setcounter{corollary}{0}
\renewcommand\thecorollary{A.\arabic{corollary}}

\setcounter{assumption}{0}
\renewcommand\theassumption{A.\arabic{assumption}}

\setcounter{definition}{0}
\renewcommand\thedefinition{A.\arabic{definition}}

\setcounter{Lemma}{0}
\renewcommand\theLemma{A.\arabic{Lemma}}

\subsection{Proof of Proposition \ref{controlfunction}}\label{cfproof}

\begin{proof}
	
	Here, we first prove that $V := F_{X\mid W,D,S}(X \mid W,D,S) = F_{\eta \mid D,S}(\eta \mid D, S)$, and then we prove that $X \independent \varepsilon \mid V, D, S$.
	
	Fix $(\overline{x}, \overline{d}, \overline{w},\overline{s}) \in \mathcal{X}\times\mathcal{D}\times\mathcal{W}\times \mathcal{S}$. Let $\overline{z} := h_0(\overline{w},\overline{s})$. We have that
	\begin{equation}\label{appV}
		\begin{split}
			F_{X\mid W,D,S}(\overline{x}\mid \overline{w},\overline{d},\overline{s}) & = \mathbb{E}[\mathds{1}\{X \leq \overline{x}\}\mid W = \overline{w}, D = \overline{d}, S = \overline{s}]\\
			& = \mathbb{E}[\mathds{1}\{h_1(\overline{z},\overline{d},\eta) \leq \overline{x}\} \mid W = \overline{w}, D = \overline{d}, S = \overline{s}]\\ 
			& = \mathbb{E}[\mathds{1}\{\eta \leq h_1^{-1}(\overline{z},\overline{d},\overline{x})\} \mid W = \overline{w}, D = \overline{d}, S = \overline{s}]\\
			& = \mathbb{E}[\mathds{1}\{\eta \leq h_1^{-1}(\overline{z},\overline{d},\overline{x})\} \mid D = \overline{d}, S = \overline{s}]\\
			& = F_{\eta \mid D,S}(\overline{\eta} \mid \overline{d}, \overline{s})
		\end{split}
	\end{equation}
	where the first equality holds from the definition of a cumulative distribution function, the second equality holds according to Equation \eqref{1st_stage}, the third equality holds according to Assumption \ref{increasing}, the fourth equality holds according to Assumption \ref{exogeneity}, the last equality holds according to the definition of a cumulative distribution function and by defining $\overline{\eta} := h_1^{-1}(\overline{z},\overline{d},\overline{x})$.
	
	Equation \eqref{appV} shows that we can identify the CDF of $\eta$ conditional on $D$ and $S$ given our knowledge of $(X,D,W,S)$. Now, define
	\begin{equation}
		V := F_{X \mid W,D,S}(X \mid W,D,S).
	\end{equation}
	
	We know that $V$ is uniformly distributed on $(0,1)$ by the Probability Integral Transformation Theorem. Now, we prove that $X \independent \varepsilon \mid V, D, S$:
	\begin{equation}
		\begin{split}
			& F_{X,\varepsilon \mid V,D,S}(\overline{x},\overline{\varepsilon}\mid \overline{v},\overline{d},\overline{s}) \\
			& \hspace{10pt} = \mathbb{E}[\mathds{1}\{X \leq \overline{x}\}\mathds{1}\{\varepsilon \leq \overline{\varepsilon}\} \mid V = \overline{v}, D = \overline{d}, S = \overline{s}]\\
			& \hspace{10pt} = \mathbb{E}[\mathds{1}\{X \leq \overline{x}\}\mathds{1}\{\varepsilon \leq \overline{\varepsilon}\} \mid \eta = \overline{\eta}, D = \overline{d}, S = \overline{s}]\\
			& \hspace{10pt} = \mathbb{E}\left[\left. \mathbb{E}[\mathds{1}\{h_1(h_0(W, \overline{s}),\overline{\eta}) \leq \overline{x}\}\mid \varepsilon, \eta = \overline{\eta}, D = \overline{d}, S = \overline{s}] \cdot \mathds{1}\{\varepsilon \leq \overline{\varepsilon}\} \right\vert \eta = \overline{\eta}, D = \overline{d}, S = \overline{s}\right]\\
			& \hspace{10pt} = \mathbb{E}[\mathbb{E}[\mathds{1}\{h_1(h_0(W, \overline{s}),\overline{\eta}) \leq \overline{x}\}\mid \eta = \overline{\eta}, D = \overline{d}, S = \overline{s}]\cdot \mathds{1}\{\varepsilon \leq \overline{\varepsilon}\} \mid \eta = \overline{\eta}, D = \overline{d}, S = \overline{s}]\\
			& \hspace{10pt} = \mathbb{E}[\mathds{1}\{X \leq \overline{x}\} \mid V = \overline{v}, D = \overline{d}, S = \overline{s}]\cdot \mathbb{E}[\mathds{1}\{\varepsilon \leq \overline{\varepsilon}\} \mid V = \overline{v}, D = \overline{d}, S = \overline{s}] 
		\end{split}
	\end{equation}
	where the first equality holds from the definition of a cumulative distribution function; the second equality holds according to Equation \eqref{appV}, Assumption \ref{increasing} and $\overline{\eta} \coloneqq F_{\eta}^{-1}\left(\left. \overline{v} \right\vert \overline{d}, \overline{s} \right)$; the third equality holds according to the Law of Iterated Expectations and Equations \eqref{1st_stage}-\eqref{EqIV}; the fourth equality holds from Assumption \ref{exogeneity}; and the last equality holds according to linearity of the expectation operator, Equation \eqref{appV} and the definition of $\overline{\eta}$.
\end{proof}

\subsection{Proof of Proposition \ref{asfidentification}}\label{asfproof}

\begin{proof}
	
	First, recall that $m(x,d,v) := \mathbb{E}[Y \mid X = x, D = d, V = v, S = \Tilde{s}]$. Fix $x \in \mathcal{X}$. Then, we have that
	\begin{equation}
		\begin{split}
			\mu(x) & = \mathbb{E}[h_2(x,D,\varepsilon) \mid S = \Tilde{s}]\\
			& = \mathbb{E}[\mathbb{E}[h_2(x,D,\varepsilon) \mid X, D, V, S = \Tilde{s}] \mid S = \Tilde{s}]\\
			& = \mathbb{E}\left[\left.\int_{\mathcal{E}_2} h_2(x, D, e) dF_{\varepsilon\mid X,D,V,S}(e \mid X,D,V,\Tilde{s}) \right\vert S = \Tilde{s}\right]\\
			& = \mathbb{E}\left[\left.\int_{\mathcal{E}_2} h_2(x, D, e) dF_{\varepsilon\mid D,V,S}(e \mid D,V,\Tilde{s}) \right\vert S = \Tilde{s}\right]\\
			& = \mathbb{E}[m(x,D,V) \mid S = \Tilde{s}]
		\end{split}
	\end{equation}
	where the first equality holds from the definition of the ASF, the second equality holds from applying the Law of Iterated Expectations, the third equality holds from the definition of expectation, the fourth equality holds from the results in Proposition \ref{controlfunction}, and the last equality holds according to Equation \eqref{mfunction}. 
\end{proof}

\subsection{Proof of Proposition \ref{propAE}}\label{aeproof}

\begin{proof}
	First, recall that $m(x,d,v) := \mathbb{E}[Y \mid X = x, D = d, V = v, S = \Tilde{s}]$. We first prove the result for the LAR parameter. We have that
	\begin{equation}
		\begin{split}
			\beta(x) & = \mathbb{E}\left[\left.\frac{\partial h_2(X,D,\varepsilon)}{\partial x}\right\vert X = x, S = \Tilde{s} \right]\\
			& = \mathbb{E}\left[\left.\mathbb{E}\left[\left.\frac{\partial h_2(X,D,\varepsilon)}{\partial x}\right \vert X,D,V, S = \Tilde{s}\right]\right\vert X = x, S = \Tilde{s} \right]\\
			& = \mathbb{E}\left[\left.\int_{\mathcal{E}_2}\frac{\partial h_2(X,D,e)}{\partial x} dF_{\varepsilon\mid X,D,V,S}(e \mid X,D,V,\Tilde{s})\right\vert X = x, S = \Tilde{s} \right]\\
			& = \mathbb{E}\left[\left.\int_{\mathcal{E}_2}\frac{\partial h_2(X,D,e)}{\partial x} dF_{\varepsilon\mid D,V,S}(e \mid D,V,\Tilde{s})\right\vert X = x, S = \Tilde{s} \right]\\
			& = \mathbb{E}\left[\left.\frac{\partial}{\partial x}\int_{\mathcal{E}_2}h_2(X,D,e) dF_{\varepsilon\mid D,V,S}(e \mid D,V,\Tilde{s})\right\vert X = x, S = \Tilde{s} \right]\\
			& = \mathbb{E}\left[\left.\frac{\partial m(X,D,V)}{\partial x}\right\vert X = x, S = \Tilde{s} \right]
		\end{split}
	\end{equation}
	where the first equality holds from the definition of the LAR, the second equality holds from applying the Law of Iterated Expectations, the third equality holds from the definition of expectation, the fourth equality holds from the results in Proposition \ref{controlfunction}, the fifth equality holds from Assumption \ref{rcond}, and the last equality holds according to Equation \eqref{mfunction}. The proof for the AD parameter is analogous:
	\begin{equation}
		\begin{split}
			\beta & = \mathbb{E}\left[\left.\frac{\partial h_2(X,D,\varepsilon)}{\partial x}\right\vert S = \Tilde{s} \right]\\
			& = \mathbb{E}\left[\left.\mathbb{E}\left[\left.\frac{\partial h_2(X,D,\varepsilon)}{\partial x}\right \vert X,D,V, S = \Tilde{s}\right]\right\vert S = \Tilde{s} \right]\\
			& = \mathbb{E}\left[\left.\int_{\mathcal{E}_2}\frac{\partial h_2(X,D,e)}{\partial x} dF_{\varepsilon\mid X,D,V,S}(e \mid X,D,V,\Tilde{s})\right\vert S = \Tilde{s} \right]\\
			& = \mathbb{E}\left[\left.\int_{\mathcal{E}_2}\frac{\partial h_2(X,D,e)}{\partial x} dF_{\varepsilon\mid D,V,S}(e \mid D,V,\Tilde{s})\right\vert S = \Tilde{s} \right]\\
			& = \mathbb{E}\left[\left.\frac{\partial}{\partial x}\int_{\mathcal{E}_2}h_2(X,D,e) dF_{\varepsilon\mid D,V,S}(e \mid D,V,\Tilde{s})\right\vert S = \Tilde{s} \right]\\
			& = \mathbb{E}\left[\left.\frac{\partial m(X,D,V)}{\partial x}\right\vert S = \Tilde{s} \right]
		\end{split}
	\end{equation}
	where the first equality holds from the definition of the AD, the second equality holds from applying the Law of Iterated Expectations, the third equality holds from the definition of expectation, the fourth equality holds from the results in Proposition \ref{controlfunction}, the fifth equality holds from Assumption \ref{rcond}, and the last equality holds according to Equation \eqref{mfunction}.
	
\end{proof}

\subsection{Proof of Proposition \ref{propPE}}\label{peproof}

\begin{proof}
	First, recall that $m(x,d,v) := \mathbb{E}[Y \mid X = x, D = d, V = v, S = \Tilde{s}]$ and that $\ell: \mathcal{X} \rightarrow \mathcal{X}$. Then, we have that
	\begin{equation}
		\begin{split}
			\gamma & = \mathbb{E}[h_2(\ell(X),D,\varepsilon) - Y \mid S = \Tilde{s}]\\
			& = \mathbb{E}[\mathbb{E}[h_2(\ell(X),D,\varepsilon) \mid X,D,V,S = \Tilde{s}] - Y \mid S = \Tilde{s}]\\
			& = \mathbb{E}\left[\left.\int_{\mathcal{E}_2} h_2(\ell(X), D, e) dF_{\varepsilon\mid X,D,V,S}(e \mid X,D,V,\Tilde{s}) - Y \right\vert  S = \Tilde{s}\right]\\
			& = \mathbb{E}\left[\left.\int_{\mathcal{E}_2} h_2(\ell(X), D, e) dF_{\varepsilon\mid D,V,S}(e \mid D,V,\Tilde{s}) - Y \right\vert  S = \Tilde{s}\right]\\
			& = \mathbb{E}[m(\ell(X),D,V) - Y \mid S = \Tilde{s}]
		\end{split}
	\end{equation}
	where the first equality holds from the definition of the Policy Effect, the second equality holds from applying the Law of Iterated Expectations, the third equality holds from the definition of expectation, the fourth equality holds from the result in Proposition \ref{controlfunction}, and the last equality holds according to Equation \eqref{mfunction}. Lastly, by Assumption 
	\ref{csupp}, we can identify $m(\ell(X),D,V)$.

\end{proof}

\subsection{Proof of Proposition \ref{thm2sls}} \label{2slsproof}

\begin{proof}

	First, define  $g(z, \eta, \varepsilon) := h_2(h_1(z, \eta), \varepsilon)$. Notice that
	\begin{equation}\label{appderivative}
		\frac{\partial g(z, \eta, \varepsilon)}{\partial z} = \frac{\partial h_2(h_1(z, \eta), \varepsilon)}{\partial z} = \frac{\partial h_2(h_1(z, \eta), \varepsilon)}{\partial x} \frac{\partial h_1(z, \eta)}{\partial z}.
	\end{equation}

	Fix $z_L, z_U \in \mathcal{Z}$, with $z_L \leq z_U$. By the Fundamental Theorem of Calculus, we have
	\begin{equation}\label{ftc}
		g(z_U, \eta, \varepsilon) - g(z_L, \eta, \varepsilon) = \int_{z_L}^{z_U} \frac{\partial h_2(h_1(\zeta, \eta), \varepsilon)}{\partial x} \frac{\partial h_1(\zeta, \eta)}{\partial z} d\zeta.
	\end{equation}
	
	Let $z_0 := \mathrm{inf}(\mathcal{Z})$. We know that 
	\begin{equation}\label{appg0}
		\mathbb{E}[g(z_0, \eta, \varepsilon)(Z - \mathbb{E}[Z]) \mid S = \Tilde{s}]  = \mathbb{E}[g(z_0, \eta, \varepsilon) \mid S = \Tilde{s}]\mathbb{E}[(Z - \mathbb{E}[Z]) \mid S = \Tilde{s}] = 0
	\end{equation}
	where the first equality holds from  Assumption \ref{exogeneity}, and the last equality holds from  the fact that $\mathbb{E}[(Z - \mathbb{E}[Z]) \mid S = \Tilde{s}] = 0$. 
	
	Recall that 
	\begin{equation}\label{app2sls}
		\beta^{\text{2SLS}} = \frac{\mathrm{Cov}(Y,Z\mid S = \Tilde{s})}{\mathrm{Cov}(X,Z\mid S = \Tilde{s})}.
	\end{equation}
	
	The numerator of Equation \eqref{app2sls} can be rewritten as
	\begin{equation}\label{appnumerator}
		\mathrm{Cov}(Y,Z\mid S = \Tilde{s}) = \mathbb{E}[Y(Z - \mathbb{E}[Z])\mid S = \Tilde{s}] = \mathbb{E}[g(Z, \eta, \varepsilon)(Z - \mathbb{E}[Z])\mid S = \Tilde{s}].
	\end{equation}
	
	From Equation \eqref{appnumerator}, we can write the numerator of the $\beta^{\text{2SLS}}$ in terms of the derivative in Equation \eqref{appderivative}:
	\begin{equation}
		\begin{split}
			\mathbb{E}[g(Z, \eta, \varepsilon)(Z - \mathbb{E}[Z]) \mid S = \Tilde{s}] & = \mathbb{E}[(g(Z, \eta, \varepsilon) - g(z_0, \eta, \varepsilon))(Z - \mathbb{E}[Z]) \mid S = \Tilde{s}]\\
			& = \int_{\mathcal{E}_2} \int_{\mathcal{E}_1} \int_{\mathcal{Z}} (g(\zeta, e_1, e_2) - g(z_0, e_1, e_2))(\zeta - \mathbb{E}[Z]) f_{Z,\eta,\varepsilon\mid S}(\zeta, e_1, e_2) d\zeta de_1 de_2\\
			& = \int_{\mathcal{E}_2} \int_{\mathcal{E}_1} \int_{\mathcal{Z}} (g(\zeta, e_1, e_2) - g(z_0, e_1, e_2))(\zeta - \mathbb{E}[Z]) f_{Z\mid S}(\zeta) f_{\eta, \varepsilon \mid S}(e_1,e_2)d\zeta de_1 de_2\\
			& = \mathbb{E}\left[\left.\int_{\mathcal{Z}} (g(\zeta, \eta, \varepsilon) - g(z_0, \eta, \varepsilon))(\zeta - \mathbb{E}[Z]) f_{Z\mid S}(\zeta) d\zeta\right \vert S = \Tilde{s}\right]\\
			& = \mathbb{E}\left[\left.\int_{\mathcal{Z}} \left(\int_{z_0}^\zeta \frac{\partial h_2(h_1(\omega, \eta), \varepsilon)}{\partial x}\frac{\partial h_1(\omega,\eta)}{\partial z} d\omega\right)(\zeta - \mathbb{E}[Z]) f_{Z\mid S}(\zeta) d\zeta\right \vert S = \Tilde{s}\right]\\
			& = \mathbb{E}\left[\left.\int_{\mathcal{Z}}  \frac{\partial h_2(h_1(\omega, \eta), \varepsilon)}{\partial x}\frac{\partial h_1(\omega,\eta)}{\partial z} \int_{z_0}^\omega(\zeta - \mathbb{E}[Z]) f_{Z\mid S}(\zeta) d\zeta d\omega\right \vert S = \Tilde{s}\right]
		\end{split}
	\end{equation}
	where the first equality holds from the result in Equation \eqref{appg0}, the second equality holds from the definition of expectation, the third equality holds from Assumption \ref{exogeneity}, the fourth equality holds from the definition of expectation, the fifth equality holds from Equation \eqref{ftc}, and the last equality holds from Fubini's Theorem.
	
	Similarly, we have that
	\begin{equation}\label{apph0}
		\mathbb{E}[h_1(z_0, \eta)(Z - \mathbb{E}[Z]) \mid S = \Tilde{s}]  =  \mathbb{E}[h_1(z_0, \eta) \mid S = \Tilde{s}] \mathbb{E}[(Z - \mathbb{E}[Z]) \mid S = \Tilde{s}] = 0
	\end{equation}
	where the first equality holds from  Assumption \ref{exogeneity}, and the last equality holds from the fact that $\mathbb{E}[(Z - \mathbb{E}[Z]) \mid S = \Tilde{s}] = 0$. The denominator in Equation \eqref{app2sls} can be rewritten as
	\begin{equation}\label{appdenominator}
		\mathrm{Cov}(X,Z\mid S = \Tilde{s}) = \mathbb{E}[X(Z - \mathbb{E}[Z])\mid S = \Tilde{s}] = \mathbb{E}[h_1(Z,\eta)(Z - \mathbb{E}[Z])\mid S = \Tilde{s}].
	\end{equation}
	
	From Equation \eqref{appdenominator}, we can write the denominator of Equation \eqref{app2sls} as 
	\begin{equation}
		\begin{split}
			\mathbb{E}[h_1(Z, \eta)(Z - \mathbb{E}[Z]) \mid S = \Tilde{s}] & = \mathbb{E}[(h_1(Z, \eta) - h_1(z_0, \eta))(Z - \mathbb{E}[Z]) \mid S = \Tilde{s}]\\
			& = \int_{\mathcal{E}_1} \int_{\mathcal{Z}} (h_1(\zeta, e_1) - h_1(z_0, e_1))(\zeta - \mathbb{E}[Z]) f_{Z,\eta\mid S}(\zeta, e_1) d\zeta de_1\\
			& = \int_{\mathcal{E}_1} \int_{\mathcal{Z}} (h_1(\zeta, e_1) - h_1(z_0, e_1))(\zeta - \mathbb{E}[Z]) f_{Z\mid S}(\zeta) f_{\eta \mid S}(e_1)d\zeta de_1\\
			& = \mathbb{E}\left[\left.\int_{\mathcal{Z}} (h_1(Z, \eta) - h_1(z_0, \eta))(\zeta - \mathbb{E}[Z]) f_{Z\mid S}(\zeta) d\zeta\right \vert S = \Tilde{s}\right]\\
			& = \mathbb{E}\left[\left.\int_{\mathcal{Z}} \left(\int_{z_0}^\zeta \frac{\partial h_1(\omega,\eta)}{\partial z} d\omega\right)(\zeta - \mathbb{E}[Z]) f_{Z\mid S}(\zeta) d\zeta\right \vert S = \Tilde{s}\right]\\
			& = \mathbb{E}\left[\left.\int_{\mathcal{Z}}  \frac{\partial h_1(\omega,\eta)}{\partial z} \int_{z_0}^\omega(\zeta - \mathbb{E}[Z]) f_{Z\mid S}(\zeta) d\zeta d\omega\right \vert S = \Tilde{s}\right]
		\end{split}
	\end{equation}
	where the first equality holds from the result in Equation \eqref{apph0}, the second equality holds from the definition of expectation, the third equality holds from Assumption \ref{exogeneity}, the fourth equality holds from the definition of expectation, the fifth equality holds from the Fundamental Theorem of Calculus, and the last equality holds from Fubini's Theorem.
	
	Now, define
	\begin{equation}
		\lambda(z,\eta) := \ffrac{\frac{\partial h_1(z,\eta)}{\partial z} \int_{z_0}^z(\zeta - \mathbb{E}[Z]) f_{Z\mid S}(\zeta) d\zeta}{\mathbb{E}\left[\left.\int_{\mathcal{Z}}  \frac{\partial h_1(\omega,\eta)}{\partial z} \int_{z_0}^\omega(\zeta - \mathbb{E}[Z]) f_{Z\mid S}(\zeta) d\zeta d\omega\,\,\right\vert S = \Tilde{s}\right]}
	\end{equation}
	
	Finally, we have that
	\begin{equation}
		\beta^{\text{2SLS}} = \mathbb{E}\left[\left.\int_{\mathcal{Z}} \frac{\partial h_2(h_1(\omega,\eta),\varepsilon)}{\partial x} \lambda(\omega,\eta) d\omega \,\,\right\vert S = \Tilde{s}\right]
	\end{equation}
\end{proof}

\subsection{First-Stage Effect}\label{fseffect}

Here, we semiparametrically identify the first-stage effect, $\sfrac{\partial h_1(z,d,\eta)}{\partial z}$. This function is a key component of the weights in Proposition \ref{thm2sls}.

\begin{proposition}\label{fseffectprop}
	Under Assumptions \ref{exogeneity}, \ref{increasing}, \ref{qrmodel}, and normalizing $\eta \sim Unif(0,1)$, we have that
	$$
	\frac{\partial K_1(z,d)}{\partial z}'\pi_1(v) = \frac{\partial h_1(z,d,v)}{\partial z}.
	$$
\end{proposition}

\begin{proof}
	First, we show that Assumption \ref{exogeneity} implies that $\eta \independent Z \mid D,S$.
	\begin{equation}
		\begin{split}
			F_{\eta,Z\mid D,S}(\overline{\eta},\overline{z}\mid\overline{d},\overline{s}) & = \mathbb{E}[\mathds{1}\{\eta \leq \overline{\eta}\}\mathds{1}\{Z \leq \overline{z}\}\mid D = \overline{d}, S = \overline{s}]\\
			& = \mathbb{E}[\mathds{1}\{\eta \leq \overline{\eta}\}\mathds{1}\{h_0(W,\overline{s}) \leq \overline{z}\}\mid D = \overline{d}, S = \overline{s}]\\
			& = \mathbb{E}[\mathds{1}\{\eta \leq \overline{\eta}\}\mathbb{E}[\mathds{1}\{h_0(W,\overline{s}) \leq \overline{z}\}\mid \eta, D = \overline{d}, S = \overline{s}]\mid D = \overline{d}, S = \overline{s}]\\
			& = \mathbb{E}[\mathds{1}\{\eta \leq \overline{\eta}\}\mathbb{E}[\mathds{1}\{h_0(W,\overline{s}) \leq \overline{z}\}\mid D = \overline{d}, S = \overline{s}]\mid D = \overline{d}, S = \overline{s}]\\
			& = \mathbb{E}[\mathds{1}\{\eta \leq \overline{\eta}\}\mid D = \overline{d}, S = \overline{s}]\mathbb{E}[\mathds{1}\{Z \leq \overline{z}\}\mid D = \overline{d}, S = \overline{s}]
		\end{split}
	\end{equation}
	
	\noindent where the first equality holds from the definition of a cumulative distribution function, the second equality holds from the definition of $Z$, the third equality holds from applying the Law of Iterated Expectations, the fourth equality holds from Assumption \ref{exogeneity}, and the last equality holds from rearranging terms. 
	
	Here, we slightly modify Assumption \ref{qrmodel}, so that
	\begin{equation}\label{appXZ}
		X = Q_X(V \mid Z, D, S) = K_1(Z,D)'\pi_1(V).
	\end{equation}
	
	From Assumption \ref{increasing}, quantile equivariance leads us to
	\begin{equation}
		Q_X(v\mid z,d,s) = h_1(z,d,Q_\eta(v\mid s)).
	\end{equation}
	
	By normalizing $\eta \sim Unif(0,1)$, we get that
	\begin{equation}
		Q_X(v\mid z,d,s) = h_1(z,d,v).
	\end{equation}
	
	From equation \eqref{appXZ}, we get that
	\begin{equation}
		\frac{\partial Q_X(v\mid z,d,s)}{\partial z} = \frac{\partial K_1(z,d)}{\partial z}'\pi_1(v) = \frac{\partial h_1(z,d,v)}{\partial z}.
	\end{equation}
	
\end{proof}

\pagebreak

\section{Details about the Dataset} \label{dataappendix}

\setcounter{table}{0}
\renewcommand\thetable{B.\arabic{table}}

\setcounter{figure}{0}
\renewcommand\thefigure{B.\arabic{figure}}

\setcounter{equation}{0}
\renewcommand\theequation{B.\arabic{equation}}

\setcounter{theorem}{0}
\renewcommand\thetheorem{B.\arabic{theorem}}

\setcounter{proposition}{0}
\renewcommand\theproposition{B.\arabic{proposition}}

\setcounter{corollary}{0}
\renewcommand\thecorollary{B.\arabic{corollary}}

\setcounter{assumption}{0}
\renewcommand\theassumption{B.\arabic{assumption}}

\setcounter{definition}{0}
\renewcommand\thedefinition{B.\arabic{definition}}

\setcounter{Lemma}{0}
\renewcommand\theLemma{B.\arabic{Lemma}}

In this section, we present descriptive statistics about the data used by \cite{autor2013}. In particular, we focus on the treatment variable $X$ of our empirical application: Exposure to Growth in Chinese Imports in each American commuting zone.

\begin{figure}[!hbt]
	\centering
	\caption{Distribution of Exposure to Growth in Chinese Imports}
	\begin{subfigure}{0.49\textwidth}
		\includegraphics[width=\textwidth]{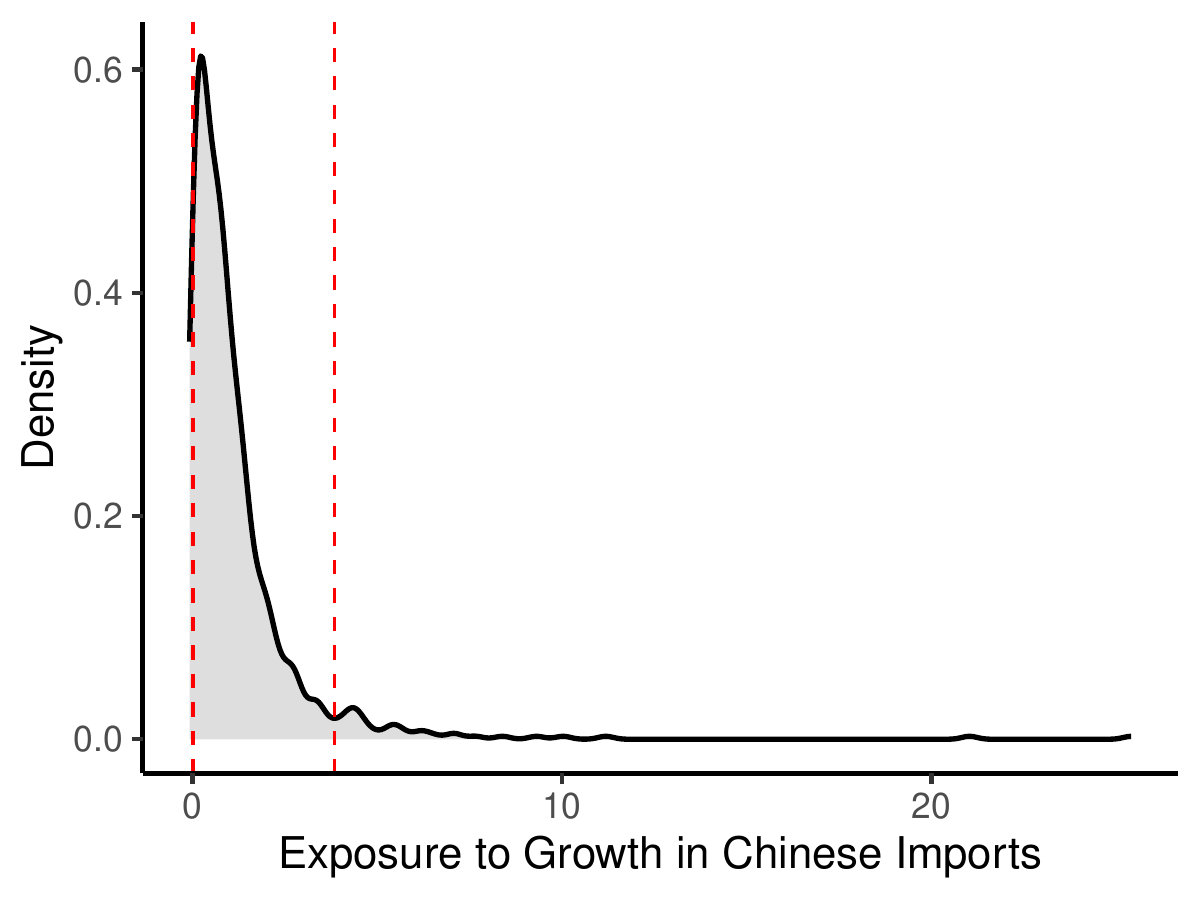}
		\caption{1990-2000}
	\end{subfigure}
	\begin{subfigure}{0.49\textwidth}
		\includegraphics[width=\textwidth]{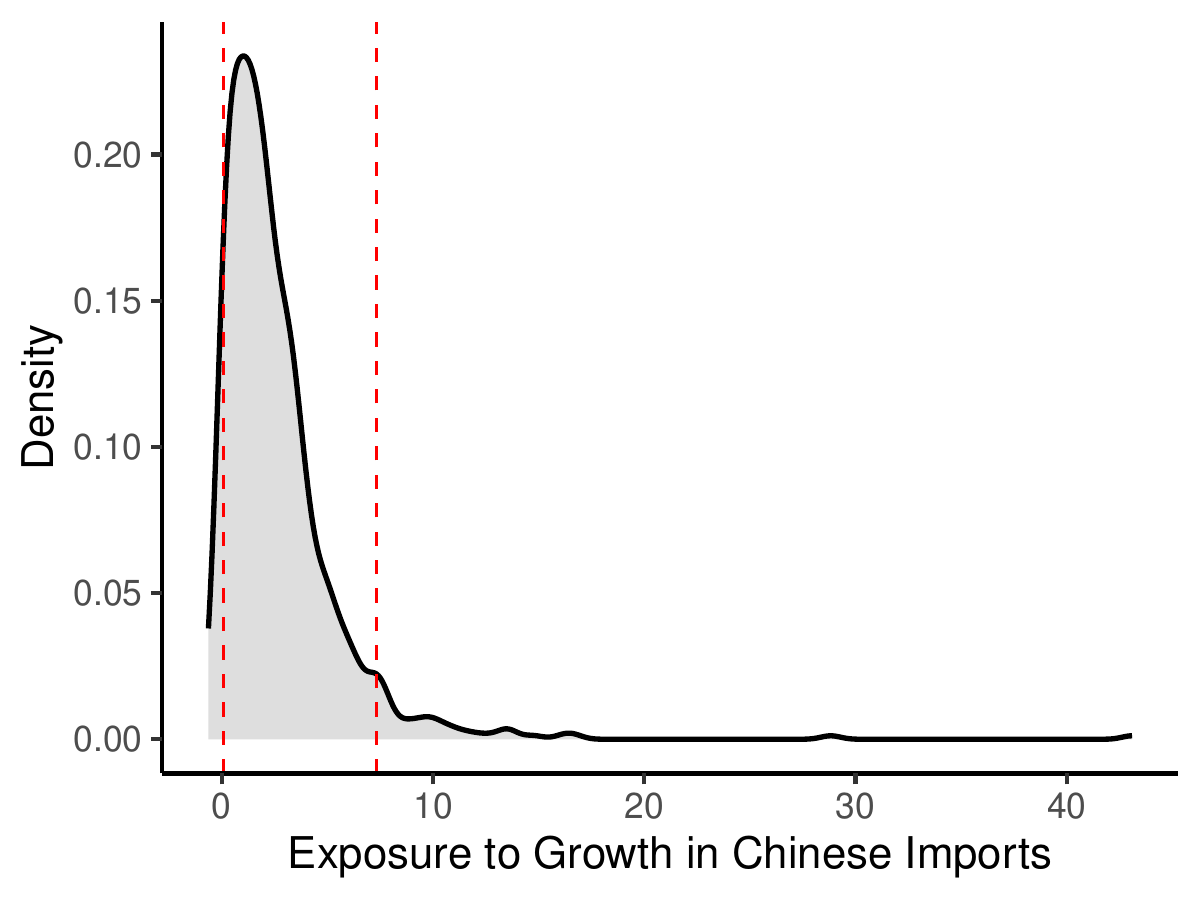}
		\caption{2000-2007}
	\end{subfigure}
	\caption*{\textit{Note:} Figure \ref{fig:dist} plots the empirical distribution of the Exposure to Growth in Chinese Imports ($X$) for the periods 1990-2000 and 2000-2007. The dashed lines in red represent the 5th and 95th quantiles of this distribution.}
	\label{fig:dist}
\end{figure}

Figure \ref{fig:dist} plots the empirical distribution of the treatment variable for the periods 1990-2000 and 2000-2007. We observe that the maximum of this distribution almost doubles over time. Moreover, the distribution for the period 1990-2000 is highly concentrated on values smaller than 4, while the distribution for the period 2000-2007 is concentrated on values smaller than 10. This result suggests that nonlinearities in treatment effects might be more relevant in the later period.

\begin{figure}[!hbt]
	\centering
	\caption{Exposure to Growth in Chinese Imports by commuting zone}
	\begin{subfigure}{0.49\textwidth}
		\includegraphics[width=\textwidth]{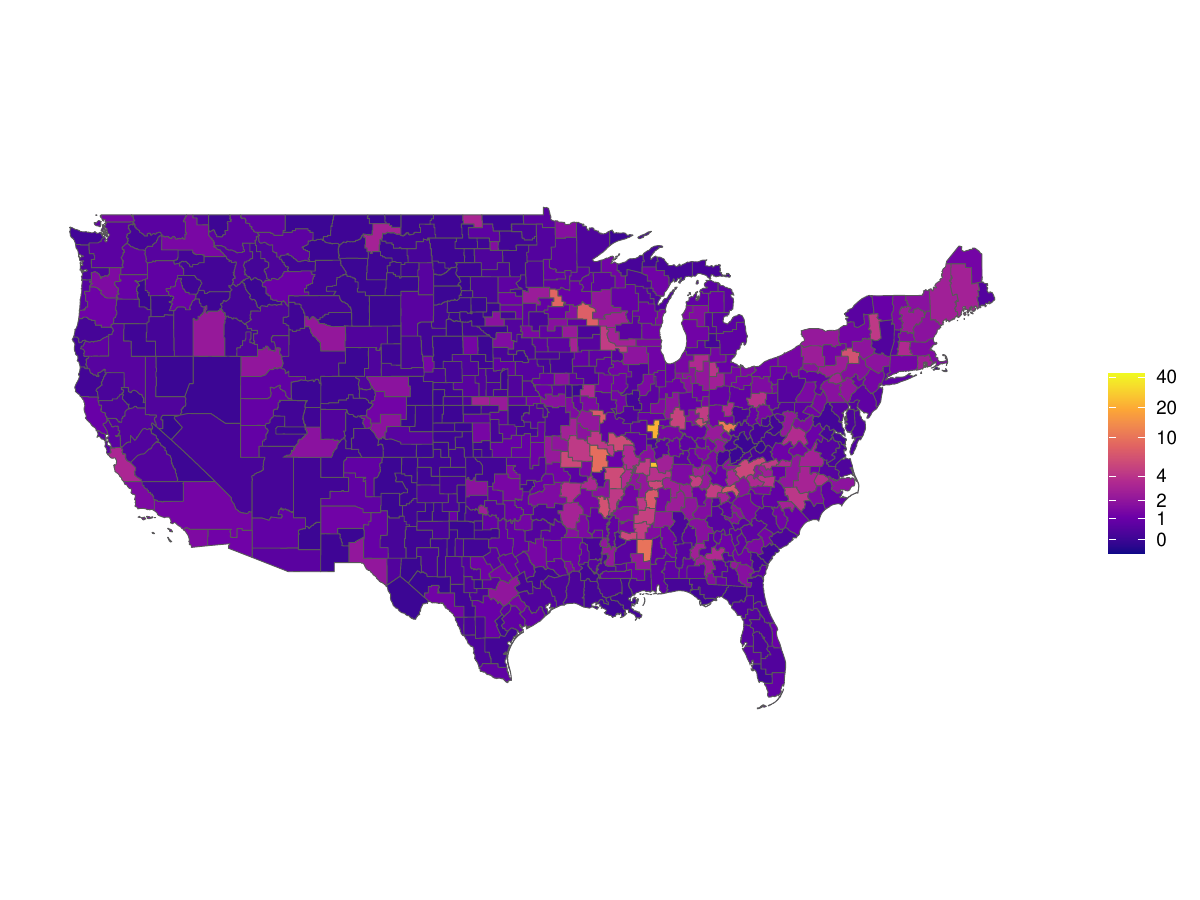}
		\caption{1990-2000}
	\end{subfigure}
	\begin{subfigure}{0.49\textwidth}
		\includegraphics[width=\textwidth]{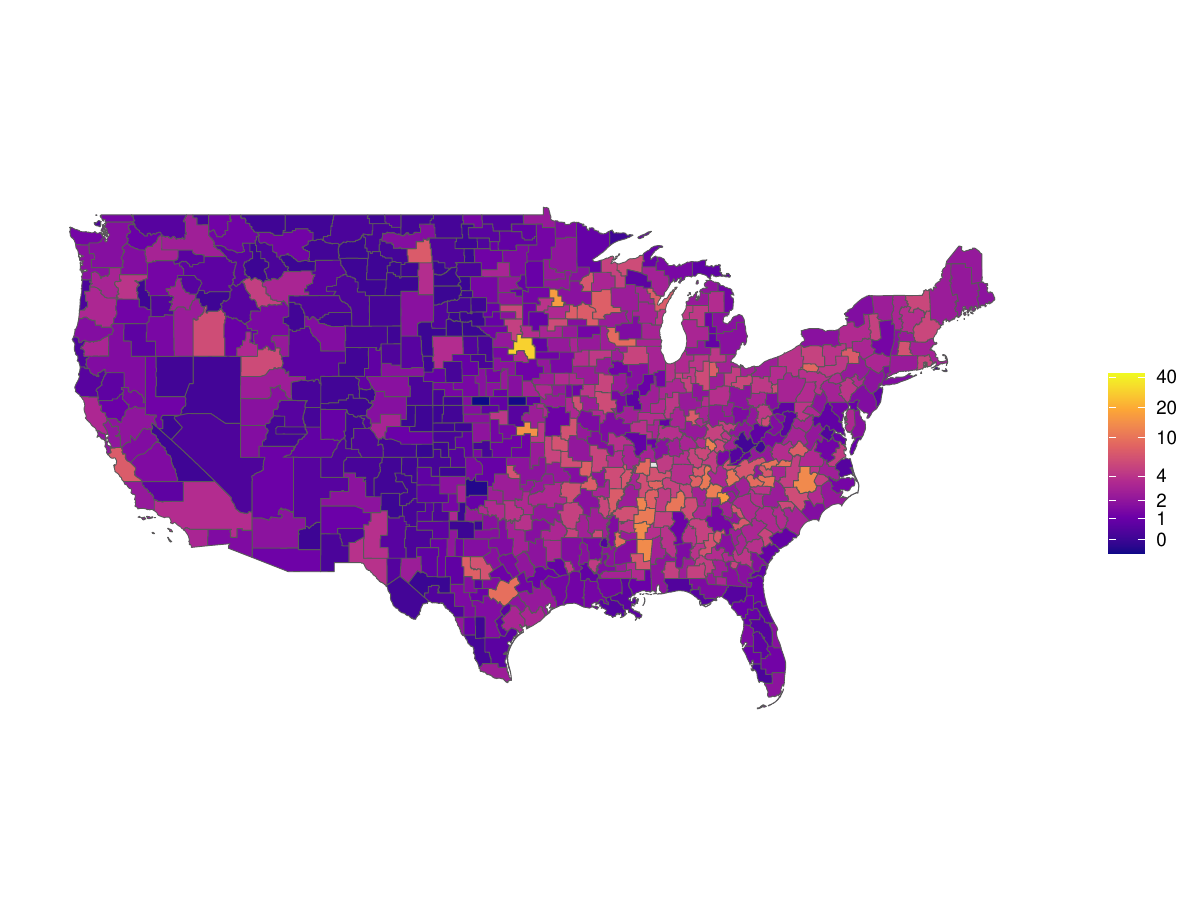}
		\caption{2000-2007}
	\end{subfigure}
	\caption*{\textit{Note:} Figure \ref{fig:map} plots the Exposure to Growth in Chinese Imports ($X$) by commuting zone in the mainland U.S. for the periods of 1990-2000 and 2000-2007.}
	\label{fig:map}
\end{figure}

Figure \ref{fig:map} plots the exposure to growth in Chinese imports by commuting zone in the mainland U.S. for the periods 1990-2000 (Panel (a)) and 2000-2007 (Panel (b)). We observe a substantial growth in exposure for commuting zones in the Rust Belt and the Southern States. These geographical disparities underscore the importance of considering nonlinear, heterogeneous treatment effects, as some regions may be considerably more affected by the China shock.

\pagebreak

\section{Robustness to the Spline Specification} \label{AppRobustness}

\setcounter{table}{0}
\renewcommand\thetable{C.\arabic{table}}

\setcounter{figure}{0}
\renewcommand\thefigure{C.\arabic{figure}}

\setcounter{equation}{0}
\renewcommand\theequation{C.\arabic{equation}}

\setcounter{theorem}{0}
\renewcommand\thetheorem{C.\arabic{theorem}}

\setcounter{proposition}{0}
\renewcommand\theproposition{C.\arabic{proposition}}

\setcounter{corollary}{0}
\renewcommand\thecorollary{C.\arabic{corollary}}

\setcounter{assumption}{0}
\renewcommand\theassumption{C.\arabic{assumption}}

\setcounter{definition}{0}
\renewcommand\thedefinition{C.\arabic{definition}}

\setcounter{Lemma}{0}
\renewcommand\theLemma{C.\arabic{Lemma}}

In this section, we present the AD, LAR, ASF, and Policy Effects for six different specifications of the splines. Our six specifications are: Degree of 2 and 3 knots; degree of 2 and 4 knots; degree of 2 and 5 knots; degree of 3 and 3 knots; degree of 3 and 4 knots (the same as in the main text), and degree of 3 and 5 knots. Section \ref{sens:ae} presents the robustness analysis for the AD and the LAR function, Section \ref{sens:asf} presents the robustness analysis for the ASF, and Section \ref{sens:pe} presents the robustness analysis for the Policy Effect.

\subsection{Robustness of Average Effects}\label{sens:ae}

\begin{figure}[!hbtp]
	\centering
	\caption{Local Average Effects with degree of 2 and 3 knots}
	\begin{subfigure}{0.49\textwidth}
		\includegraphics[width=\textwidth]{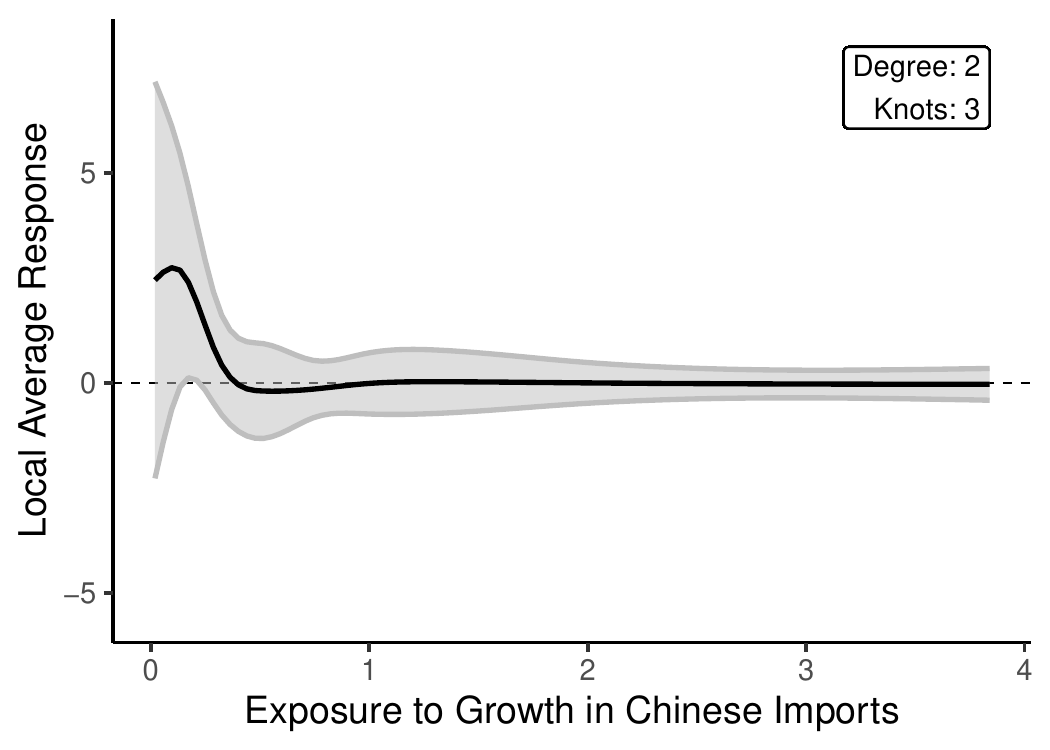}
		\caption{LAR for 1990-2000}
	\end{subfigure}
	\begin{subfigure}{0.49\textwidth}
		\includegraphics[width=\textwidth]{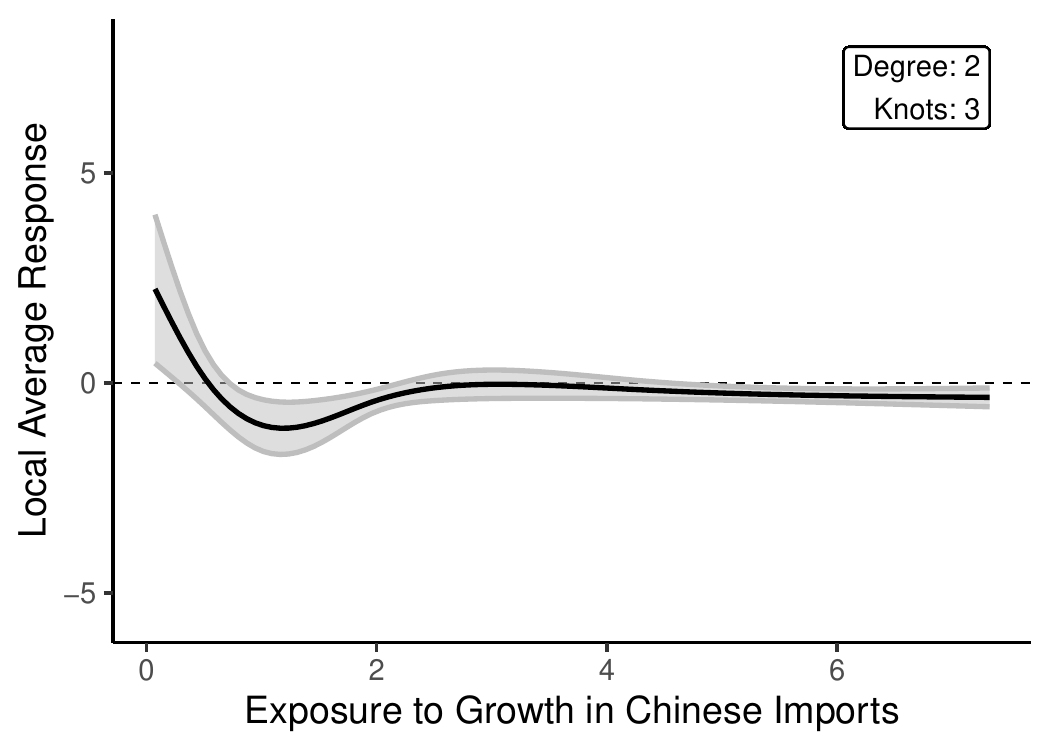}
		\caption{LAR for 2000-2007}
	\end{subfigure}
	\caption*{\textit{Note:} This figure plots the estimates for the Local Average Effect for the periods of 1990-2000 and 2000-2007. Both estimates were produced according to the specification in Section \ref{spec}, but using a degree of 2 and 3 knots in the specification of the spline. The values displayed on the x-axes represent the 5th to the 95th quantiles of the empirical distribution of the Exposure to Growth in Chinese Imports. }
	\label{fig:lar23}
\end{figure}

\begin{figure}[!hbtp]
	\centering
	\caption{Local Average Effects with degree of 2 and 4 knots}
	\begin{subfigure}{0.49\textwidth}
		\includegraphics[width=\textwidth]{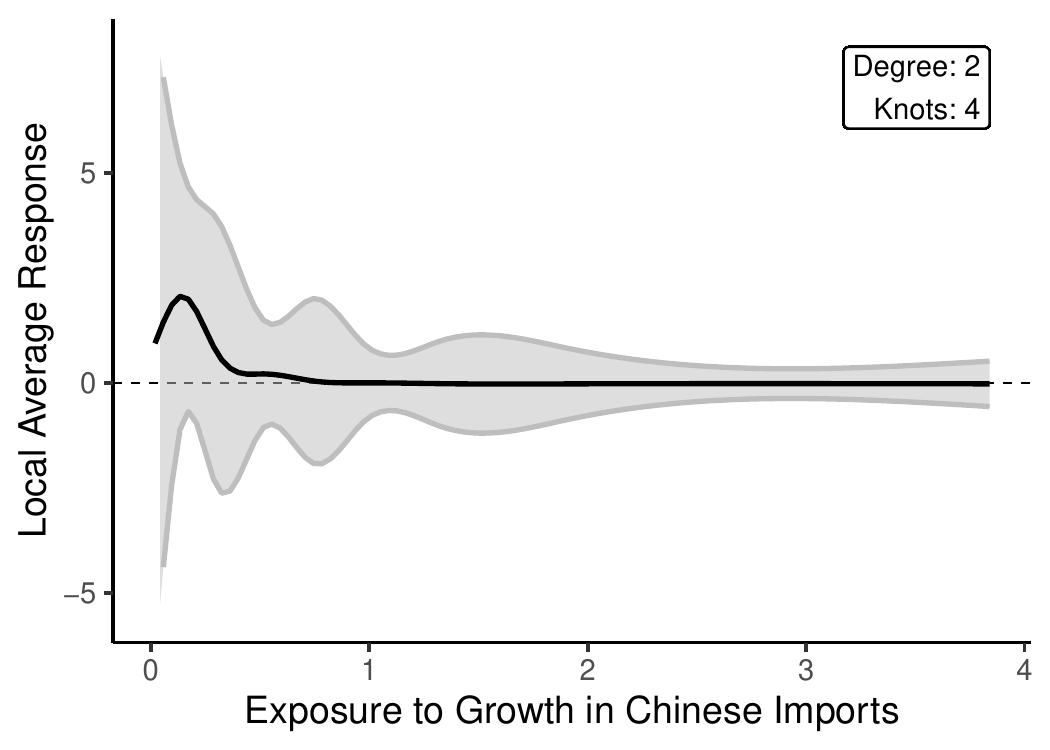}
		\caption{LAR for 1990-2000}
	\end{subfigure}
	\begin{subfigure}{0.49\textwidth}
		\includegraphics[width=\textwidth]{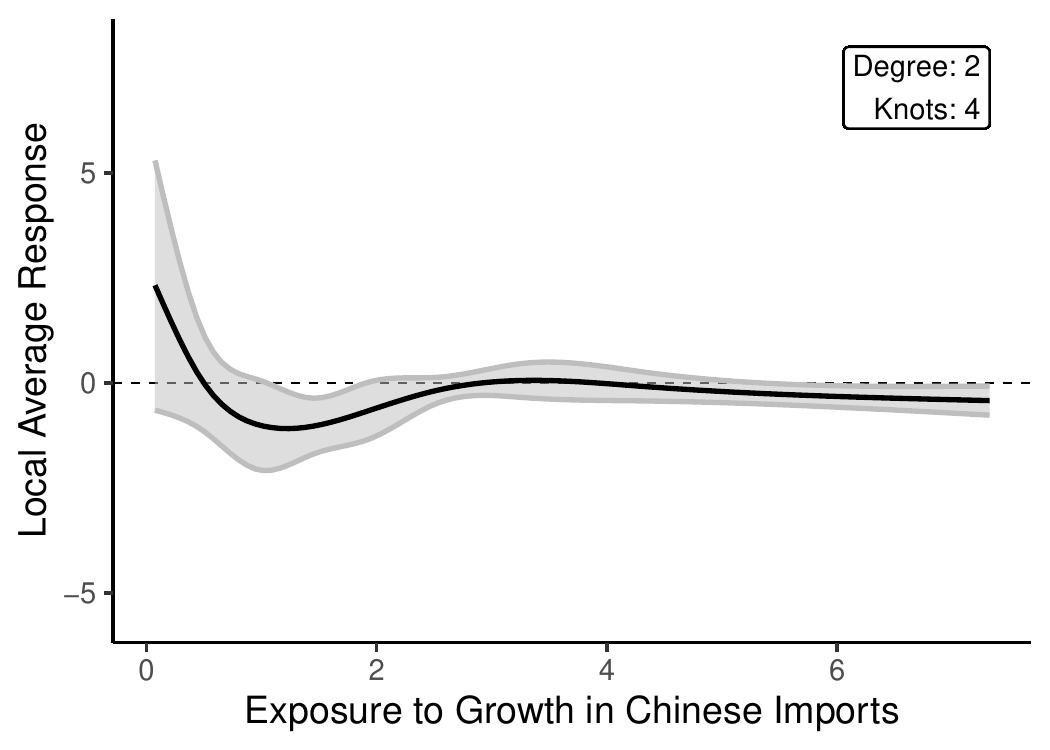}
		\caption{LAR for 2000-2007}
	\end{subfigure}
	\caption*{\textit{Note:} This figure plots the estimates for the Local Average Effect for the periods of 1990-2000 and 2000-2007. Both estimates were produced according to the specification in Section \ref{spec}, but using a degree of 2 and 4 knots in the specification of the spline. The values displayed on the x-axes represent the 5th to the 95th quantiles of the empirical distribution of the Exposure to Growth in Chinese Imports. }
	\label{fig:lar24}
\end{figure}

\begin{figure}[!hbtp]
	\centering
	\caption{Local Average Effects with degree of 2 and 5 knots}
	\begin{subfigure}{0.49\textwidth}
		\includegraphics[width=\textwidth]{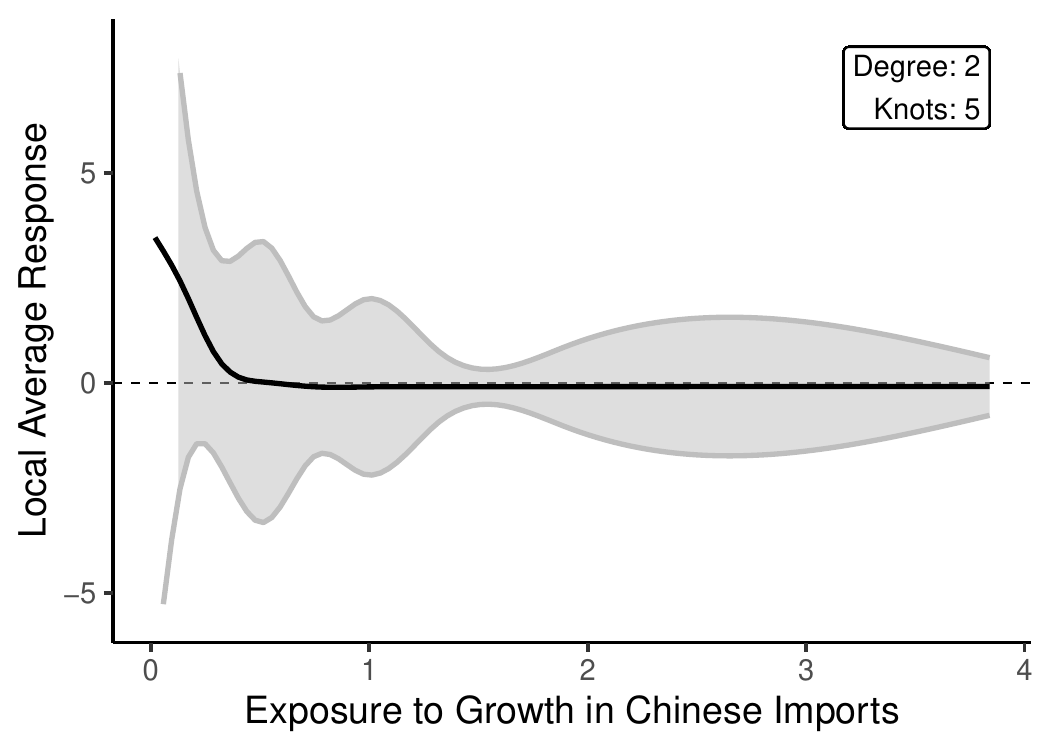}
		\caption{LAR for 1990-2000}
	\end{subfigure}
	\begin{subfigure}{0.49\textwidth}
		\includegraphics[width=\textwidth]{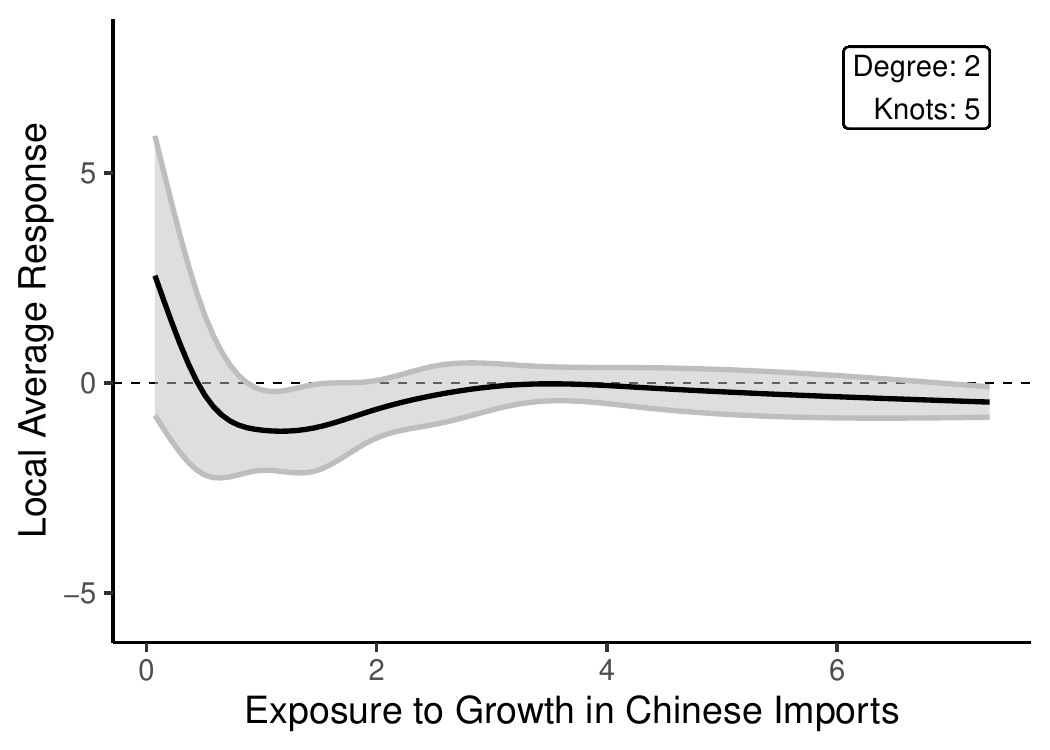}
		\caption{LAR for 2000-2007}
	\end{subfigure}
	\caption*{\textit{Note:} This figure plots the estimates for the Local Average Effect for the periods of 1990-2000 and 2000-2007. Both estimates were produced according to the specification in Section \ref{spec}, but using a degree of 2 and 5 knots in the specification of the spline. The values displayed on the x-axes represent the 5th to the 95th quantiles of the empirical distribution of the Exposure to Growth in Chinese Imports. }
	\label{fig:lar25}
\end{figure}

\begin{figure}[!hbtp]
	\centering
	\caption{Local Average Effects with degree of 3 and 3 knots}
	\begin{subfigure}{0.49\textwidth}
		\includegraphics[width=\textwidth]{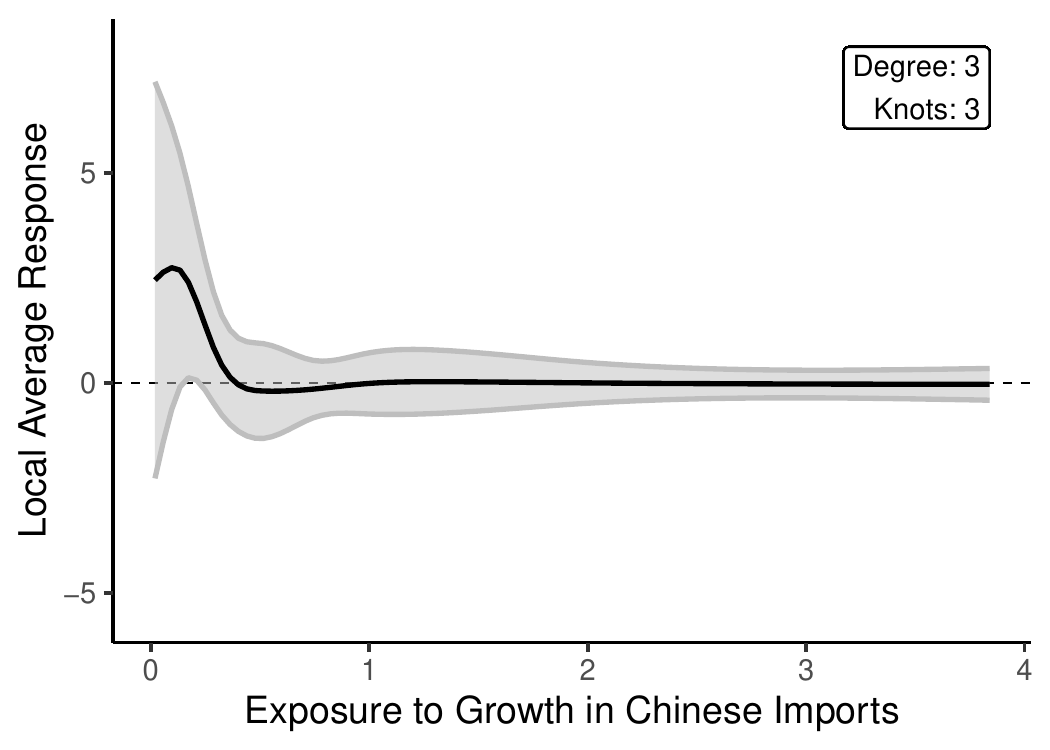}
		\caption{LAR for 1990-2000}
	\end{subfigure}
	\begin{subfigure}{0.49\textwidth}
		\includegraphics[width=\textwidth]{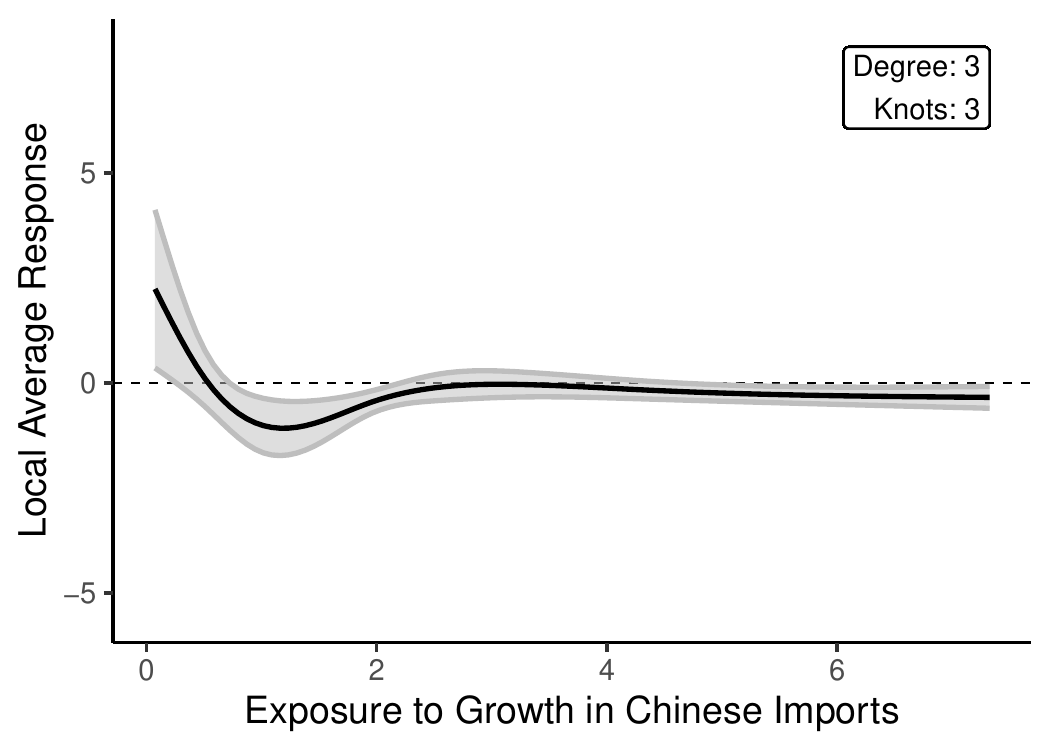}
		\caption{LAR for 2000-2007}
	\end{subfigure}
	\caption*{\textit{Note:} This figure plots the estimates for the Local Average Effect for the periods of 1990-2000 and 2000-2007. Both estimates were produced according to the specification in Section \ref{spec}, but using a degree of 3 and 3 knots in the specification of the spline. The values displayed in the x-axes represents the 5th to the 95th quantiles of the empirical distribution of the Exposure to Growth in Chinese Imports. }
	\label{fig:lar33}
\end{figure}

\begin{figure}[!hbtp]
	\centering
	\caption{Local Average Effects with degree of 3 and 4 knots}
	\begin{subfigure}{0.49\textwidth}
		\includegraphics[width=\textwidth]{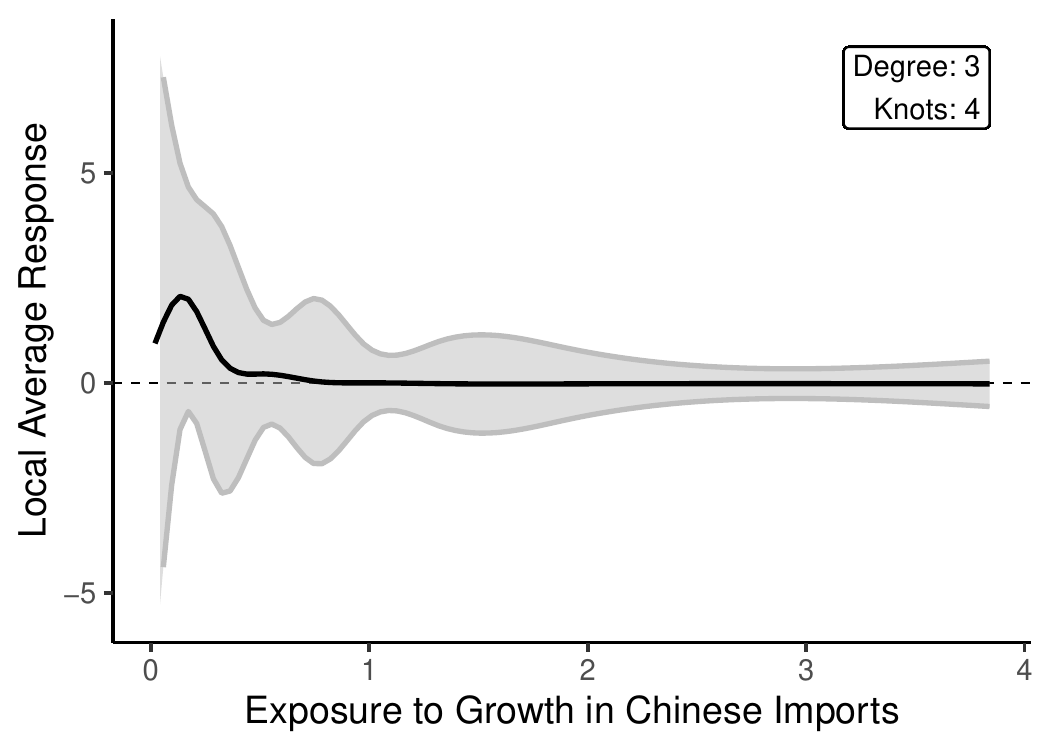}
		\caption{LAR for 1990-2000}
	\end{subfigure}
	\begin{subfigure}{0.49\textwidth}
		\includegraphics[width=\textwidth]{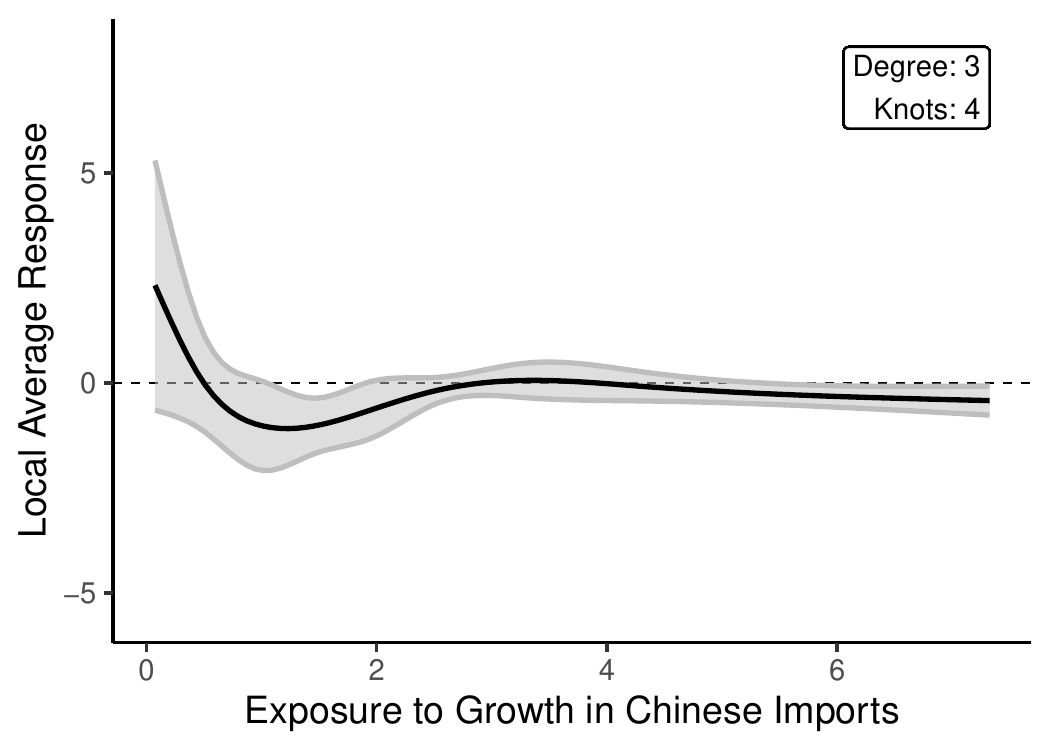}
		\caption{LAR for 2000-2007}
	\end{subfigure}
	\caption*{\textit{Note:} This figure plots the estimates for the Local Average Effect for the periods of 1990-2000 and 2000-2007. Both estimates were produced according to the specification in Section \ref{spec}, but using a degree of 3 and 4 knots in the specification of the spline. The values displayed in the x-axes represents the 5th to the 95th quantiles of the empirical distribution of the Exposure to Growth in Chinese Imports. }
	\label{fig:lar34}
\end{figure}

\begin{figure}[!hbtp]
	\centering
	\caption{Local Average Effects with degree of 3 and 5 knots}
	\begin{subfigure}{0.49\textwidth}
		\includegraphics[width=\textwidth]{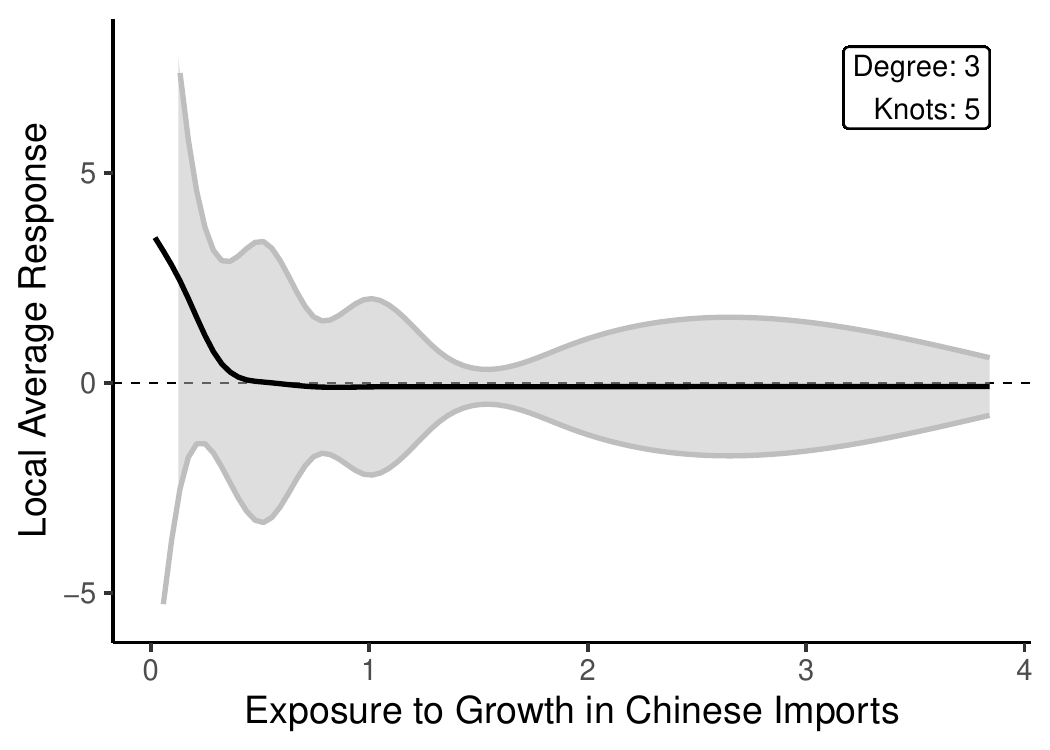}
		\caption{LAR for 1990-2000}
	\end{subfigure}
	\begin{subfigure}{0.49\textwidth}
		\includegraphics[width=\textwidth]{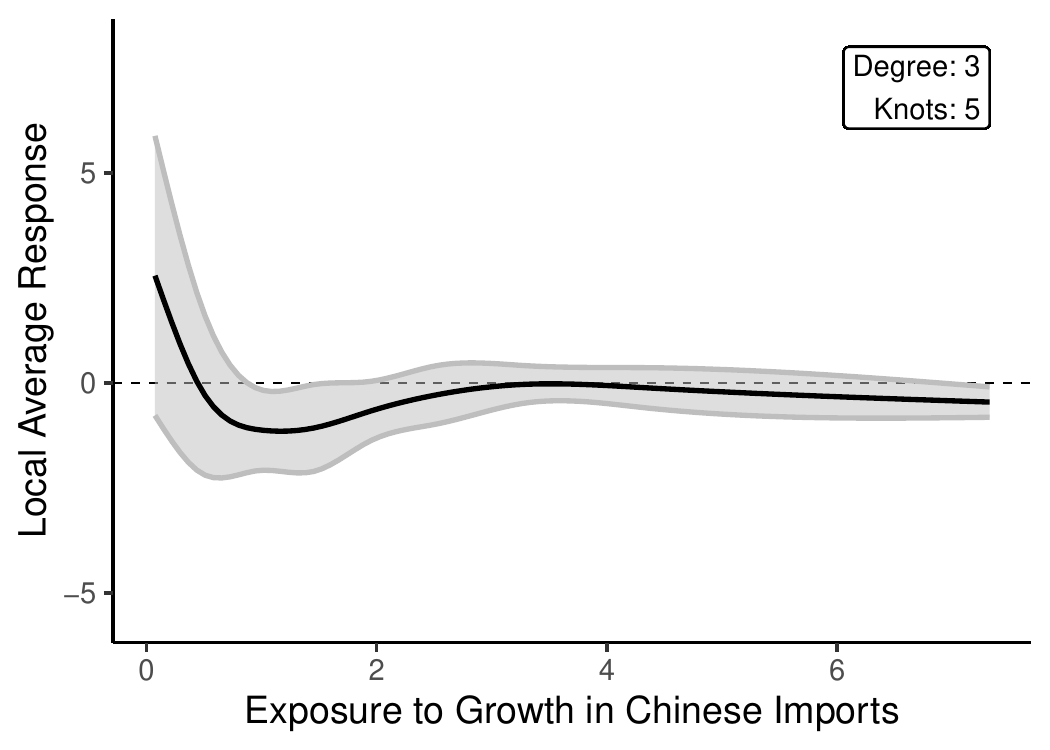}
		\caption{LAR for 2000-2007}
	\end{subfigure}
	\caption*{\textit{Note:} This figure plots the estimates for the Local Average Effect for the periods of 1990-2000 and 2000-2007. Both estimates were produced according to the specification in Section \ref{spec}, but using a degree of 3 and 5 knots in the specification of the spline. The values displayed in the x-axes represents the 5th to the 95th quantiles of the empirical distribution of the Exposure to Growth in Chinese Imports. }
	\label{fig:lar35}
\end{figure}

In this section, we present the AD (Table \ref{tab:ad_sens}) and LAR (Figures \ref{fig:lar23}-\ref{fig:lar35}) estimates for six different specifications.

\begin{table}[!htbp]
	\centering
	\caption{Average Derivative}
	\begin{tabular}{lcc}
		\toprule
		& \multicolumn{2}{c}{Period}\\
		\cmidrule{2-3}
		& 1990-2000 & 2000-2007 \\
		\textit{Specification} & (1) & (2) \\
		\midrule
		Degree of 2 and 3 knots & 0.538 & 0.072 \\
		& [-0.204, 1.281] & [-0.189, 0.333]  \\
		& &  \\  
		Degree of 2 and 4 knots & 0.106 & 0.021 \\
		& [-0.972, 1.185] & [-0.337, 0.378]  \\
		& &  \\  
		Degree of 2 and 5 knots & 0.379 & -0.014 \\
		& [-0.957, 1.715] & [-0.502, 0.475]  \\
		& &  \\  
		Degree of 3 and 3 knots & 0.538 & 0.072 \\
		& [-0.204, 1.281] & [-0.189, 0.333]  \\
		& &  \\  
		Degree of 3 and 4 knots & 0.106 & 0.021 \\
		& [-0.972, 1.185] & [-0.337, 0.378]  \\
		& &  \\  
		Degree of 3 and 5 knots  & 0.379 & -0.014 \\
		& [-0.957, 1.715] & [-0.502, 0.475]  \\
		& &  \\  
		\midrule
		Observations & 722 & 722  \\
		\bottomrule
	\end{tabular}
	\caption*{\textit{Note}: Table \ref{tab:ad_sens} reports results for the Average Derivative for the six specifications. Column (1) reports results for the period of 1990-2000, column (2) reports results for the period 2000-2007, and column (3) reports results pooling both periods. Confidence intervals for the AD estimates are reported between brackets below each estimate.}
	\label{tab:ad_sens}
\end{table}

\subsection{Robustness of the Average Structural Function}\label{sens:asf}

\begin{figure}[!hbtp]
	\centering
	\caption{Average Structural Functions with degree of 2 and 3 knots}
	\begin{subfigure}{0.49\textwidth}
		\includegraphics[width=\textwidth]{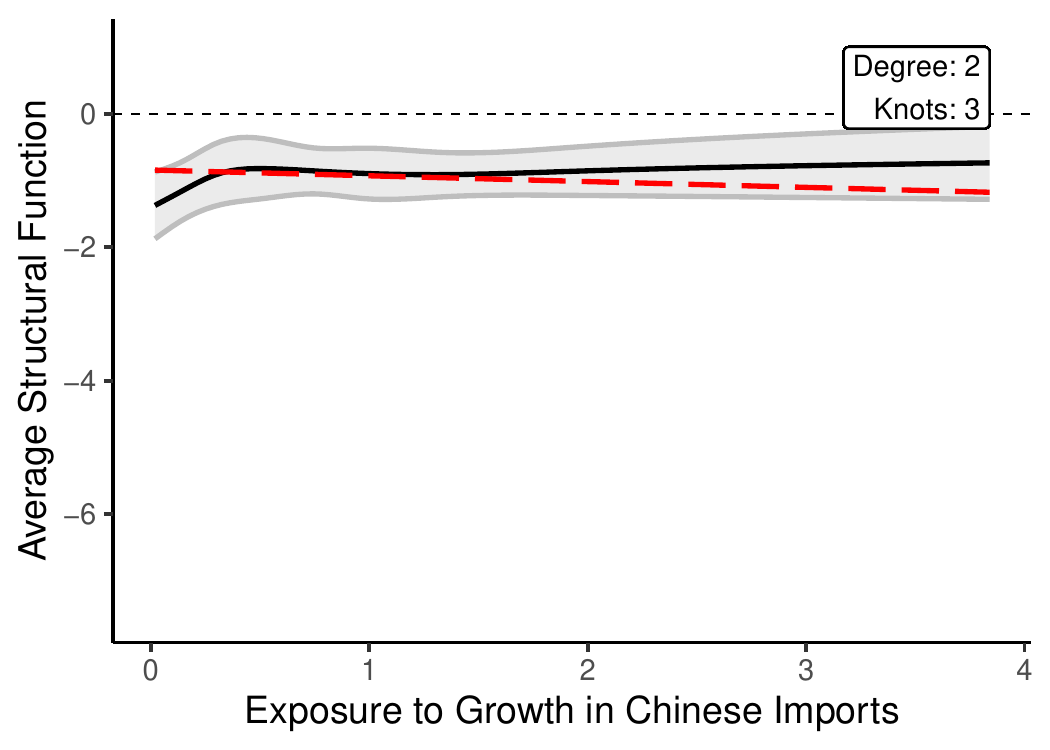}
		\caption{ASF for 1990-2000}
	\end{subfigure}
	\begin{subfigure}{0.49\textwidth}
		\includegraphics[width=\textwidth]{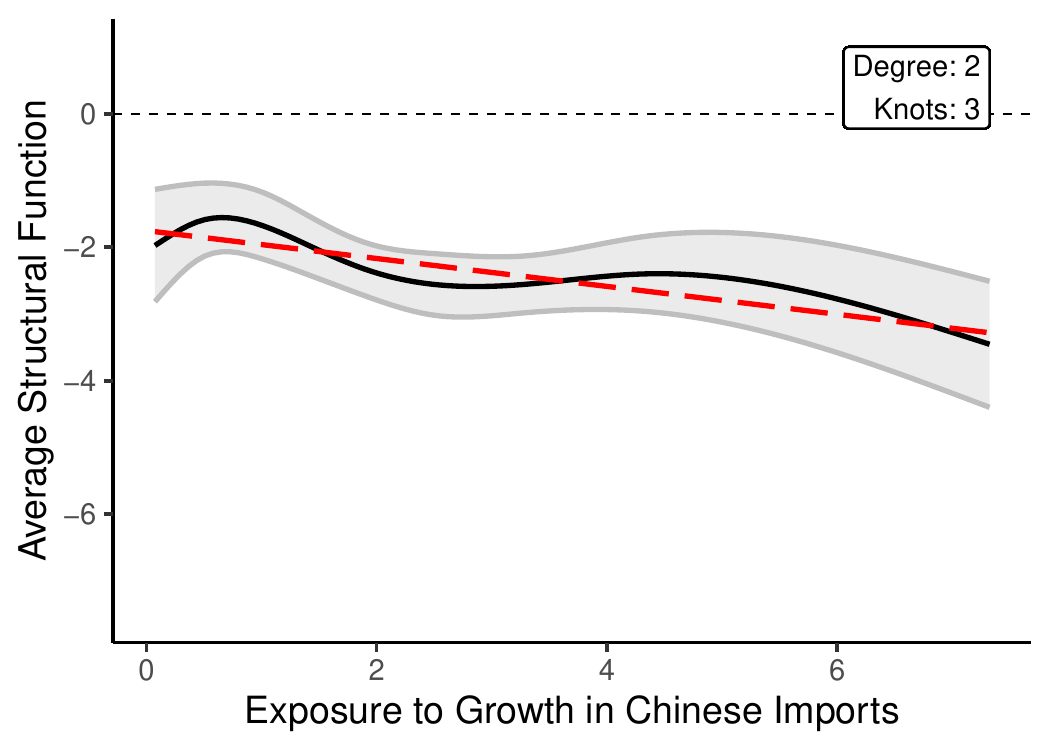}
		\caption{ASF for 2000-2007}
	\end{subfigure}
	\caption*{\textit{Note:} This figure plots the estimates for the Average Structural Function for the periods of 1990-2000 and 2000-2007. Both estimates were produced according to the specification in Section \ref{spec}, but using a degree of 2 and 3 knots in the specification of the spline. The red long-dashed lines represent the estimates from a 2SLS specification. The values displayed in the x-axes represents the 5th to the 95th quantiles of the empirical distribution of the Exposure to Growth in Chinese Imports. }
	\label{fig:asf23}
\end{figure}

\begin{figure}[!hbtp]
	\centering
	\caption{Average Structural Functions with degree of 2 and 4 knots}
	\begin{subfigure}{0.49\textwidth}
		\includegraphics[width=\textwidth]{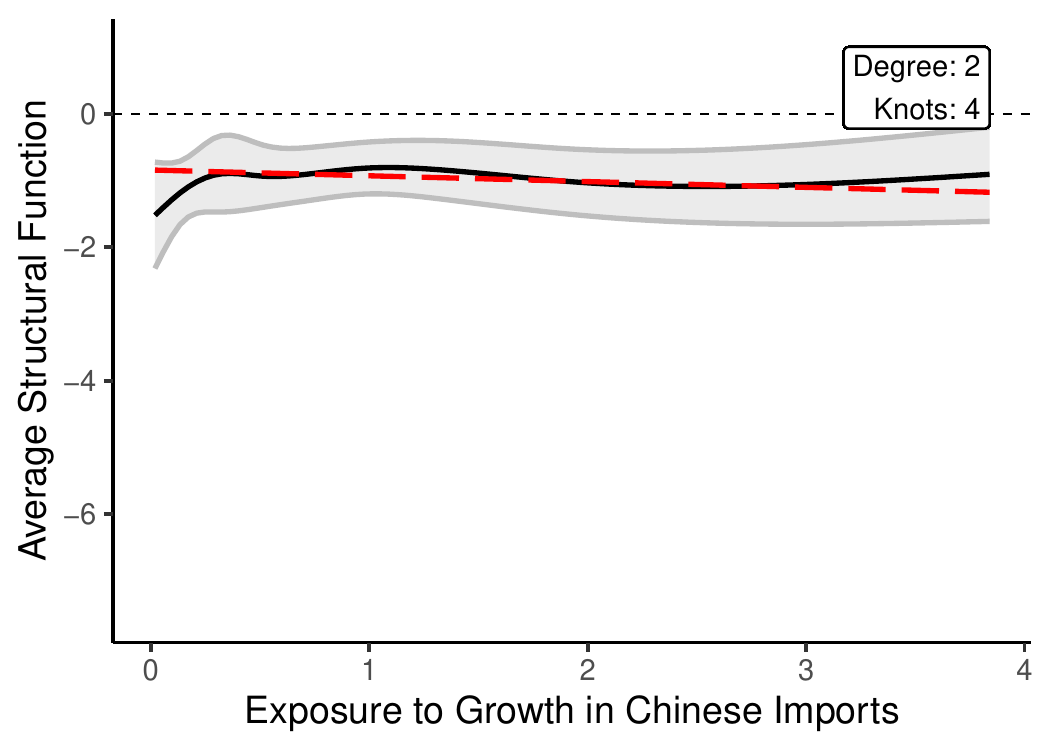}
		\caption{ASF for 1990-2000}
	\end{subfigure}
	\begin{subfigure}{0.49\textwidth}
		\includegraphics[width=\textwidth]{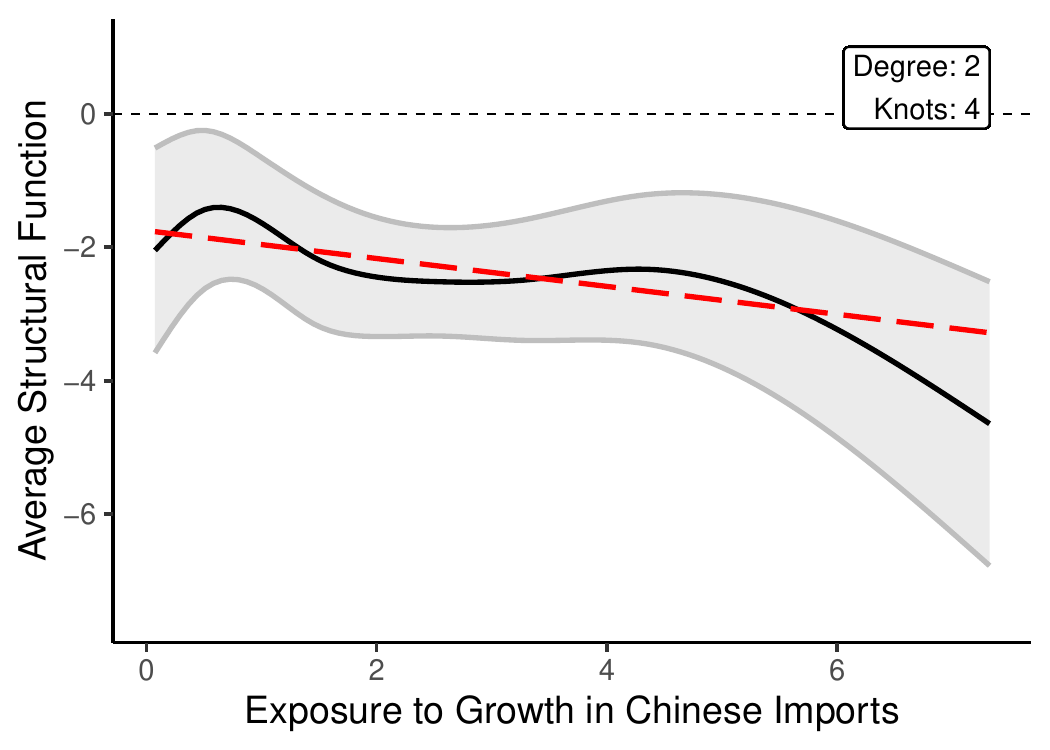}
		\caption{ASF for 2000-2007}
	\end{subfigure}
	\caption*{\textit{Note:} This figure plots the estimates for the Average Structural Function for the periods of 1990-2000 and 2000-2007. Both estimates were produced according to the specification in Section \ref{spec}, but using a degree of 2 and 4 knots in the specification of the spline. The red long-dashed lines represent the estimates from a 2SLS specification. The values displayed in the x-axes represents the 5th to the 95th quantiles of the empirical distribution of the Exposure to Growth in Chinese Imports. }
	\label{fig:asf24}
\end{figure}

\begin{figure}[!hbtp]
	\centering
	\caption{Average Structural Functions with degree of 2 and 5 knots}
	\begin{subfigure}{0.49\textwidth}
		\includegraphics[width=\textwidth]{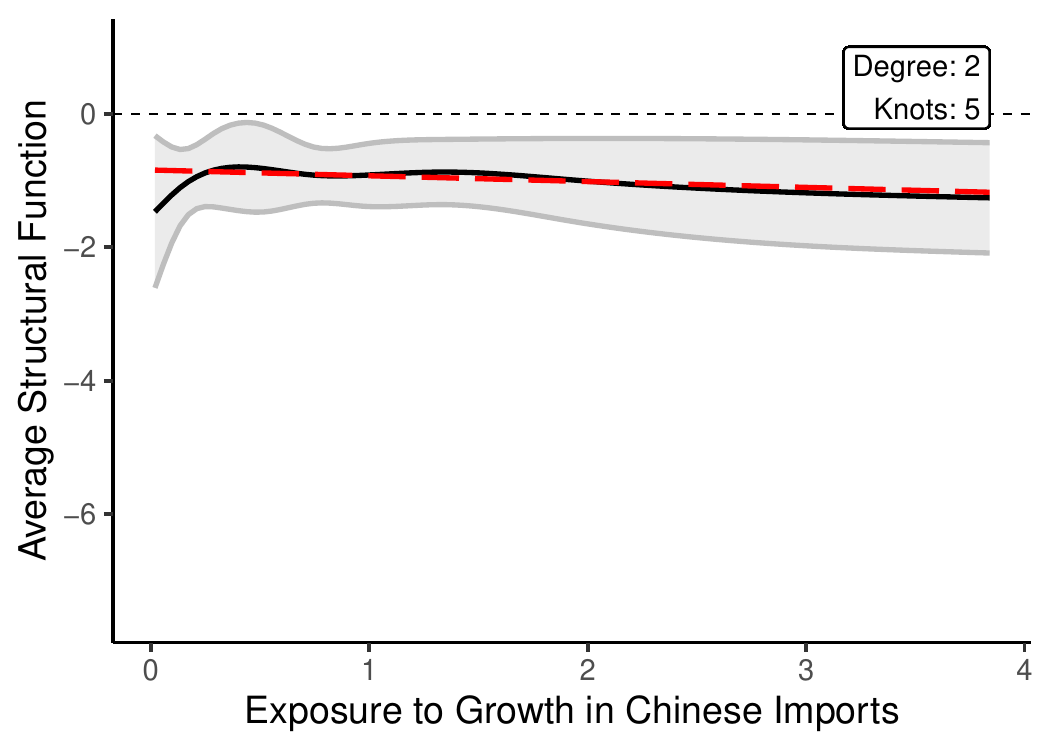}
		\caption{ASF for 1990-2000}
	\end{subfigure}
	\begin{subfigure}{0.49\textwidth}
		\includegraphics[width=\textwidth]{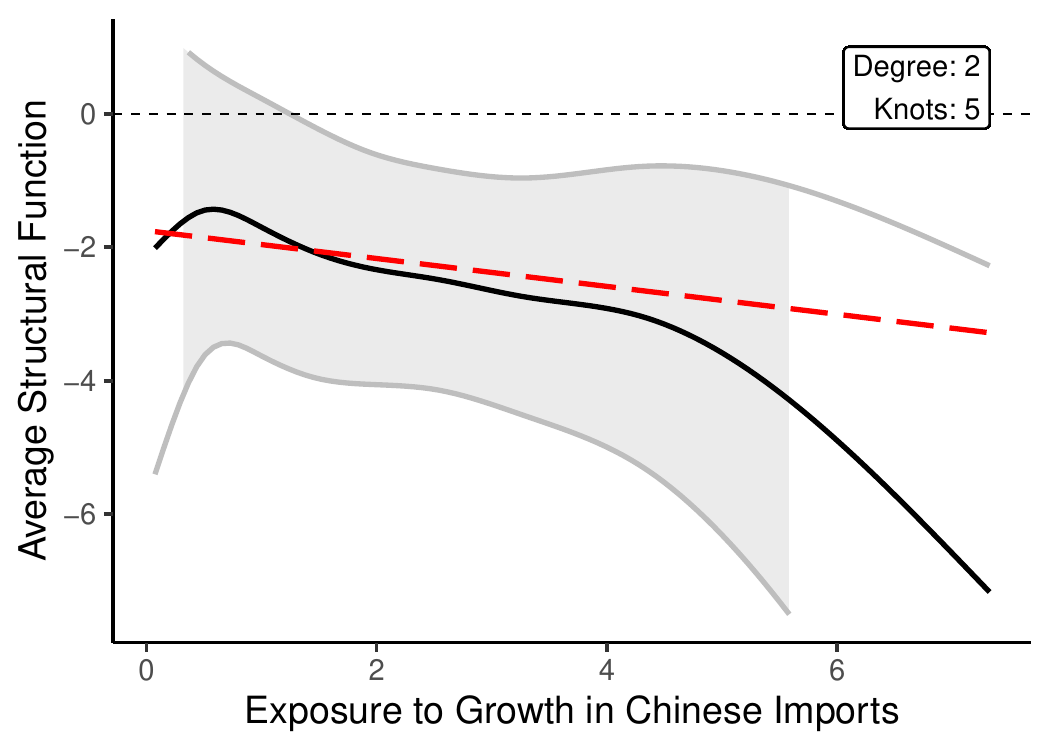}
		\caption{ASF for 2000-2007}
	\end{subfigure}
	\caption*{\textit{Note:} This figure plots the estimates for the Average Structural Function for the periods of 1990-2000 and 2000-2007. Both estimates were produced according to the specification in Section \ref{spec}, but using a degree of 2 and 5 knots in the specification of the spline. The red long-dashed lines represent the estimates from a 2SLS specification. The values displayed in the x-axes represents the 5th to the 95th quantiles of the empirical distribution of the Exposure to Growth in Chinese Imports. }
	\label{fig:asf25}
\end{figure}

\begin{figure}[!hbtp]
	\centering
	\caption{Average Structural Functions with degree of 3 and 3 knots}
	\begin{subfigure}{0.49\textwidth}
		\includegraphics[width=\textwidth]{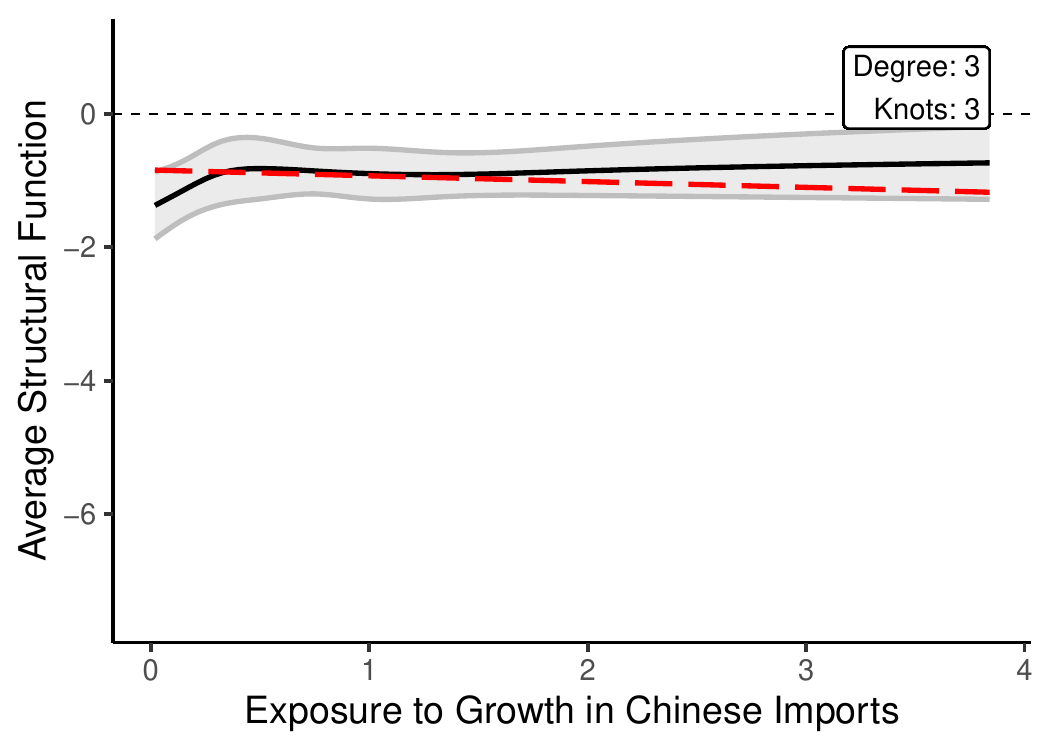}
		\caption{ASF for 1990-2000}
	\end{subfigure}
	\begin{subfigure}{0.49\textwidth}
		\includegraphics[width=\textwidth]{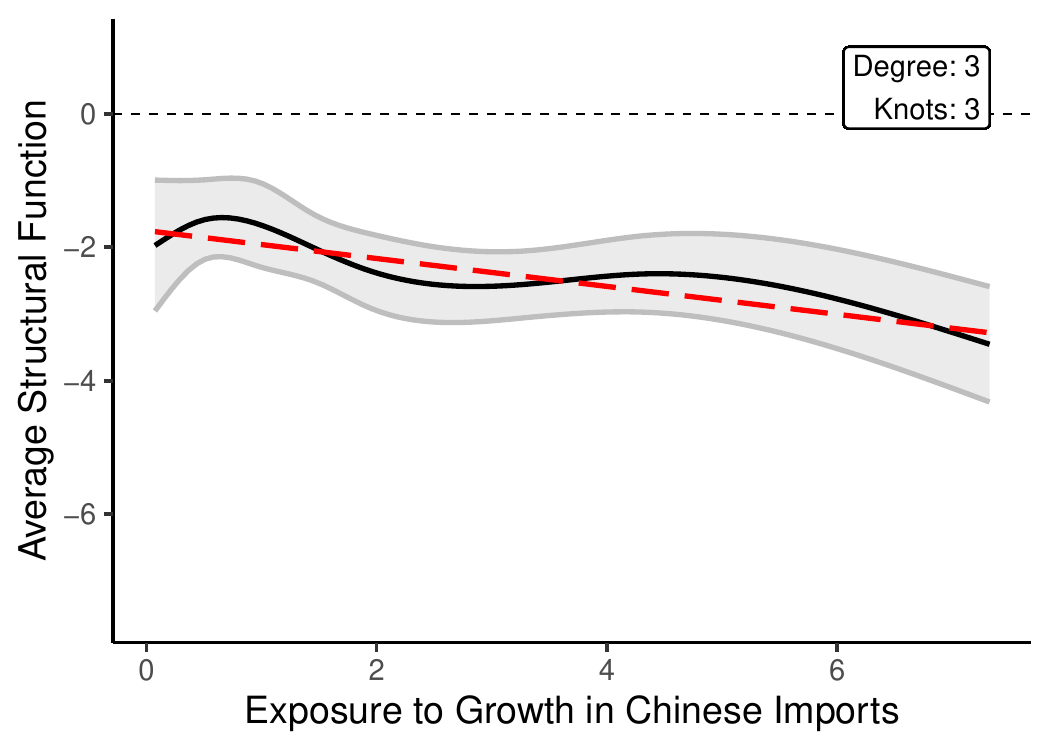}
		\caption{ASF for 2000-2007}
	\end{subfigure}
	\caption*{\textit{Note:} This figure plots the estimates for the Average Structural Function for the periods of 1990-2000 and 2000-2007. Both estimates were produced according to the specification in Section \ref{spec}, but using a degree of 3 and 3 knots in the specification of the spline. The red long-dashed lines represent the estimates from a 2SLS specification. The values displayed in the x-axes represents the 5th to the 95th quantiles of the empirical distribution of the Exposure to Growth in Chinese Imports. }
	\label{fig:asf33}
\end{figure}

\begin{figure}[!hbtp]
	\centering
	\caption{Average Structural Functions with degree of 3 and 4 knots}
	\begin{subfigure}{0.49\textwidth}
		\includegraphics[width=\textwidth]{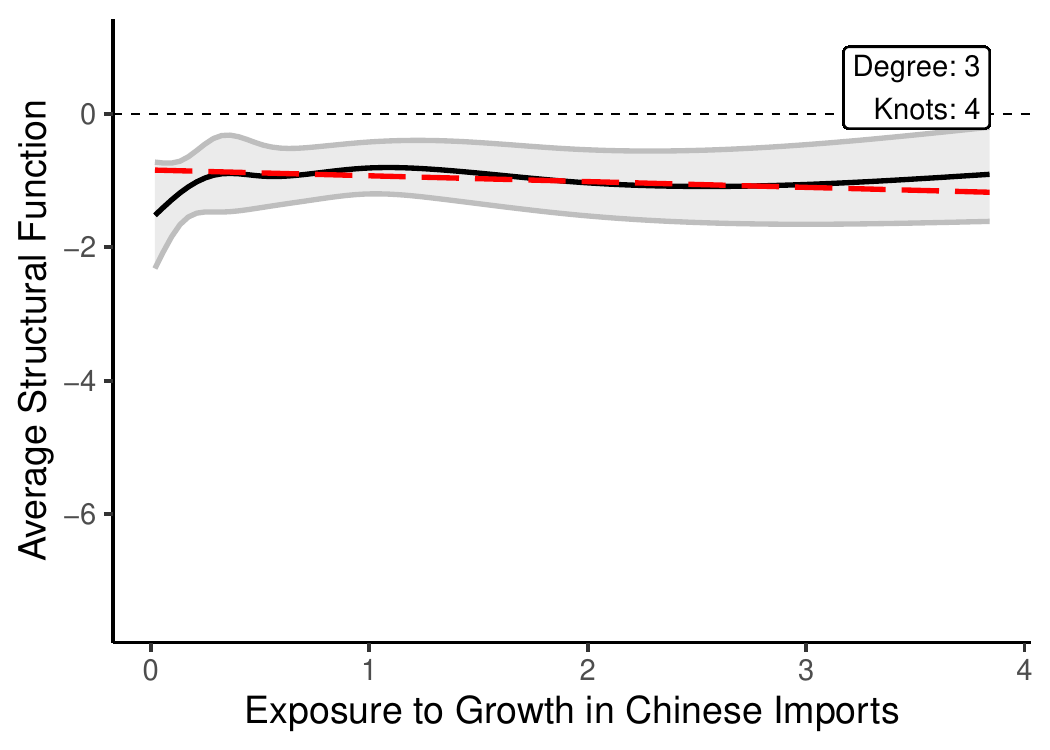}
		\caption{ASF for 1990-2000}
	\end{subfigure}
	\begin{subfigure}{0.49\textwidth}
		\includegraphics[width=\textwidth]{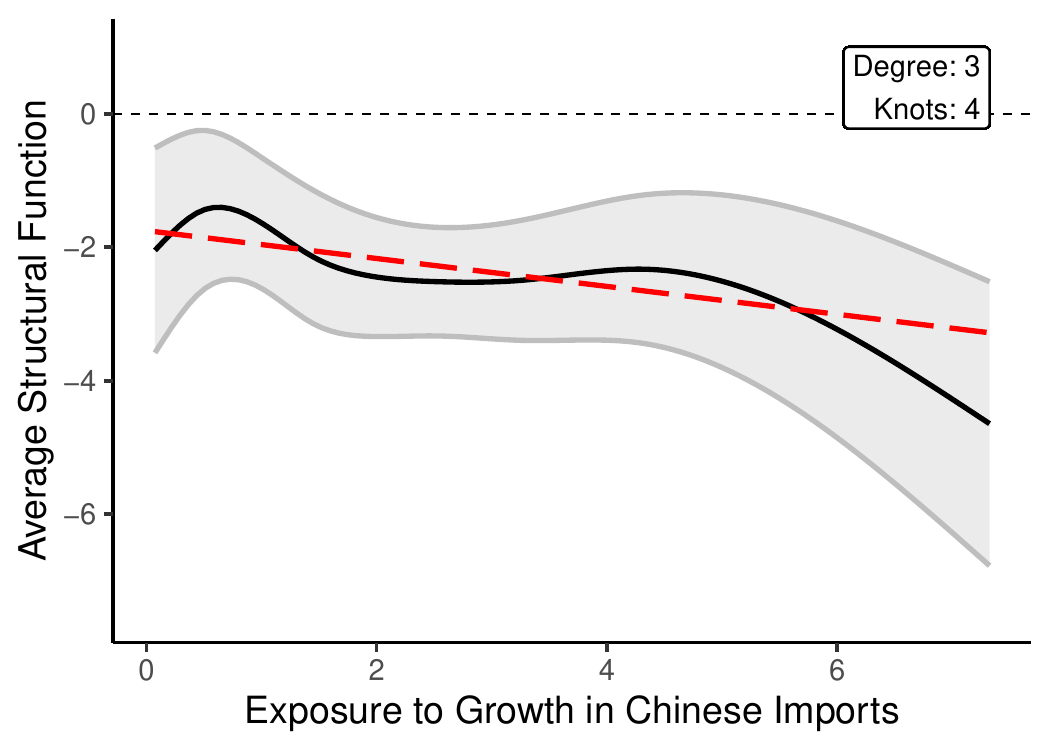}
		\caption{ASF for 2000-2007}
	\end{subfigure}
	\caption*{\textit{Note:} This figure plots the estimates for the Average Structural Function for the periods of 1990-2000 and 2000-2007. Both estimates were produced according to the specification in Section \ref{spec}, but using a degree of 3 and 4 knots in the specification of the spline. The red long-dashed lines represent the estimates from a 2SLS specification. The values displayed in the x-axes represents the 5th to the 95th quantiles of the empirical distribution of the Exposure to Growth in Chinese Imports. }
	\label{fig:asf34}
\end{figure}

\begin{figure}[!hbtp]
	\centering
	\caption{Average Structural Functions with degree of 3 and 5 knots}
	\begin{subfigure}{0.49\textwidth}
		\includegraphics[width=\textwidth]{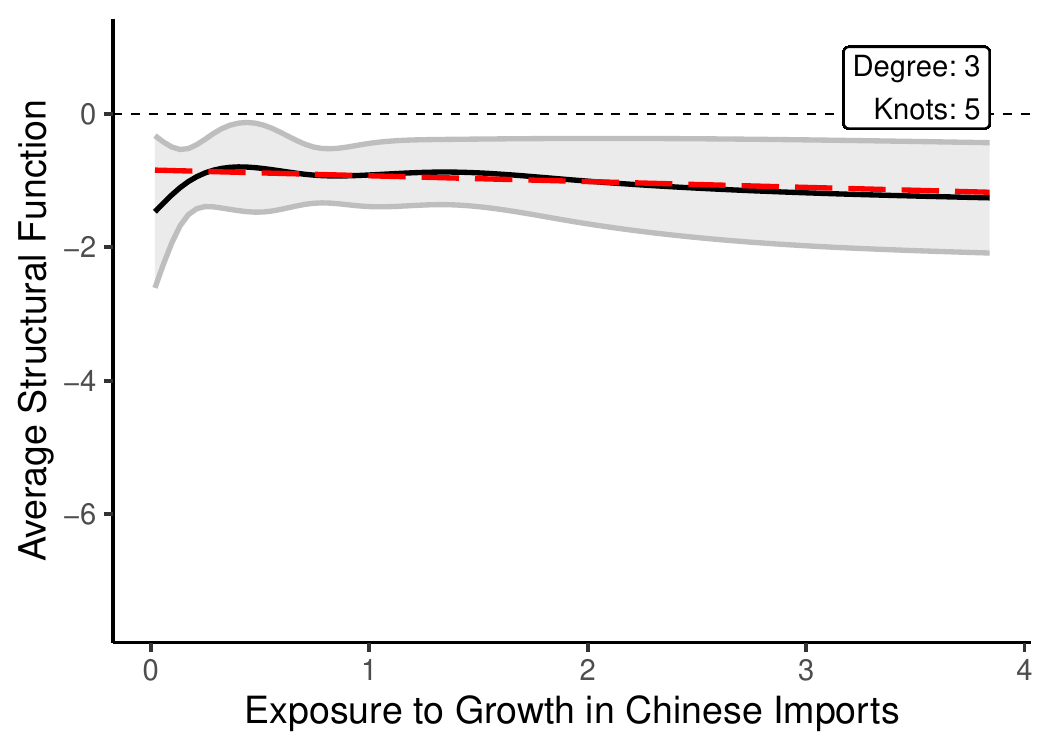}
		\caption{ASF for 1990-2000}
	\end{subfigure}
	\begin{subfigure}{0.49\textwidth}
		\includegraphics[width=\textwidth]{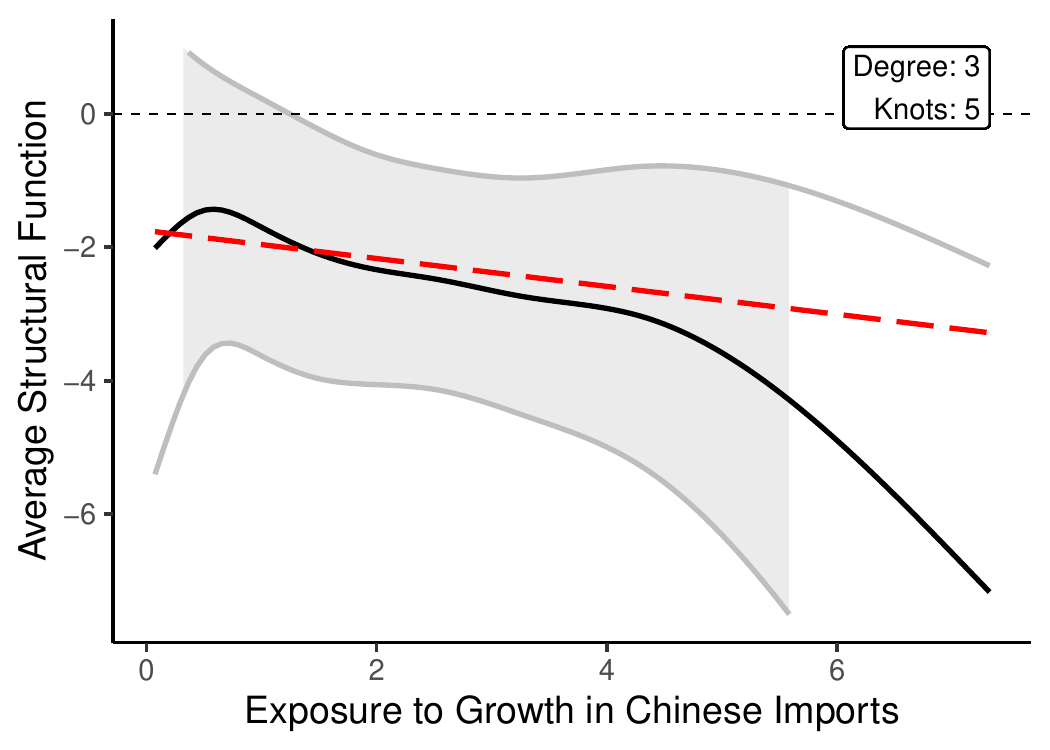}
		\caption{ASF for 2000-2007}
	\end{subfigure}
	\caption*{\textit{Note:} This figure plots the estimates for the Average Structural Function for the periods of 1990-2000 and 2000-2007. Both estimates were produced according to the specification in Section \ref{spec}, but using a degree of 3 and 5 knots in the specification of the spline. The red long-dashed lines represent the estimates from a 2SLS specification. The values displayed in the x-axes represents the 5th to the 95th quantiles of the empirical distribution of the Exposure to Growth in Chinese Imports. }
	\label{fig:asf35}
\end{figure}

In this section, we present the ASF estimates for six different specifications (Figures \ref{fig:asf23}-\ref{fig:asf35}).

\subsection{Robustness of the Policy Effect}\label{sens:pe}

\begin{figure}[!hbtp]
	\centering
	\caption{Policy Effects with degree of 2 and 3 knots}
	\begin{subfigure}{0.49\textwidth}
		\includegraphics[width=\textwidth]{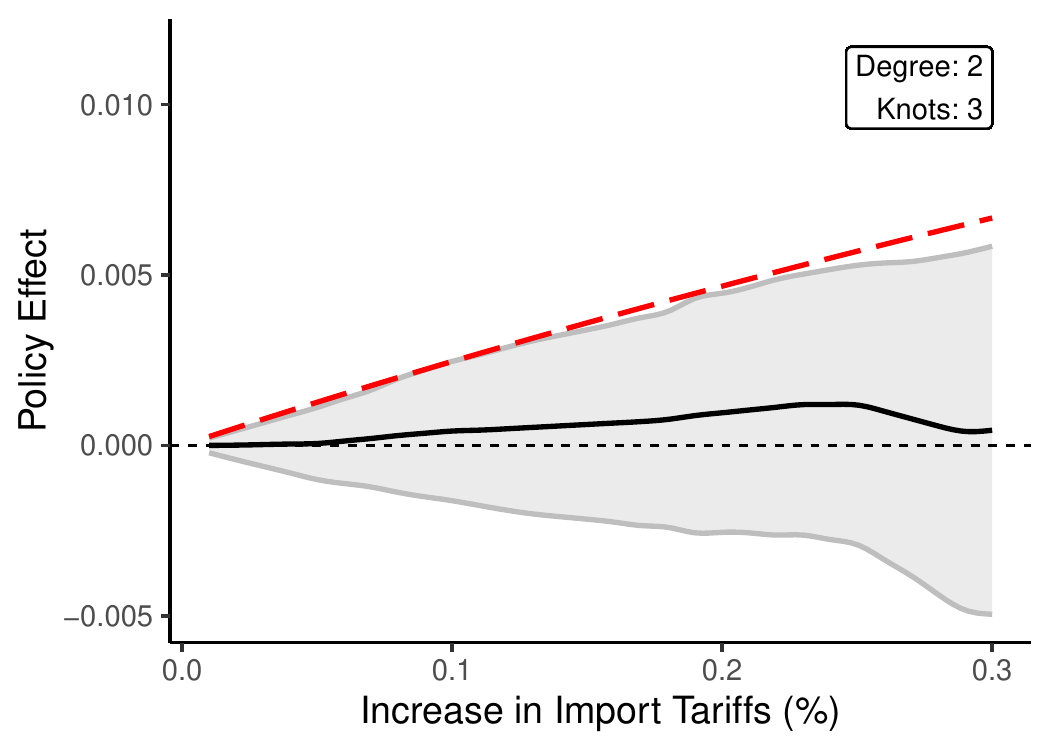}
		\caption{Policy Effect for 1990-2000}
	\end{subfigure}
	\begin{subfigure}{0.49\textwidth}
		\includegraphics[width=\textwidth]{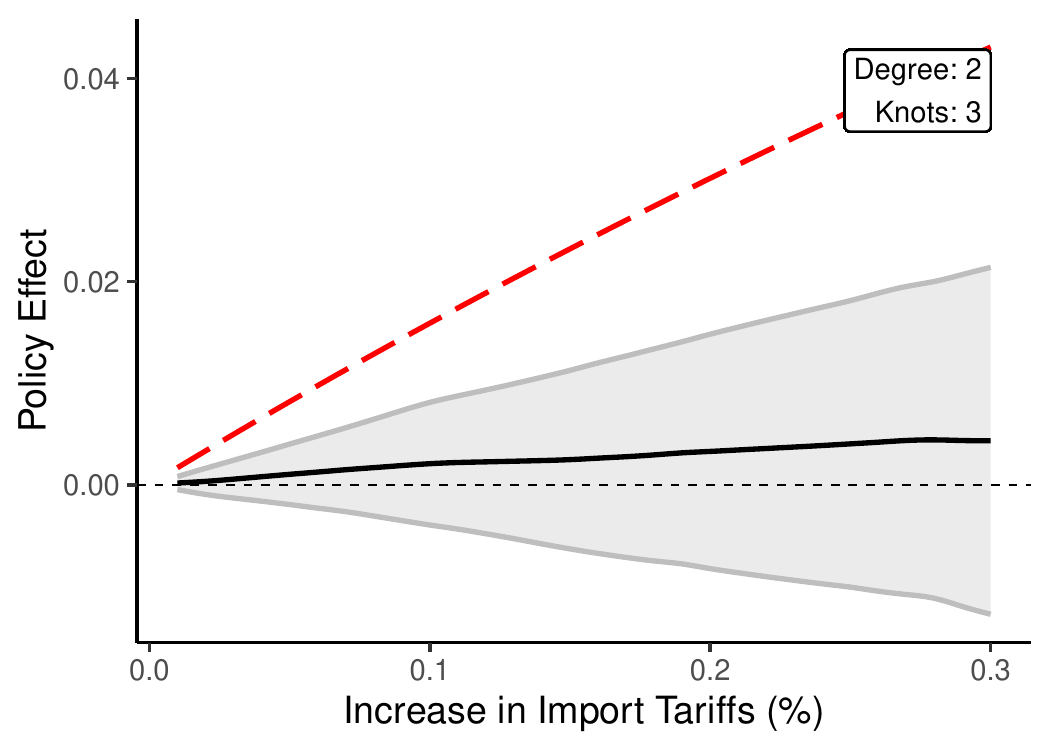}
		\caption{Policy Effect for 2000-2007}
	\end{subfigure}
	\caption*{\textit{Note:} This figure plots the estimates for the Policy Effect for the periods of 1990-2000 and 2000-2007. The red long-dashed lines represent the estimates from a 2SLS specification. Both estimates were produced according to the specification in Section \ref{spec}, but using a degree of 2 and 3 knots in the specification of the spline. The values displayed on the x-axes represent an increase in import tariff, going from a 1\% increase to a 30\% increase. }
	\label{fig:pe23}
\end{figure}

\begin{figure}[!hbtp]
	\centering
	\caption{Policy Effects with degree of 2 and 4 knots}
	\begin{subfigure}{0.49\textwidth}
		\includegraphics[width=\textwidth]{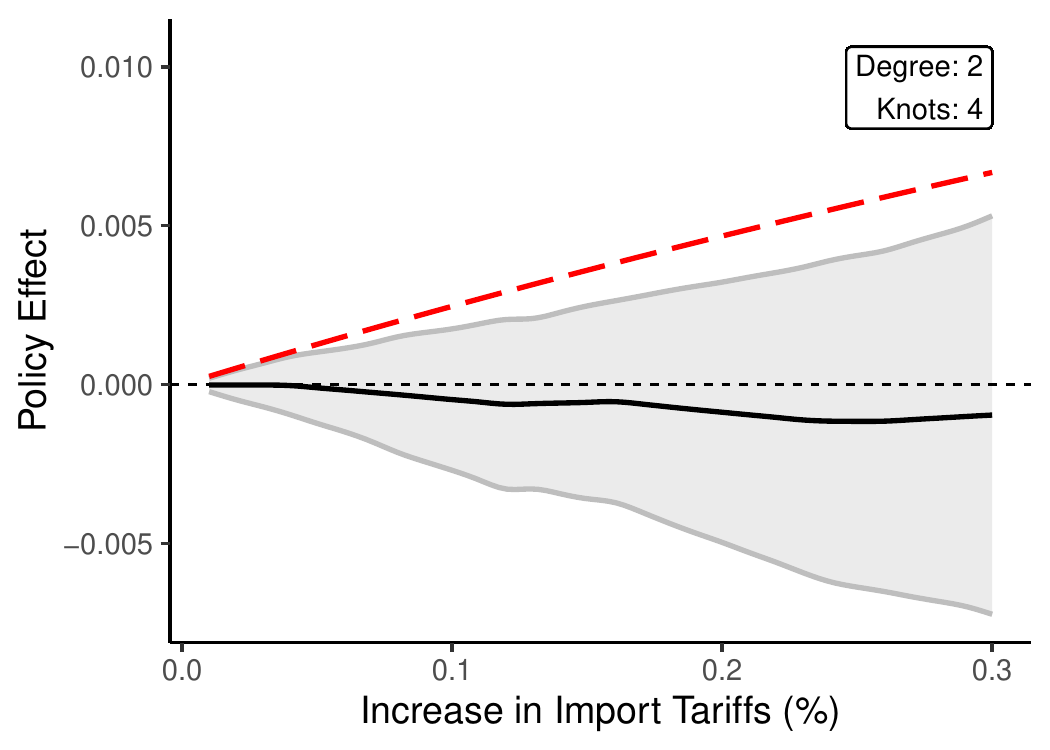}
		\caption{Policy Effect for 1990-2000}
	\end{subfigure}
	\begin{subfigure}{0.49\textwidth}
		\includegraphics[width=\textwidth]{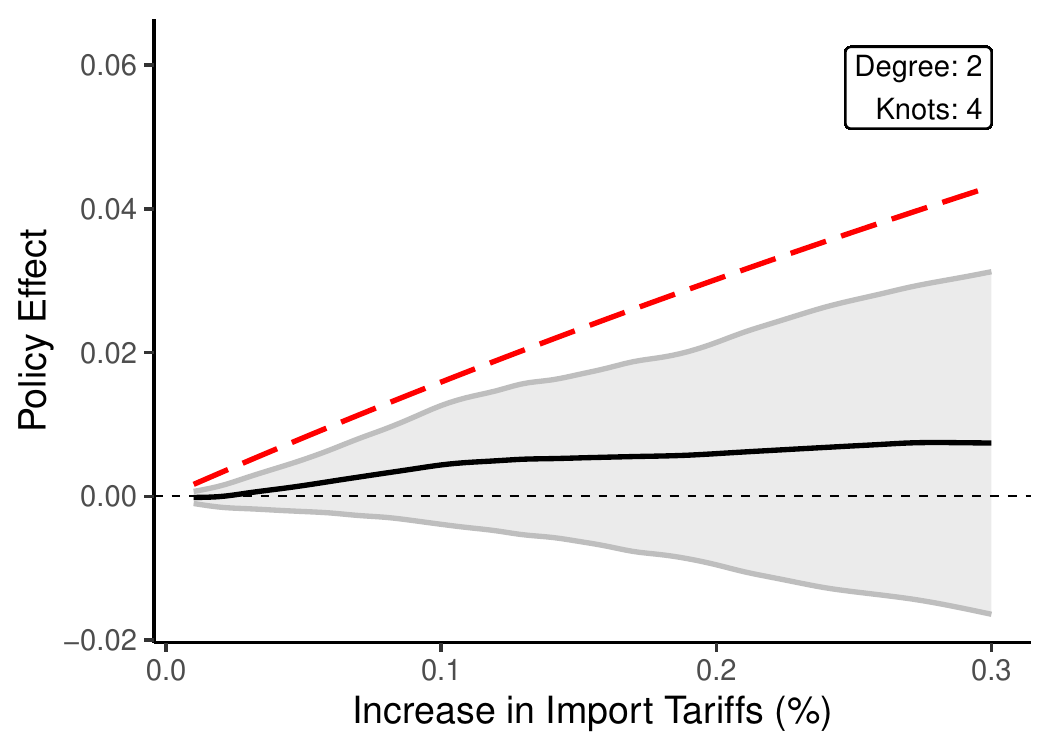}
		\caption{Policy Effect for 2000-2007}
	\end{subfigure}
	\caption*{\textit{Note:} This figure plots the estimates for the Policy Effect for the periods of 1990-2000 and 2000-2007. The red long-dashed lines represent the estimates from a 2SLS specification. Both estimates were produced according to the specification in Section \ref{spec}, but using a degree of 2 and 4 knots in the specification of the spline. The values displayed on the x-axes represent an increase in import tariff, going from a 1\% increase to a 30\% increase. }
	\label{fig:pe24}
\end{figure}

\begin{figure}[!hbtp]
	\centering
	\caption{Policy Effects with degree of 2 and 5 knots}
	\begin{subfigure}{0.49\textwidth}
		\includegraphics[width=\textwidth]{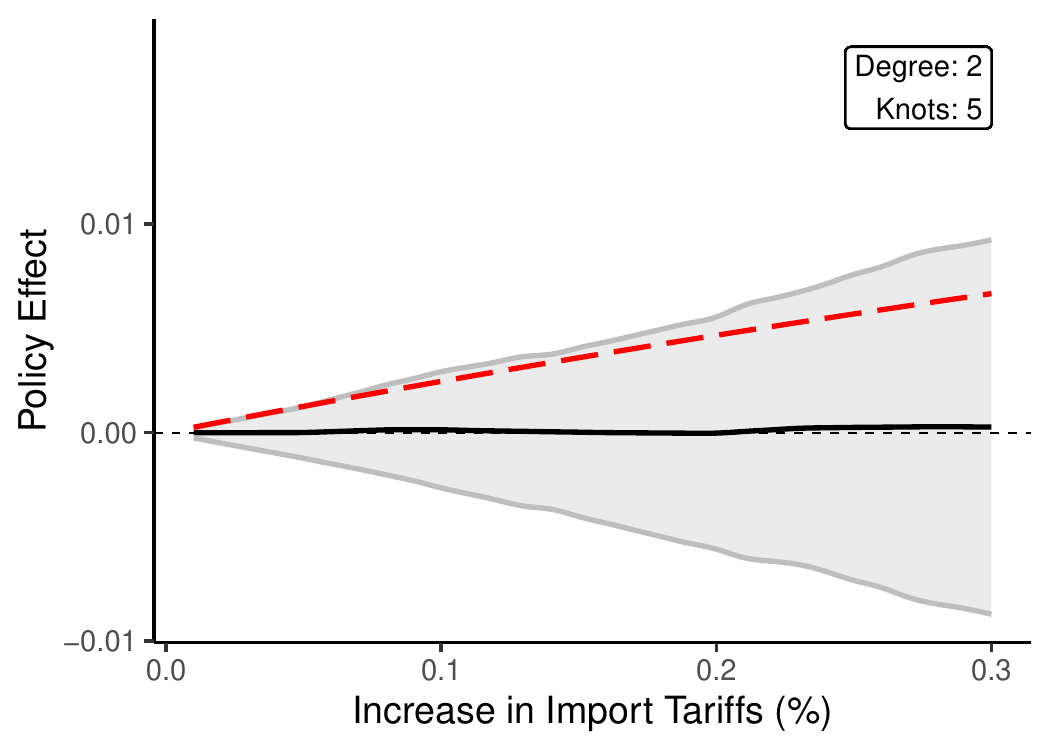}
		\caption{Policy Effect for 1990-2000}
	\end{subfigure}
	\begin{subfigure}{0.49\textwidth}
		\includegraphics[width=\textwidth]{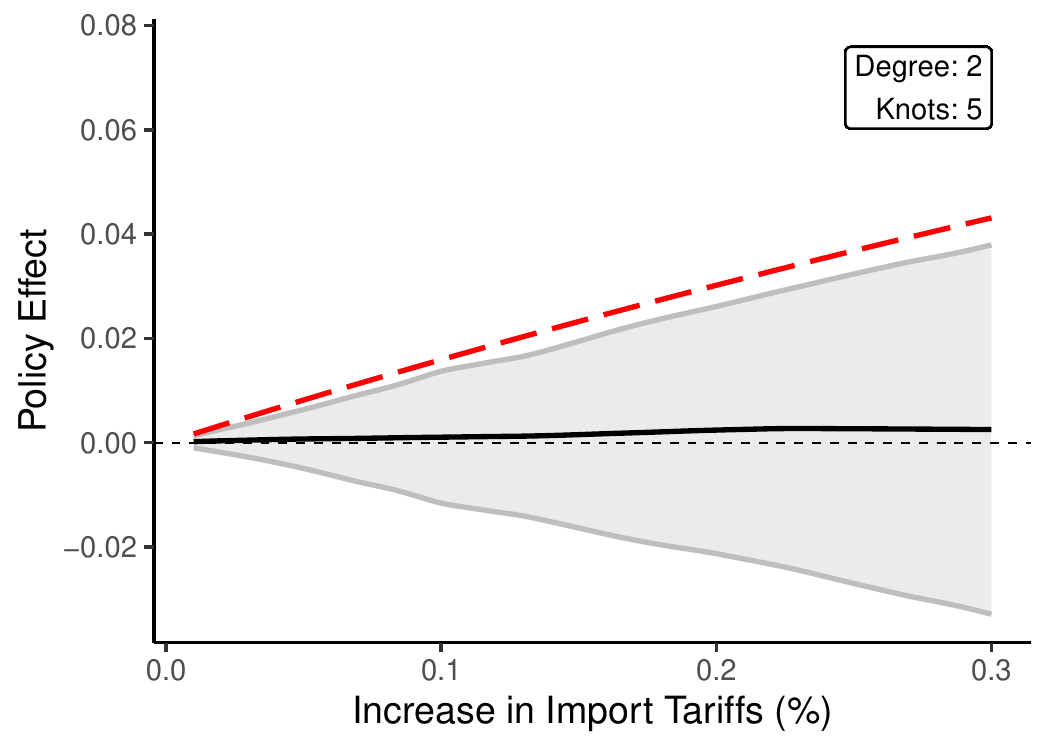}
		\caption{Policy Effect for 2000-2007}
	\end{subfigure}
	\caption*{\textit{Note:} This figure plots the estimates for the Policy Effect for the periods of 1990-2000 and 2000-2007. The red long-dashed lines represent the estimates from a 2SLS specification. Both estimates were produced according to the specification in Section \ref{spec}, but using a degree of 2 and 5 knots in the specification of the spline. The values displayed on the x-axes represent an increase in import tariff, going from a 1\% increase to a 30\% increase. }
	\label{fig:pe25}
\end{figure}

\begin{figure}[!hbtp]
	\centering
	\caption{Policy Effects with degree of 3 and 3 knots}
	\begin{subfigure}{0.49\textwidth}
		\includegraphics[width=\textwidth]{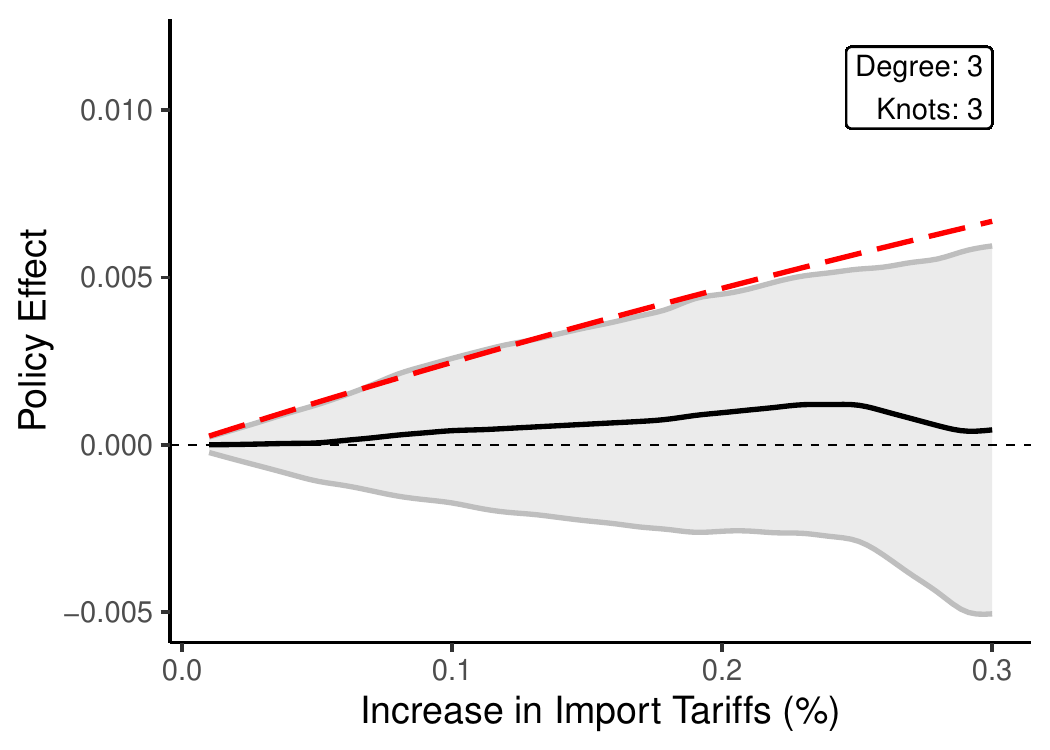}
		\caption{Policy Effect for 1990-2000}
	\end{subfigure}
	\begin{subfigure}{0.49\textwidth}
		\includegraphics[width=\textwidth]{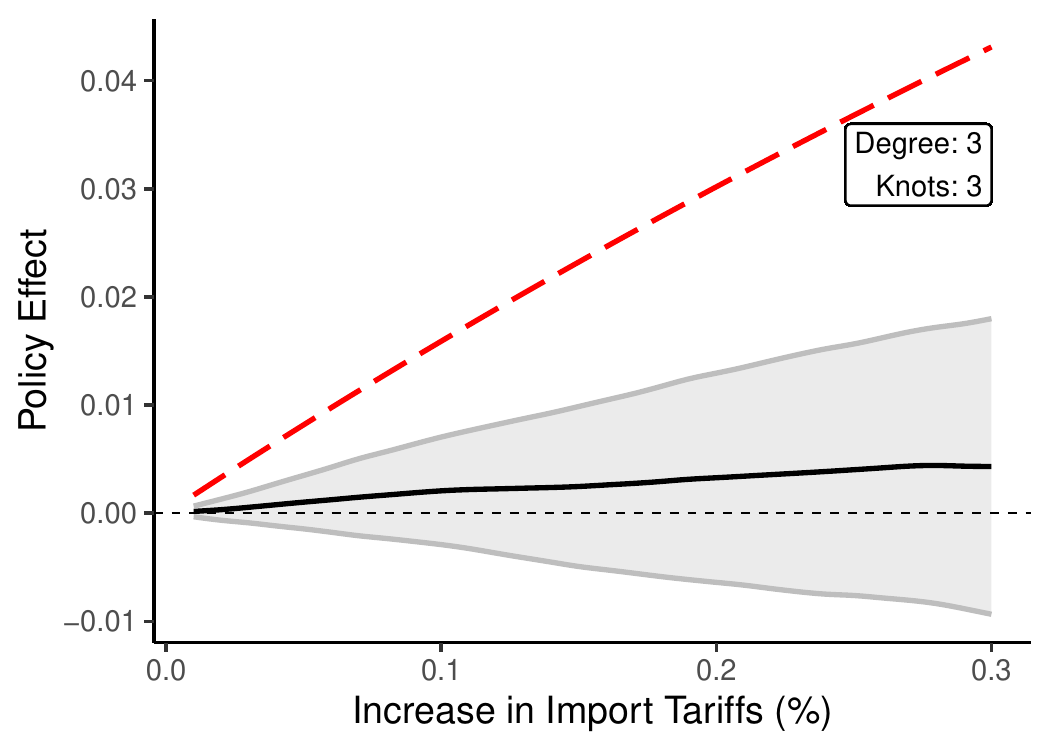}
		\caption{Policy Effect for 2000-2007}
	\end{subfigure}
	\caption*{\textit{Note:} This figure plots the estimates for the Policy Effect for the periods of 1990-2000 and 2000-2007. The red long-dashed lines represent the estimates from a 2SLS specification. Both estimates were produced according to the specification in Section \ref{spec}, but using a degree of 3 and 3 knots in the specification of the spline. The values displayed on the x-axes represent an increase in import tariff, going from a 1\% increase to a 30\% increase. }
	\label{fig:pe33}
\end{figure}

\begin{figure}[!hbtp]
	\centering
	\caption{Policy Effects with degree of 3 and 4 knots}
	\begin{subfigure}{0.49\textwidth}
		\includegraphics[width=\textwidth]{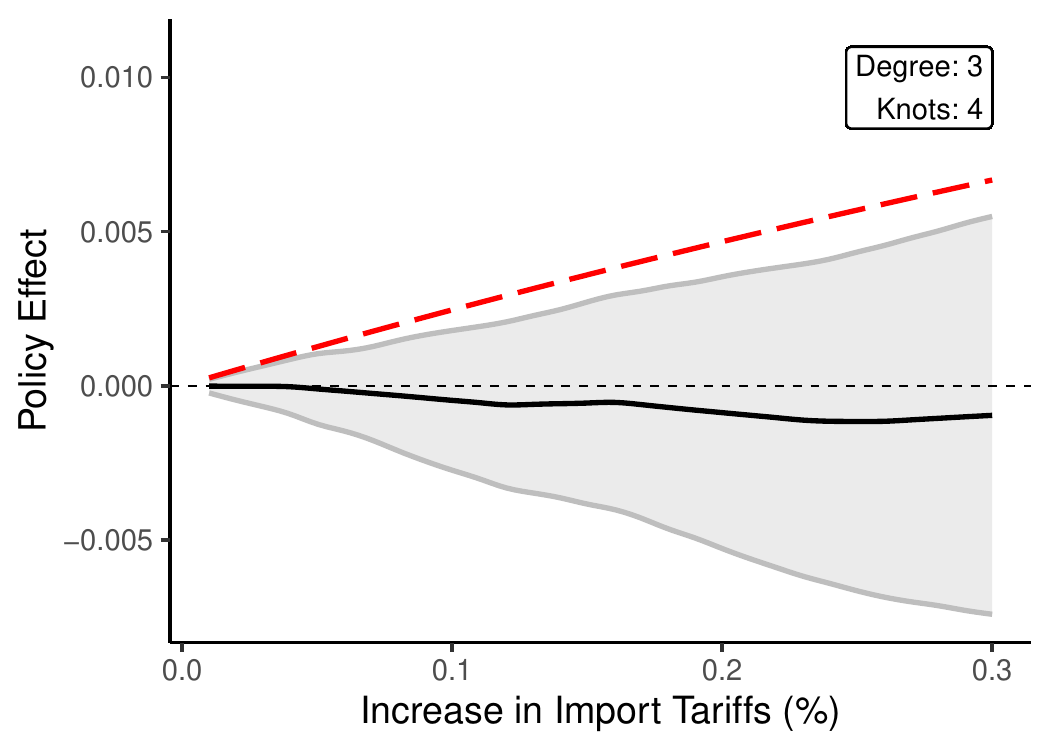}
		\caption{Policy Effect for 1990-2000}
	\end{subfigure}
	\begin{subfigure}{0.49\textwidth}
		\includegraphics[width=\textwidth]{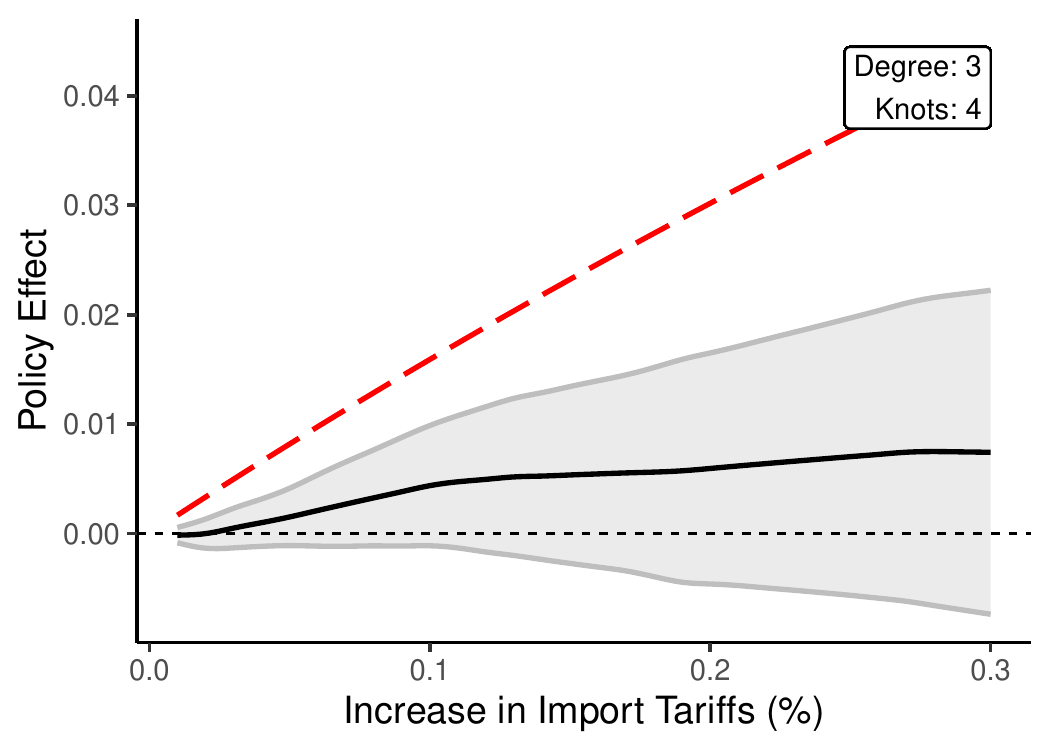}
		\caption{Policy Effect for 2000-2007}
	\end{subfigure}
	\caption*{\textit{Note:} This figure plots the estimates for the Policy Effect for the periods of 1990-2000 and 2000-2007. The red long-dashed lines represent the estimates from a 2SLS specification. Both estimates were produced according to the specification in Section \ref{spec}, but using a degree of 3 and 4 knots in the specification of the spline. The values displayed on the x-axes represent an increase in import tariff, going from a 1\% increase to a 30\% increase. }
	\label{fig:pe34}
\end{figure}

\begin{figure}[!hbtp]
	\centering
	\caption{Policy Effects with degree of 3 and 5 knots}
	\begin{subfigure}{0.49\textwidth}
		\includegraphics[width=\textwidth]{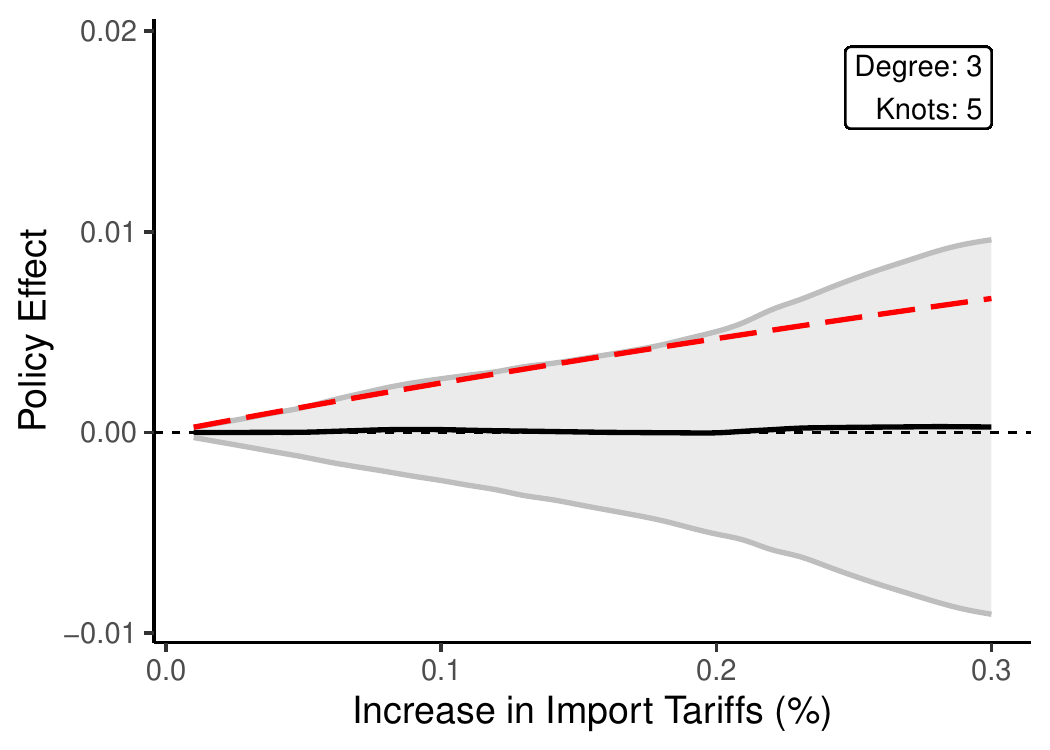}
		\caption{Policy Effect for 1990-2000}
	\end{subfigure}
	\begin{subfigure}{0.49\textwidth}
		\includegraphics[width=\textwidth]{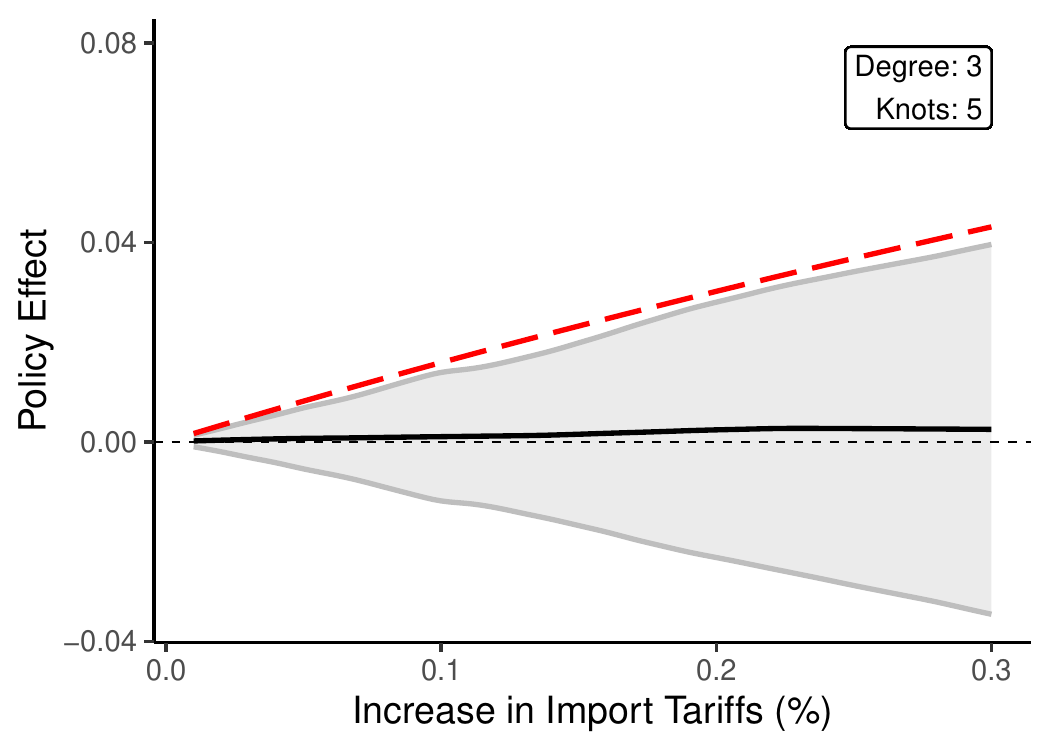}
		\caption{Policy Effect for 2000-2007}
	\end{subfigure}
	\caption*{\textit{Note:} This figure plots the estimates for the Policy Effect for the periods of 1990-2000 and 2000-2007. The red long-dashed lines represent the estimates from a 2SLS specification. Both estimates were produced according to the specification in Section \ref{spec}, but using a degree of 3 and 5 knots in the specification of the spline. The values displayed on the x-axes represent an increase in import tariff, going from a 1\% increase to a 30\% increase. }
	\label{fig:pe35}
\end{figure}

In this section, we present the Policy Effects estimates for six different specifications (Figures \ref{fig:pe23}-\ref{fig:pe35}).

\end{document}